\documentclass[12pt]{article}
\usepackage{graphicx}
\usepackage{epsf}
\usepackage{amsmath}
\usepackage{here}
\usepackage{setspace}
\usepackage{natbib}
\usepackage{xcolor}
\usepackage{varioref}
 \usepackage{wrapfig}
 \usepackage{threeparttable}
 \usepackage{dcolumn}
  \newcolumntype{d}{D{.}{.}{-1}}
 \usepackage{nomencl}
  \makeglossary
  \usepackage{subfigure}
 \usepackage{subfigmat}
 \usepackage{fancyvrb}
  \fvset{fontsize=\footnotesize,xleftmargin=2em}
 \usepackage{lettrine}

\usepackage{caption}
\usepackage{epstopdf}
\usepackage{lineno}

\begin{document}
\title{ Inward Swirling Flamelet Model}
 \author{William A. Sirignano \\
  {\normalsize\itshape  Department of Mechanical and Aerospace Engineering }\\
   {\normalsize\itshape  University of California, Irvine, CA 92697}  \\
 }

\newcommand{\hs}{\mbox{\hspace{0.10in}}}
\newcommand{\hsm}{\mbox{\hspace{-0.06in}}}
\newcommand{\be}{\begin{equation}}
\newcommand{\ee}{\end{equation}}
\newcommand{\bea}{\begin{eqnarray}}
\newcommand{\eea}{\end{eqnarray}}

\maketitle

\begin{abstract}
A new rotational flamelet model with inward swirling flow through a stretched vortex tube is developed for sub-grid modeling to be coupled with the resolved flow for turbulent combustion. The model has critical new features compared to existing models. (i) Non-premixed flames, premixed flames, or multi-branched flame structures are determined rather than prescribed. (ii) The effects of  vorticity  and the related centifugal acceleration are determined. (iii) The strain rates and vorticity applied at the sub-grid level can be directly determined from the resolved-scale strain rates and vorticity without a contrived progress variable. (iv) The flamelet model is three-dimensional. (v) The effect of variable density is addressed. (vi) The inward swirl is created by vorticity combined with  two compressive normal strain components; this feature distinguishes the model from counterflow flamelet models.  Solutions to the multicomponent Navier-Stokes equations governing the flamelet model are obtained. By coordinate transformation, a similar solution is found for the model, through a system of ordinary differential equations. Vorticity  creates a centrifugal force on the sub-grid counterflow that modifies the molecular transport rates, burning rates, and flammability limits.  Sample computations of the inward swirling rotational flamelet model without coupling to the resolved flow are presented  to demonstrate the importance of the new features. Premixed, nonpremixed, and multi-branched flame structures are examined. Parameter surveys are made with rate of normal strain, vorticity, Damk\"{o}hler number, and Prandtl number.  The centrifugal effect has interesting consequences when combined with the variable-density field. Flow direction can reverse; burning rates can be modified; flammability limits can be extended.

\end{abstract}



\linespread{1.0}

\section{Introduction}

Combustion in high mass-flux chambers is the practical and major method for energy conversion for mechanical power and heating. Inherently, the high mass-flow rate leads to turbulent flow. Thereby, many length and time scales appear in the physics making serious challenges for both computational and experimental analyses. For computations where the smallest scales typically cannot be resolved, the method of large-eddy simulations (LES) is employed wherein the smaller scales are filtered via integration over a window size commensurate with the computational mesh size that allows affordable computations. Consequently, the essential, rate-controlling, physical and chemical processes that occur on shorter scales than the filter size must be modelled.  Those sub-grid models must be properly coupled to the resolved LES flow field.

Current flamelet models that are used for LES or Reynolds-averaged Navier-Stokes (RANS) methods have some advantages. Typically, the flamelet equations are a  system of ordinary differential equations (ODEs) that can be solved offline with solutions available in tabular form or through neural networks (NN). The flamelet models can handle multi-species, multi-step oxidation kinetics without requiring small time steps during the solution of the resolved-scale fluid dynamics. Thus, for several reasons, savings of computational resources can be huge compared to direct numerical simulation. We aim here to retain these very attractive features while removing some less desirable features. Already, some progress has been made in extending the fundamental flamelet theory beyond its long-term limitation of a single-flame structure, two-dimensional (or axisymmetric) configuration, and use of the uniform-density assumption. However, those advances still must be applied to LES or RANS. In addition, the flamelet theory must be advanced to consider shear strain and vorticity at the small scale of the flamelet; these are the vital forgotten physics in current flamelet modelling. Furthermore, the strain rates in the flamelet model are far from properly connected to the strain rates at the resolved scale. Attempts at corrections of some of these weaknesses are made here.

The goals in this paper are to improve the flamelet model by including  several important physical effects that are commonly neglected in present models and to identify other issues, related to the coupling between the sub-grid-scale physics and the resolved-scale (or time-averaged) physics, that require further study.

\subsection{Existing Flamelet Theory}

 The laminar mixing and combustion that commonly occur within the smallest turbulent eddies is important in determining the performance for many power and propulsion applications. These laminar flamelet sub-domains experience significant strain of all types, shear, tensile, and compressive. Some important works exist here but typically for either counterflows with only normal strain or simple vortex structures in planar or axisymmetric geometry and often with a constant-density approximation. See \cite{Linan}, \cite{Williams1975}, \cite{Marble}, \cite{Karagozian}), \cite{Cetegen1}, \cite{Cetegen2}, \cite{Peters}, and \cite{Pierce}.  An interesting review of the early flamelet theory is given by \cite{Williams2000}.  \cite{Williams1975} first established the concept of laminar flamelets in the turbulent diffusion flame structure.    Flamelet studies have focused on either premixed or nonpremixed flames; a unifying approach to premixed, nonpremixed, and multi-branched flames has not been developed until the recent counterflow-based rotational flamelet study by \cite{Sirignano2022}.       Here, we attempt a unification for the vortex-tube-based flamelet.

 Most flamelet studies have not directly considered vorticity interaction with the flamelet. See,  for example, \cite{Linan, Peters, Williams2000, Pierce}. \cite{Williams1975} first recognized the advantage of separating rotation (due to vorticity) and stretching by transformation to a rotating, non-Newtonian reference frame. However,  the momentum consequences in the new reference frame were not examined.  Some other works that have examined vortex-flame interaction have not separated the effects of stretching and rotation. See \cite{Marble, Cetegen1, Cetegen2, Meneveau-Poinsot}.
\cite{Karagozian} examined a three-dimensional flow with radial inward velocity, axial jetting, and a vortex centered on the axis. The flame sheet wrapped around the axis due to the vorticity; an incompressible-flow  velocity field was used with an ad hoc adjustment for the variable density effect.

 The two-dimensional planar or axisymmetric counterflow configuration is generally a foundation for a flamelet model. Local conversion to a coordinate system based on the principal strain-rate  directions can provide the counterflow configuration in a general flow.  Furthermore, the quasi-steady counterflow can be analyzed by ordinary differential equations because the dependence on the transverse coordinate is either constant or linear, depending on the variable.   \cite{Pierce} modified the nonpremixed-flamelet counterflow configuration by fixing domain size and forcing flux to zero at the boundaries. Flamelet theory as a closure model for turbulent combustion is typically based on the tracking of two variables: a normalized conserved scalar and the strain rate; the latter is generally given indirectly through a progress variable. Mixture fraction is traditionally used for the conserved scalar.
 The flamelet model for LES developed by \cite{Pierce} was a substantial advancement through the introduction of the flamelet progress variable (FPV). Their approach has also been used by \cite{Ihme2009}, \cite{Tuan1, Tuan2, Tuan3}, and others.       \cite{Ihme2009, Shadram2021a, Shadram2021b} introduced the use of neural networks in place of the look-up table.          \cite{Mueller2020} presented the flamelet model in a somewhat different mathematical framework but without the addition of new physical description.

There are concerns about incompleteness and contradiction in the above models.  (i) The models are  designed  specifically for non-premixed flames or premixed flames. The flame structure should be determined rather than prescribed. Multi-branched flames should be allowed. (ii) The effects of vorticity  are commonly neglected with a very few exceptions identified above.  Yet, the models are applied to turbulent flows where the strain rates and vorticity magnitudes are known to be larger at the small scales than at the large scales. (iii) The above flamelet models are two-dimensional or axisymmetric although key three-dimensional behavior  can be shown to exist. (iv) The effect of variable density is not thoroughly addressed. (v) Clear connections are not given between the strain rates and vorticity at the flamelet level and those variables at the resolved scale of the combustor.


\subsection{Stretched Vortex with Inward Swirl}

In one part of their paper, \cite{Karagozian} treated a flame within a stretched vortex tube with an inward swirling flow. Their incompressible velocity field was defined by \cite{Burgers1948} and \cite{Rott} and is commonly known as Burgers vortex. In particular,
with $u_r, u_z, $ and $u_{\theta}$ as the velocity components in cylindrical coordinates and parameter $a$ and kinematic viscosity $\nu$ taken as constants, we have
\begin{eqnarray}
u_r = ar   \;\;\;  ;  \;\;\;  u_z = 2az  \nonumber  \\
u_{\theta} =\frac{\Gamma }{4 \pi r}\big[ 1 - exp\big(-\frac{ar^2}{2\nu}\big)\big]
\end{eqnarray}
where $\Gamma$ is the circulation taken through the far field surrounding the vortex tube. Note that $u_{\theta} \rightarrow \Gamma/(2 \pi r) $ as $r \rightarrow  \infty$, yielding potential flow for the far field.   This description gives an exact steady-state solution to the incompressible Navier-stokes equations. Although the tube is being stretched in the $z$ direction, diffusion of momentum and vorticity in the $r$ direction allows a balance with radial advection that results in a steady solution.

\cite{Karagozian} made an ad hoc adjustment to correct for expansion by variable density; however, by not accounting for spatial variation of density, the effect of centrifugal acceleration was not considered. They also focused on diffusion flames. Here, a stretched vortex will be considered but with full account of variable density and allowance for a premixed flame, multibranched flame, or diffusion flame as determined by the boundary conditions. We will not use the incompressible Burgers vortex velocity field; however, it does provide useful guidance. In particular, note that for small $r$ values, the Burgers vortex gives wheel motion for the fluid, i.e, $u_{\theta} \approx (\Gamma a r)/(4 \pi \nu)$.

\subsection{Relative Orientations of Principal Strain Axes, Vorticity, and Scalar Gradients}\label{orientation}

Both normal strain rate and shear strain rate are imposed on the flamelet and are important.   Shear strain can, in general, be decomposed into a normal strain and a rotation (whose rate is half of the vorticity magnitude).   The magnitudes of strain rate and vorticity increase as the eddy size  decreases in the turbulence energy cascade.  The  strain and rotation become especially important on the smallest scales of turbulence where mixing and chemical reaction occur.   The smallest (i.e., Kolmogorov) scale  size is determined by the dissipation rate of turbulence kinetic energy. The final molecular mixing and chemical reaction  occur on this smaller scale, where  there will be an axis (or direction) of principal compressive normal strain and an orthogonal axis for principal tensile strain, the third orthogonal axis could be either tensile or compressive.  These axes  rotate due to  vorticity. Thereby, the direction of the scalar gradient rotates .  A useful flamelet model must have a statistically accurate representation of the relative orientations on this smallest scale of the vorticity vector, scalar gradients, and the directions of the three principal axes for strain rate.  Several studies exist that are helpful in understanding this important alignment issue.

Generally (and always for incompressible flow), one principal strain rate $\gamma$ locally will be compressive (corresponding to inflow in a counterflow configuration), another principal strain rate $\alpha$ will be tensile (also named extensional and corresponding to outflow), and the third can be either extensional or compressive and will have an intermediate  strain rate $\beta$ of lower magnitude than the other like strain rate. Specifically, $\alpha > \beta > \gamma , \;\alpha > 0 , \; \gamma <0,$ and, for incompressible flow, $\alpha + \beta + \gamma = 0$.  If the intermediate strain rate $\beta < 0$, there is inflow from two directions with outflow in one direction; a contracting jet flow occurs locally. Conversely, with $\beta > 0$, there is outflow in two directions and inflow in one direction; a counterflow  or, in other words, the head-on collision of two opposed jets occurs.

Several interesting findings result from direct numerical simulations (DNS) for incompressible flows. Both \cite{Ashurstetal} and \cite{Nomura1992} compared a case of homogeneous sheared turbulence with a case of isotropic turbulence. They report that the vorticity alignment with the intermediate strain direction is most probable in both cases but especially in the case with shear. Furthermore, the intermediate strain rate is most likely to be extensive (positive) implying a counterflow configuration.

\cite{Nomura1993} studied reacting flow and show that in regions of exothermic reaction and variable density, alignment of the vorticity with the most tensile strain direction can occur. Still though as the strain rates increase, the intermediate direction becomes more favored for alignment with vorticity; that direction is also preferred in regions where mixing occurs without substantial divergence of the velocity due to chemical  reaction.

A material interface most probably aligns to be normal to the direction of the compressive normal strain. That is, the scalar gradient and the direction of compressive strain are aligned. See \cite{Ashurstetal, Nomura1992, Nomura1993} and \cite{Boratav1996, Boratav1998}. Authors agree that the most common intermittent vortex structures in regions of high strain rate are sheets or ribbons rather than tubes.  Nevertheless, vortex tubes can exist in a combustor and can be relevant.

Based on those understandings concerning vector orientations, \cite{Sirignano2020}, extended flamelet theory in a second significant aspect beyond the inclusion of both premixed and non-premixed flame structures; namely, a model was created of a three-dimensional field with both shear and normal strains.  The three-dimensional problem is reduced to a two-dimensional form and then, for the counterflow or mixing-layer flow, to a one-dimensional similar form.  The system of ordinary differential equations (ODEs) is presented for the thermo-chemical variables and the velocity components. Conserved scalars are determined and can become the independent variable if they behave in a monotonic fashion.   These new findings are very helpful in improving the foundations for flamelet theory and its use in sub-grid modeling for turbulent combustion.

Based on the observations of the needed improvements, \cite{Sirignano2022}  has developed a rotational flamelet model based on a counterflow with rotation. The model  (i) determines rather than prescribes the existence of non-premixed flames, premixed flames, or multi-branched flame structures; (ii) determines directly the the effect of shear strain and vorticity on the  flames; (iii) applies directly the resolved-scale strain rates and vorticity to the sub-grid level without the use of a contrived progress variable; (iv) employs a three-dimensional flamelet model; (v) considers the effect of variable density.  The analysis uses one-step kinetics to avoid complications in this initial study; however, a clear template will exist for the employment of multi-step kinetics. The goal with the new inward-swirl flamelet   model presented here is to extend the rotational flamelet concept to a vortex-tube configuration with inward spiralling flow rather than the traditional counterflow. Elements of the counterflow character will remain because fluid of differing compositions will be strained to move towards each other enhancing transport and reaction.

  Section \ref{flamelet} presents the analysis supporting  a new sub-grid flame model that better addresses effects of rotation, variable density,  three-dimensional character, and multibranched flame structure for the stretched vortex tube Computational results are discussed in Section \ref{flamelet}. Results and the related discussion are presented in Section \ref{results}.   Concluding comments are made in Section \ref{conclusions}.

\section{Sub-grid Flamelet Analysis}\label{flamelet}

The problem is stated here in a quasi-steady, three-dimensional form where variable density is allowed. These assumed orientations are consistent with the statistical findings of \cite{Nomura1993}. The direction of major compressive principal strain aligns with the scalar gradient and is orthogonal to the vorticity vector direction;  an extensional principal strain direction is aligned with the vorticity. The stretched vortex character is created by imposing compressive normal strain in both coordinate direction that are orthogonal to the vorticity vector. Thus, we have a stretched vortex tube that qualitatively relates to the incompressible-flow  configurations of \cite{Burgers1948}, \cite{Rott}, and \cite{Karagozian}.

\subsection{Coordinate Transformation} \label{transformation}

\begin{figure}
  \centering
  \includegraphics[height = 7.0cm, width=0.8\linewidth]{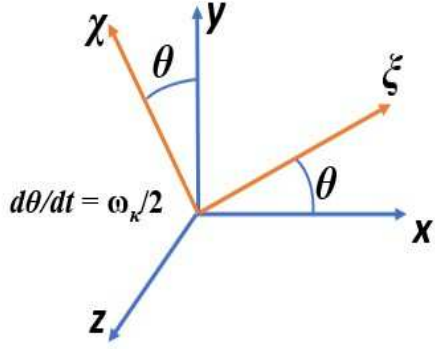}
  \caption{Transformation to $\xi, \chi, z'$ rotating coordinate system from $x', y', z'$ Newtonian system. $\theta$ increases in the counterclockwise direction.}
  \label{Coordinate}
\end{figure}

 In Figure \ref{Coordinate}, the Newtonian frame is transformed to a rotating, non-Newtonian frame  where the curl of the velocity is zero.   The vorticity aligns with the $z'$ direction. $\omega_{\kappa}$ is the vorticity magnitude on this sub-grid (Kolmogorov) scale. $x, y, z$ are transformed to $\xi, \chi, z$ wherein the material rotation is removed from the $\xi, \chi$ plane by having it rotate at angular velocity $d\theta/dt = \omega_{\kappa}/2$ relative to $x, y$. Here, $\theta$ is the angle between the $x$ and $\xi$ axes and simultaneously the angle between the $y$ and $\chi$ axes. The sub-grid domain is sufficiently small to consider a uniform value of $\omega$ across it, consistent with the truncated Taylor series expansion used elsewhere in the flamelet analysis.
 The scalar gradients align with the major principal axis for compressive strain.  In many of our calculations, the two compressive normal strains will have equal magnitude; so, the choice of the normal strain direction which aligns with the scalar gradient is arbitrary. The scalar gradient is always aligned with the $\chi$ direction in the analysis here.

\;
 Thereby,
\begin{eqnarray}
\xi &=& x\; cos \theta + y sin\theta \;\; ; \;\;
\chi = y \;cos \theta - x \;sin\theta \nonumber \\
\frac{\partial \xi}{\partial x} &=&  cos \theta \;\; ; \;\; \frac{\partial \xi}{\partial y}= sin\theta \;\; ; \;\;
\frac{\partial \chi}{\partial x} = - sin \theta \;\; ; \;\; \frac{\partial \chi}{\partial y} = cos \theta \nonumber \\
u_{\xi} &=& u \;cos \theta + v \; sin \theta  + \chi\frac{\omega_{\kappa}}{2}   \;\; ; \;\;
u_{\chi} = v\; cos \theta -u \;sin \theta  - \xi\frac{\omega_{\kappa}}{2}  \nonumber \\
\frac{\partial u}{\partial x} &=& \frac{\partial u}{\partial \xi}cos \theta
- \frac{\partial u}{\partial \chi}sin \theta  \;\; ; \;\;
\frac{\partial u}{\partial y} = \frac{\partial u}{\partial \xi}sin \theta
+ \frac{\partial u}{\partial \chi}cos \theta   \nonumber \\
\frac{\partial v}{\partial x} &=& \frac{\partial v}{\partial \xi}cos \theta
- \frac{\partial v}{\partial \chi}sin \theta  \;\; ; \;\;
\frac{\partial v}{\partial y} = \frac{\partial v}{\partial \xi}sin \theta
+ \frac{\partial v}{\partial \chi}cos \theta
\end{eqnarray}
Since
\begin{eqnarray}
\frac{\partial v}{\partial x}  - \frac{\partial u}{\partial y} = \omega_{\kappa}
\end{eqnarray}
it follows that
\begin{eqnarray}
\frac{\partial u_{\chi}}{\partial \xi}  - \frac{\partial u_{\xi}}{\partial \chi} = 0
\end{eqnarray}
Thus, the rotating frame of reference does not have vorticity appearing explicitly.  However,  the frame is not Newtonian and a reversed (centrifugal) force is imposed. The expansions due to energy release  produce new vorticity but it will integrate to zero globally; the flow will be  antisymmetric and have zero circulation in the new reference frame.

In the new reference frame, the normal rates of strain, imposed in the far field, in the $\xi, \chi, $ and $z$ directions are $S_1, -(S_1 + S_2), $ and $S_2$, respectively. \cite{Sirignano2022}, for the rotational flamelet with counterflow, considered both $S_1$ and $S_2$ to be positive. Here, with the inward swirl flamelet, $S_1 < 0, \; S_2 >0, $ and $S_1 +S_2>0$. In the next sub-section, these strain rates will be non-dimensionalized.

\subsection{Governing Equations} \label{equations}

Quasi-steady behavior is considered. The governing equations for steady 3D flow in the non-Newtonian frame can be written with $u_i = u_{\xi}, u_{\chi}, w \; ; \; x_i = \xi, \chi, z$. The centrifugal acceleration $a_i = \xi \omega_{\kappa}^2/4 , \; \chi \omega_{\kappa}^2 /4, \; 0$.  The quantities $p, \rho, h, h_m, Y_m, \dot{\omega},  \mu, \lambda, D,$ and $ c_p  $ are pressure, density, specific enthalpy, heat of formation of species $m$, mass fraction of species $m$, chemical reaction rate of species $m$, dynamic viscosity, thermal conductivity, mass diffusivity, and specific heat, respectively. Furthermore, the Newtonian viscous stress tensor with the Stokes hypothesis is considered. The boundary-layer approximation is not needed and the Navier-Stokes equations for a multicomponent field will be solved. The  system is described as
\begin{eqnarray}
\frac{\partial (\rho u_j)}{\partial x_j} =0
\label{cont}
\end{eqnarray}
\begin{eqnarray}
 \rho u_j\frac{\partial u_i}{\partial x_j} +\frac{\partial p}{\partial x_i} = \frac{\partial}{\partial x_j}\Big(\mu \Big[ \frac{\partial u_i}{\partial x_j} + \frac{\partial u_j}{\partial x_i}
			-\frac{2}{3}\delta_{ij} \frac{\partial u_k}{\partial x_k}\Big]\Big) + \rho a_i
			\label{momentum}
		\end{eqnarray}	
		\begin{eqnarray}
			\rho u_j\frac{\partial h}{\partial x_j}
			=\frac{\partial}{\partial x_j} \Big( \frac{\lambda}{c_p} \frac{\partial h}{\partial x_j}  \Big)
			+ \frac{\partial}{\partial x_j} \Big( \rho D (1 - Le)  \Sigma^N_{m=1}h_m \frac{\partial Y_m}{\partial x_j}  \Big)
\nonumber \\
			-\rho \Sigma^N_{m=1}h_{f,m} \dot{\omega}_m	
			\label{energy}
		\end{eqnarray}
		\begin{eqnarray}
			\rho u_j\frac{\partial Y_m}{\partial x_j}  = \frac{\partial }{\partial x_j}\Big(\rho D \frac{\partial Y_m}{\partial x_j}\Big) + \rho \dot{\omega}_m   \;\;;\;\; m=1, 2, ...., N
			\label{species}
		\end{eqnarray}	
The viscous dissipation, the energy source term $\rho u_j a_j = \rho (\omega_{\kappa}/2)^2(\xi u_{\xi} + \chi u_{\chi})$ in the new reference frame, and other terms of the  order of the kinetic energy per mass have been neglected.

Here, we define the non-dimensional  Prandtl, Schmidt, and Lewis numbers:
$ Pr \equiv c_p \mu / \lambda$ ; $Sc \equiv \mu / (\rho D)$  ; and $ Le \equiv Sc/ Pr$.  These numbers will be assumed to be constants. Furthermore, $Pr = Sc$ (i.e., $Le =1$).

		The non-dimensional forms of the above equations remain identical to the above forms if we choose certain reference values for normalization. In the remainder of this article, the non-dimensional forms of the above equations will be considered. The superscript $^*$ is used here to designate a dimensional property. The variables $u_i^*, t^*, x_i^*, \rho^*, h^*, p^*,$ and $\dot{\omega}_m^* ,$ and properties $\mu^*, \lambda^*/ c_p^*,$ and $ D^*$ are normalized respectively by
		$ [(S_1^* +S_2^*) \mu_{\infty}^*/ \rho_{\infty}^*]^{1/2}, (S_1^* + S_2^*)^{-1},
 [ \mu_{\infty}^*/ (\rho_{\infty}^*(S_1^* +S_2^*))]^{1/2},
		\rho_{\infty}^*, (S_1^* +S_2^*) \mu_{\infty}^*/ \rho_{\infty}^*, (S_1^* + S_2^*)\mu_{\infty}^*, (S_1^* + S_2^*), \mu_{\infty}^*, $ \;$ \mu_{\infty}^*,$ and $\mu_{\infty}^*/ \rho_{\infty}^*$. The dimensional strain rates $S^*_1$ and $S^*_2$ and  vorticity $\omega_{\kappa}^*$ are normalized by $S^*_1 + S^*_2$.  The reference values for strain rates and far-stream variables and properties used for normalization  will be constants. The reference length $[ \mu_{\infty}^*/ (\rho_{\infty}^*(S_1^* +S_2^*))]^{1/2}$ is the estimate for the magnitude of the viscous-layer thickness.  In the following flamelet analysis, the vorticity $\omega_{\kappa}$ and the velocity derivatives $\partial u_i/ \partial x_j$ are non-dimensional quantities; their dimensional values can be obtained through multiplication by $S^*_1 + S^*_2$. Now, $S_1 + S_2 =1$.

In the rotating reference frame, two different gaseous mixtures exist the far field, one for large positive values of $\chi$ and another for large negative values of $\chi$. They both advect and diffuse towards each other. For the study of a diffusion (nonpremixed)  flame, one mixture is fuel and the other is an oxidizer. With a premixed flame, one far field has a combustible mixture of fuel and oxidizer while the other has a hot inert gas (e.g., combustion products). In another case where multiple flame branches may occur, both streams can be combustible; one can be fuel rich while the other is fuel lean.
		
\subsection{Similar Form for the Equations} \label{similar}

		The stagnation point  is taken as the origin $\xi = \chi =z=0$. Along the line $\xi=z=0$ normal to the interface , we can expect the first derivatives of $u_{\chi}, \rho, h, T,$ and $Y_m$ with respect to either $\xi$ or $z$ to be zero-valued.   The velocity components $u_{\xi}$ and $w$ will be odd functions of $\xi$ and $z$, respectively, going through zero and changing sign at that line. $v $ also changes sign, going through zero at the origin;  however, generally the reaction zone will be offset and an odd function does not result for $v$.  Upon neglect of terms of $O(\xi^2)$ and $O(z^2)$, the variables $u_{\chi}, \rho, h, T,$ and $Y_m$ can be considered to be functions only of $t$ and $\chi$. The density-weighted \cite{Illingworth} transformation of $\chi$  replaces $\chi$ with $\eta \equiv \int^y_0 \rho (\chi') d\chi'$. Neglect of the same order of terms implies that $u_{\xi}= S_1 \xi(df_1/d\eta)$ and $w= S_2 z(df_2/d\eta)$. Note $u_{\xi}$ is independent of $z$ and $w$ is independent of $\xi$ ). At the edge of the viscous layer at large positive $\eta$,
		$df_1/d \eta \rightarrow 1, df_2/d \eta \rightarrow 1, f_1 \rightarrow \eta$, and $f_2 \rightarrow \eta$  .
		Ordinary differential equations are created here through the variable $\eta$ and the convenient notation is used so that  $( )' \equiv d( )/d\eta$.

		In the non-dimensional form given by Equations (\ref{cont}) through (\ref{species}), the dimensional strain rates $S_1^*$ and $S_2^*$ are each normalized by the dimensional sum $S* = S_1^* + S_2^*$. If the far field has uniform density, $S*$  is the magnitude of the major compressive normal strain. Thus, the non-dimensional relation is $S_2 = 1 - S_1$ and only one independent non-dimensional strain-rate parameter is needed. Two strain rates are presented above and in the following analysis with the understanding that one depends on the other such that $S_1 + S_2 = 1$.  $S_1 + S_2$ will be explicitly stated in our analysis without substitution of the unity value in order to emphasize the summation which is consequential in the dimensional formulation.   This choice clarifies whether a particular  term when converted to a dimensional form depends on $S_1^*, S_2^*$, or the sum of the two strain rates.
		
		For steady state, the continuity equation (\ref{cont}) is readily integrated to give
		\begin{eqnarray}
			\rho u_{\chi} = -S_1 f_1(\eta) - S_2 f_2(\eta)
			\label{cont2}
		\end{eqnarray}
		and then
		\begin{eqnarray}
			u'_{\chi} =  \frac{S_1 f_1(\eta) + S_2 f_2(\eta)}{\rho^2 }
			\rho' -\frac{S_1 f_1'(\eta)+ S_2 f_2'(\eta)}{\rho}
			\label{dv}
		\end{eqnarray}
		Thus, the incoming inviscid flow outside the boundary layer is described by $u_{\chi}=-(S_1 + S_2)\eta$ for positive $\eta$ and $u_{\chi}=-(S_1 + S_2)\eta/ \rho_{-\infty}$ for negative $\eta$.
	
 At $\eta =\infty$, $f_1' =f_2'=1$ and $f_1'' = f_2'' = f_1''' = f_2''' =0$ which allows the two constants to be determined.
 A perfect gas with $\rho\mu = 1$ is assumed. The perfect gas law and the assumption of constant specific heat $c_p$ will give the relation that $1/ \rho = h$.  Specifically, we obtain
		\begin{eqnarray}
			f_1'''  +f f_1'' + S_1[ h -(f_1')^2]    &+& \frac{\omega_{\kappa}^2}{4S_1}(1 - h)
=0 \nonumber \\
		f_2'''  +f f_2'' + S_2[h-(f_2')^2] &=& 0
			\label{ODEs}
		\end{eqnarray}
The boundary conditions  use the assumption that two velocity components asymptote to
the constant values
$ u_{\xi}(\infty), u_{\xi}(-\infty), w(\infty),$ and $w(-\infty)$ at large magnitudes of $\eta$.   The stream function bounding the two incoming streams is arbitrarily given a zero value and placed at $\eta =0$.	
\begin{eqnarray}
f_1'(\infty)&=&  1 \;\; ;\;\; f_1'(-\infty)  = \sqrt{h_{-\infty}+  \bigg(\frac{\omega_{\kappa}}{2 S_1}\bigg)^2
(  1 -   h_{-\infty})  }       \;\;   ; \; \;  	f_1(0)  =0 \;\;  ; \;\;  \nonumber \\	
f_2' (\infty)&=& 1 \;\; ;\;\;  f_2'(-\infty) = \sqrt{h_{-\infty}}  \;\;\; ; f_2(0)  =0
\label{BCs}
\end{eqnarray}

When density varies through the flow because of heating or variation of composition, $u_{\xi}$ and $w$ vary with $\chi$, thereby creating a shear stress and vorticity albeit that the frame  transformation removed vorticity and shear  from the incoming flow.

The dependence of $u_{\chi}$ on $f\equiv S_1f_1 + S_2 f_2$ is shown by Equation (\ref{cont2}). Thus, the function $f$ will be important in determining both the field for $u_{\chi}$ and the scalar fields.

		Consequently,  $f$ as well as $f_1$ and $f_2$ depend on both $S_1$ and $S_2$, not merely on $S_1 + S_2$. That is, the particular distribution of the normal strain rate between the two transverse direction  matters. $f$ and $f_1$  also depend directly on $\omega_{\kappa}$ (unless $ S_1 = 0$). $f_2$ depends on $\omega_{\kappa}$ indirectly through its coupling with $f_1$.
	
Here, an exact solution of the variable-density Navier-Stokes equation is obtained subject to determination of $h$  through solutions of the energy and species equations as discussed below. Thus, the solution here is the natural solution, subject to neglect of terms of $O(\xi^2)$ and $O(z^2)$.

	 The similar form of the scalar equations becomes
\begin{eqnarray}
 Y''_m  + Pr f Y'_m &=& - Pr\dot{\omega}_m   \;\;;\;\; m=1, 2, ...., N  \nonumber  \\
 h'' + Pr f h' + (Pr - Sc)  \Sigma^N_{m=1}h_m Y''_m
&=& Pr  \Sigma^N_{m=1}h_{f,m} \dot{\omega}_m
\label{ODEs3}
\end{eqnarray}

The boundary conditions are
\begin{eqnarray}
h(\infty) &=& 1  \;\; ; \; \; h(-\infty)= \frac{1}{\rho_{-\infty}}\;\; ;\;\; \nonumber \\
Y_m (\infty) &=& Y_{m,\infty} \;\; ; \;\; Y_m (-\infty) = Y_{m,-\infty} \;\; ;\;\;\nonumber \\
		\label{BCs3}
	\end{eqnarray}

	Equations (\ref{ODEs3})  indicate a dependence of the heat and mass transport on $f\equiv S_1f_1 + S_2f_2$. Manipulation of the first two equations of (\ref{ODEs3}) leads to an  ODE for $f$ with $S_1
S_2$ and $S_1S_2f_1'f_2'$ as parameters, clearly indicating that generally $f$ will have a dependence on $S_1S_2$. Thus, the behavior for the counterflow can vary from the planar value of $S_1 =1, S_2 =0$ (or vice versa) or from the case $S_1 = S_2 =1/2$.  This clearly shows that distinctions must be made amongst the various possibilities for three-dimensional strain fields as $S_1S_2$ varies between large negative numbers and $1/4$. An exception is the incompressible case with constant properties where the $S_1S_2$ terms cancel in the equation for $f$.

	The vorticity $\omega_{\kappa}$  impacts directly $f_1$ and $f$; thereby, it is affecting the velocity field. Then, through the advection of the scalar properties, there is impact on mass fractions and enthalpy.   If the vorticity $\omega_{\kappa} =0$, a simple inspection of the governing ODEs leads to the conclusion that the values for $f_1, f_1', f_2, f_2',u/x,$ and $w/z$ can be interchanged with the values for $f_2, f_2', f_1, f_1',w/z,$ and $u/x$, respectively, when $S_1$ and $S_2$ are replaced by $1-S_1$ and $1-S_2$, respectively.

The analysis is formulated in identical fashion to the approach of \cite{Sirignano2022}. However, there, with two directions for extensional strain rate in the rotating frame of reference, both $S_1$ and $S_2$ are positive numbers. However, in the computations here, we  consider the vortex tube with inward swirl so that, in the far field,  there are two directions of compressive strain and only one direction of extensional strain. That extensional strain is aligned with the vorticity vector. Thus, here, $S_1 \leq 0$ and $S_2\geq 1$.  The basic case  takes the two compressive strain rates to be equal; thereby, $S_1 = -1.0$ and $S_2 = 2.0$.

Consider the production or consumption rate of a particular species over the counterflow volume. We can either integrate over a volume using the original form in Equation (\ref{species}) or, more conveniently, using Equation (\ref{ODEs3}) to get exactly the same result. Consider the volume $-a < \xi < a, -b < y < b, -c <z < c$. The choices of lengths $a$ and $c$ do not matter on a per-unit-volume basis  since mass fraction $Y_m$ and reaction rate $\dot{\omega}_m$ do not vary with $x$ or $z$. $c$ is chosen to be of the order of the Kolmogorov scale. Volume $V = 8abc$ and $ \widetilde{\rho\dot{\omega}_m} $  is the average mass production rate over the volume. From integration of the Equations (\ref{ODEs3}) after multiplication by density $\rho$ and division by $Pr V$,
\begin{eqnarray}
\int^a_{-a}\int^b_{-b}\int^c_{-c}  \frac{\rho}{Pr V} [Y''_m  &+&  Pr f Y'_m + Pr\dot{\omega}_m]dx dy dz = 0   \;\;;\;\; m=1, 2, ...., N  \nonumber  \\
\widetilde{\rho\dot{\omega}_m} &\equiv& \frac{1}{V}\int_V \rho \dot{\omega}_m dV= - \frac{1}{2b} \int^{\eta(b)}_{\eta(-b)}fY'_m d\eta \;\;;\;\; m=1, 2, ...., N  \nonumber  \\
\int^a_{-a}\int^b_{-b}\int^c_{-c}  \frac{\rho}{Pr V}  [ h'' &+& Pr f h'
- Pr  \Sigma^N_{m=1}h_{f,m} \dot{\omega}_m ] dx dy dz =0
\nonumber \\
\Sigma^N_{m=1}h_{f,m} \widetilde{\rho \dot{\omega}_m } &=& \frac{1}{2b}\int^{\eta(b)}_{\eta(-b)} fh' d\eta
\label{rateintegral}
\end{eqnarray}
Here, $b$ is considered large enough so that $Y'_m =0$ and $h'=0$ at those boundaries are good approximations.  However, the value for $ \widetilde{\rho \dot{\omega}_m }$ depends strongly on the chosen domain size $2b$, which has  a value of $O(10)$ typically in our analysis.

Consider a species $m$ that is flowing inward away from $\eta = \infty$ towards $\eta = 0$. If it is being produced (consumed), the derivative $Y'_m$ in Equation (\ref{rateintegral}) will be negative (positive) for $\eta > 0$ where velocity $v<0$ and $f>0$. The signs are opposite for a species flowing inward away from $\eta = -\infty$ and towards $\eta = 0$.  The equation provides two ways to evaluate the average production (consumption) rate for species $m$. The volume integral of the reaction rate has highly nonuniform integrand values over the space while the outflow integral over $\eta$ has a smoother variation of the integrand.

\subsection{Chemical Kinetics Model}

The above analysis applies for both diffusion-flame  and partially-premixed-flame configurations . Multi-branched flames can also be described. While the analysis allows for the use of detailed chemical kinetics, we focus here on propane-oxygen flows with one-step kinetics.  \cite{Westbrook_Dryer:1984}   kinetics are used; they were developed for premixed flames but any error for nonpremixed flames is often viewed as tolerable  here because diffusion  generally is rate-controlling.  Using astericks to denote dimensional quantities,
\begin{eqnarray}
\dot{\omega}^*_F = - A^* {\rho^*}^{0.75} Y_F^{0.1} Y_O^{1.65} e^{-50.237/\tilde{h}}
\label{onestep}
\end{eqnarray}
where the ambient temperature is set at 300 K and density $\rho^*$ is to be given in units of kilograms per cubic meter. $A^* =4.79\times10^8 (kg/m^3)^{-0.75}/s$. The dimensional strain rate $S^*_1 + S^*_2$ (at the sub-grid scale) is used to normalize time and reaction rate.
In non-dimensional terms,
\begin{eqnarray}
\dot{\omega}_F &=& - \frac{A^* {\rho_{\infty}^*}^{0.75}}{S_1^* + S^*_2}\tilde{h}^{-0.75} Y_F^{0.1} Y_O^{1.65} e^{-50.237/\tilde{h}}   \nonumber  \\
\dot{\omega}_F &=& - \frac{Da}{\tilde{h}^{0.75}} Y_F^{0.1} Y_O^{1.65} e^{-50.237/\tilde{h}}
\label{Damkohler}
\end{eqnarray}
The  equation defines the Damk\"{o}hler number $Da$. We define $K$ so that $Da \equiv K Da_{ref}$ where
\begin{eqnarray}
Da_{ref} \equiv \frac{ \tilde{A}(10 kg/m^3)^{0.75}}{(10^4/s)} =2.693 \times 10^5 \;\;  ;  \;\;
 K \equiv \Big[\frac{\rho_{\infty}^*}{10 kg/m^3}\Big]^{0.75}\frac{10^4/s}{S_1^* + S_2^*}
\end{eqnarray}
10 $kg/m^3$ and 10,000$/s$ are conveniently chosen as reference values for density and strain rate, respectively.

 It is not necessary to set pressure (or its proxy, density) and the strain rate separately for a one-step reaction. For propane and oxygen, the mass stoichiometric ratio $\nu =0.275$.    The non-dimensional parameter $K$ will increase (decrease) as the strain rate decreases (increases) and/ or the pressure increases (decreases). $K =1$ is our reference case. The value of $K$ will be varied as  needed to address the variations  in strain rate and pressure that affect premixed flamelets, diffusion flamelets, and multi-branched flamelets.

\section{Computational Results and Discussion} \label{results}

The ordinary differential equations are solved numerically  using a relaxation method with a pseudo-time variable and central differences.  The parameters that are varied are $ K, Pr,\omega_{\kappa}$ and $S_1$ (and thereby $S_2 = 1 -S_1$). Here, calculations have $Sc = Pr = 1$ with emphasis on the effect of variation in $K$, i.e., pressure and strain rate.

Here, the effects of vorticity on three types of oxygen-propane flame structure will be examined.  In Subsection \ref{DiffFlame}, a diffusion flame near its flammability is considered. The premixed flame is discussed in Subsection \ref{PremixedFlame} while the multi-branched flame calculations are shown in Subsection \ref{MultiFlame}. The basic calculations pertain to the inward swirling flow with $Pr = Sc =1$ and equal compressive strain rate in the far field from two directions in the rotating frame of reference, i.e., $S_1 = -1.0$ and $S_2 =2.0.$   Values for $K =  Da / Da_{ref}$ and thereby for the Damk\"{o}hler number $Da$ are deliberately chosen in the vicinity of the flammability limit where vorticity and its centrifugal effect can have a significant role.  In addition to boundary values at  $\eta=\infty$ and $\eta = - \infty$, the system of equations has four independent, non-dimensional parameters as inputs: $\omega_{\kappa}, S_1 = 1 - S_2, Pr = Sc$, and $Da = K Da_{ref}$.

\subsection{Diffusion Flamelet Calculations}\label{DiffFlame}
		
\begin{figure}[thbp]
  \centering
 \subfigure[enthalpy, $h/h_{\infty}$ ]{
  \includegraphics[height = 4.6cm, width=0.45\linewidth]{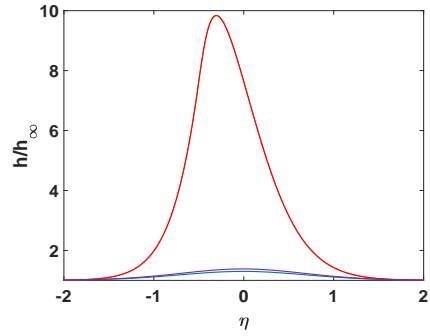}}
  \subfigure[fuel mass fraction, $Y_F$]{
  \includegraphics[height = 4.6cm, width=0.45\linewidth]{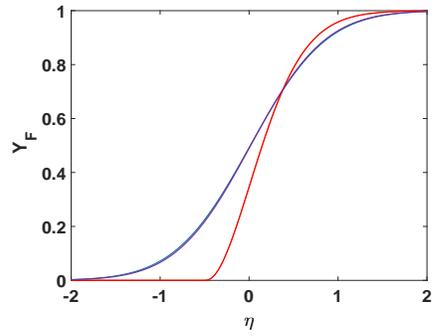}}     \\
  \vspace{0.2cm}
  \subfigure[ mass ratio x oxygen mass fraction, $\nu Y_O$]{
  \includegraphics[height = 4.6cm, width=0.45\linewidth]{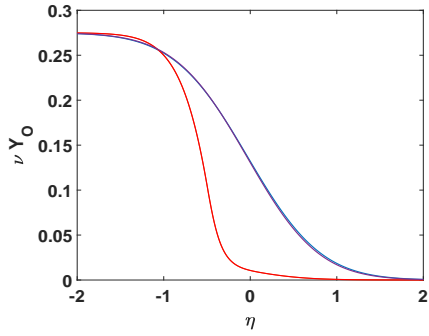}}
  \subfigure[integral of reaction rate, $\int \dot{\omega}_F d \eta$]{
  \includegraphics[height = 4.6cm, width=0.45\linewidth]{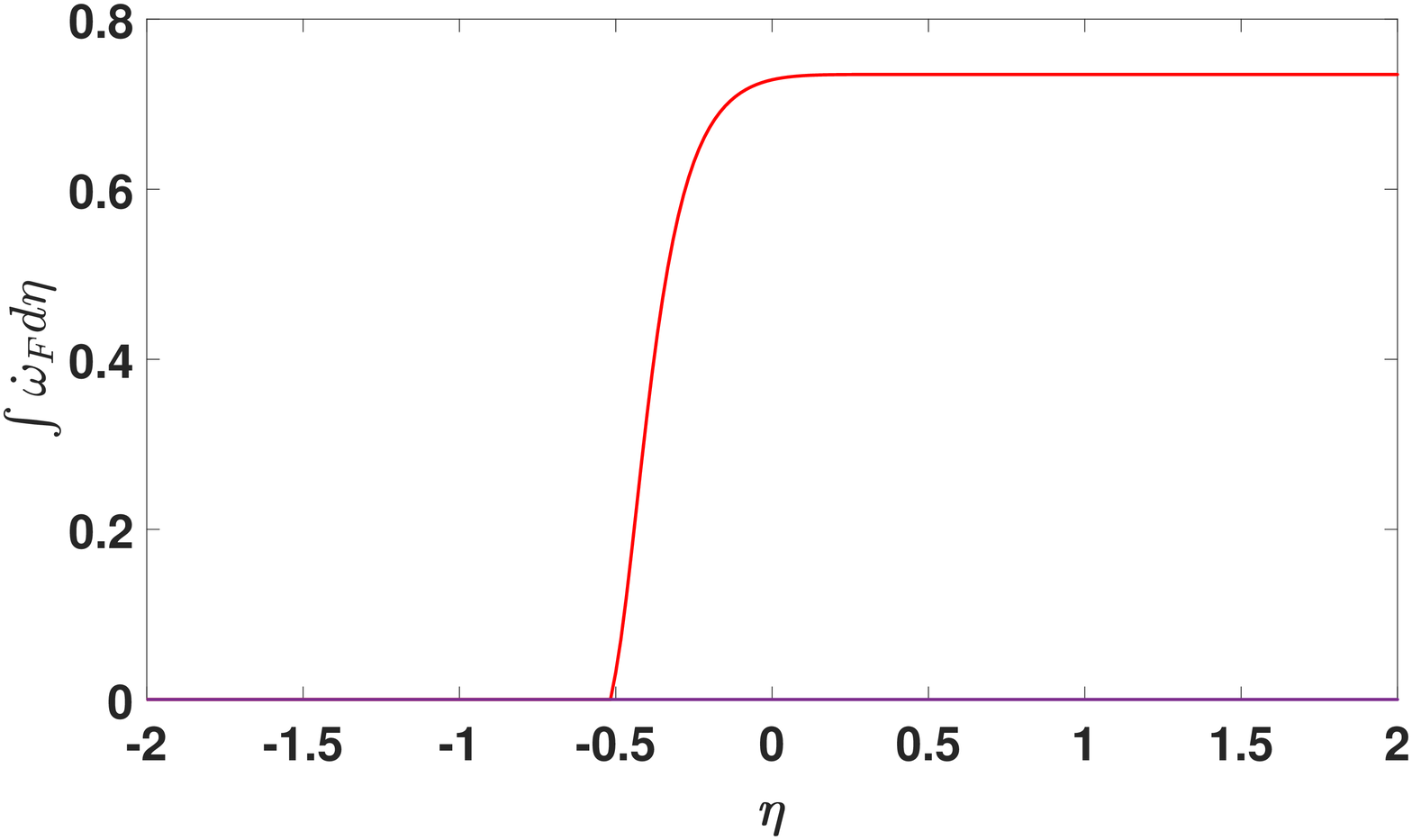}}     \\
  \vspace{-0.1cm}
  \caption{Scalar properties for diffusion flame with varying vorticity. \\
  $S_1 =-1.00 ;  \; S_2 = 2.00; \; K= 0.275.  \;\;  \omega_{\kappa} = 0,$   \;  \;  blue, \; no flame; \; $\omega_{\kappa} = 0.5,$   \;   purple,  \;no flame;  \; $\omega_{\kappa} = 1.0$,  \;  red, \; flame. }
  \label{WeakDiffFlame1}
\end{figure}

\begin{figure}[thbp]
  \centering
 \subfigure[mass flux per area, $f=\rho u_{\chi}$ ]{
  \includegraphics[height = 4.8cm, width=0.45\linewidth]{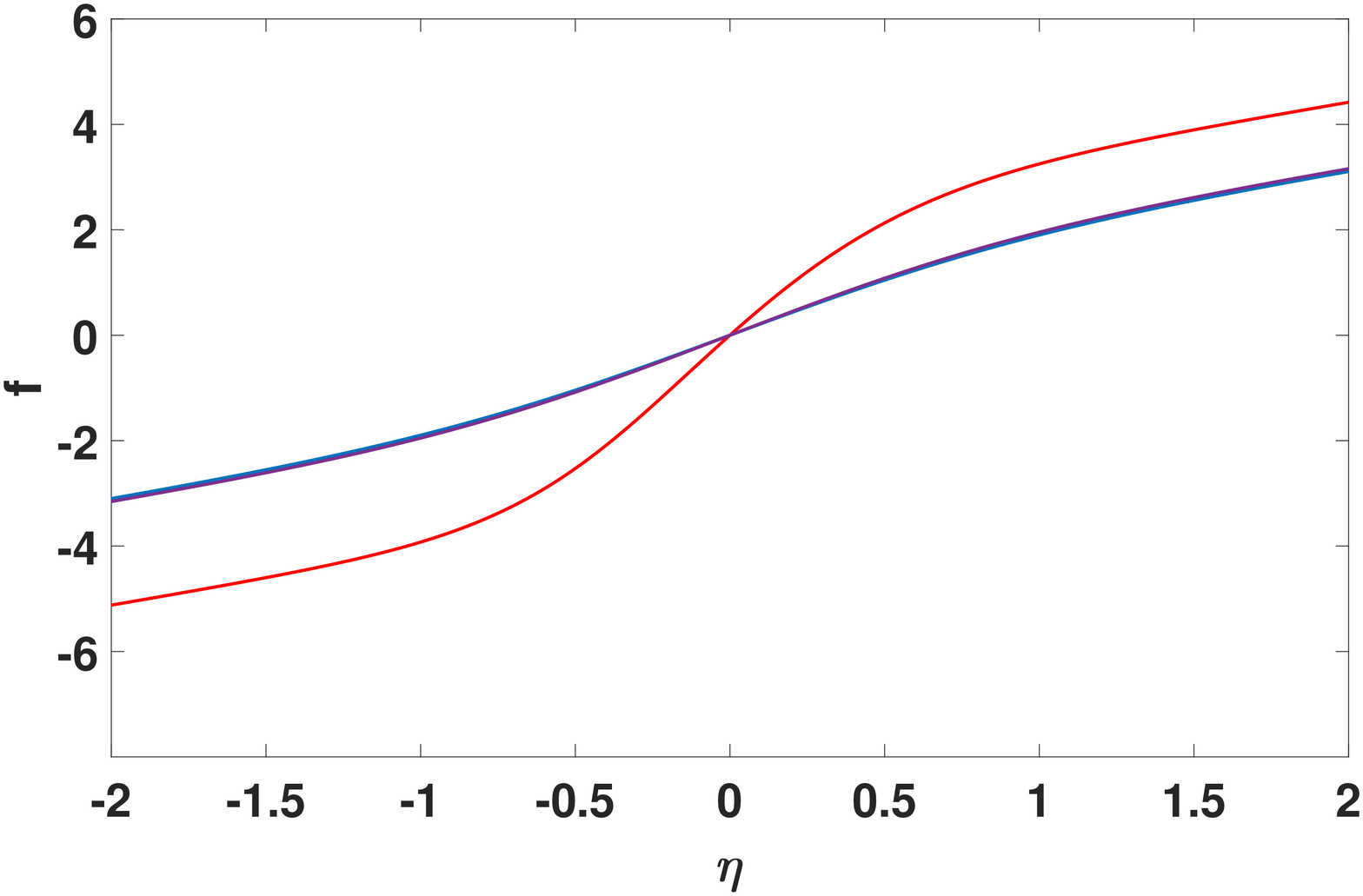}}
  \subfigure[velocity component, $f_1' = u_{\xi}/(S_1 \xi)$]{
  \includegraphics[height = 4.8cm, width=0.45\linewidth]{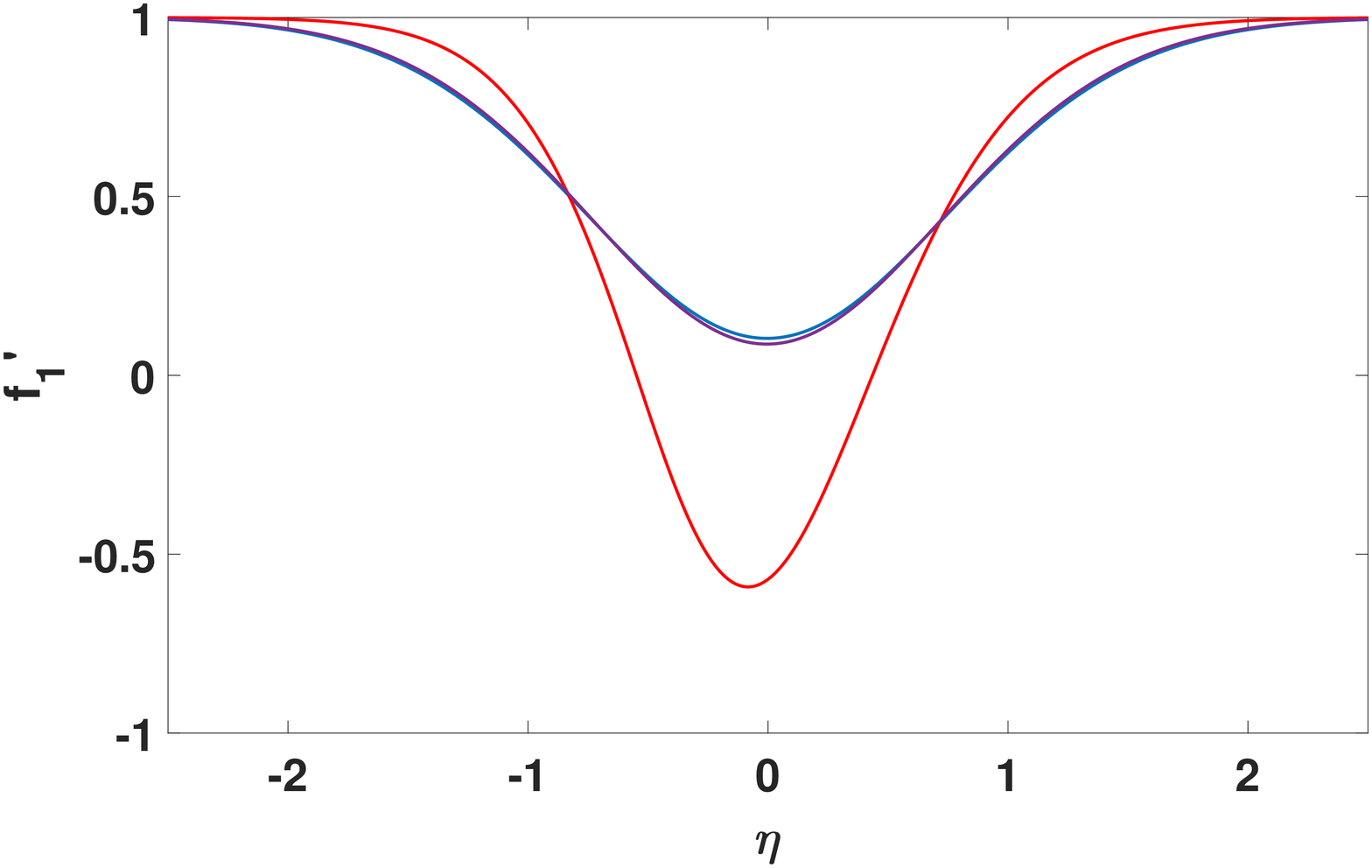}}     \\
  \vspace{0.2cm}
  \subfigure[ velocity component, $f_2' = w/(S_2z)$]{
  \includegraphics[height = 4.8cm, width=0.45\linewidth]{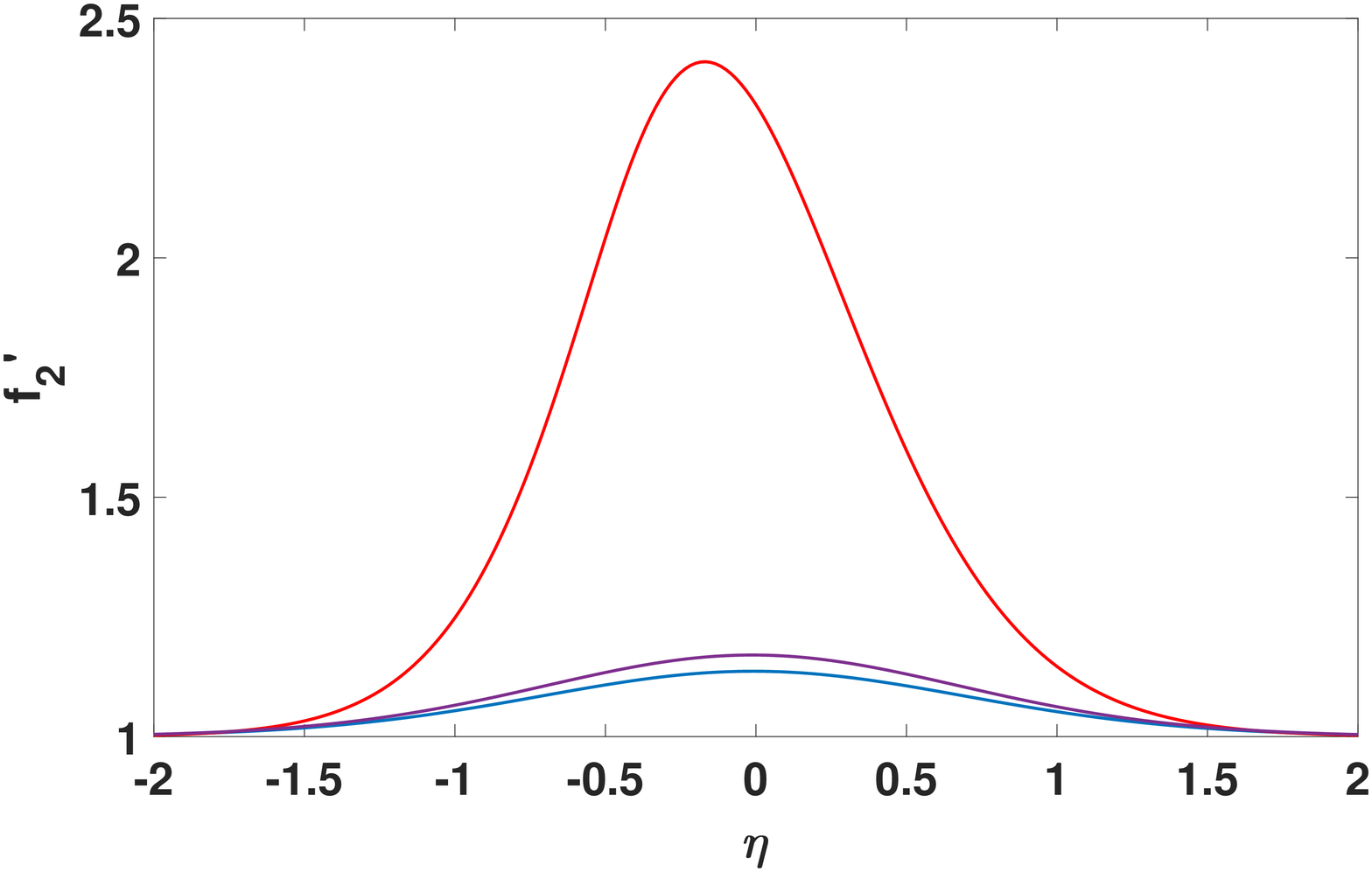}}
  \subfigure[velocity component, $u_{\chi}$]{
  \includegraphics[height = 4.8cm, width=0.45\linewidth]{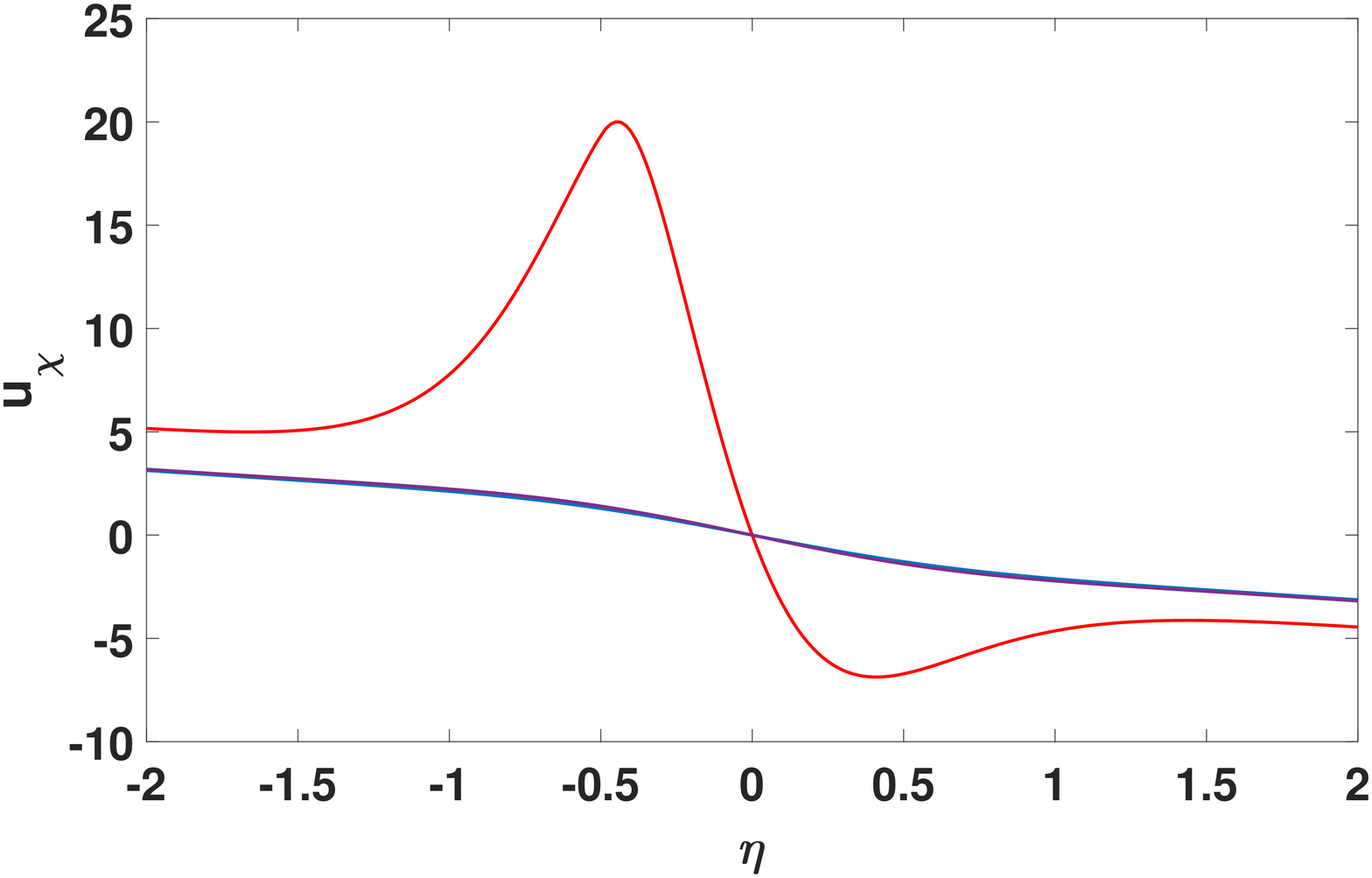}}     \\
  \vspace{0.2cm}
   \caption{Velocity behavior for diffusion flame with varying vorticity. \\  $S_1 =-1.00;  \; S_2 = 2.00; \; \; K= 0.275. \; \; \; \omega_{\kappa} = 0$, \;  blue, \; no flame ; \; $\omega_{\kappa} = 0.5,$   \;   purple,  \; no flame;  \; $\omega_{\kappa} = 1.0$,  \;  red, \;  flame, \;  flow reversal. }
  \label{WeakDiffFlame2}
\end{figure}

First, we treat a situation with a  three-dimensional diffusion-flame structure. Figures \ref{WeakDiffFlame1} and  \ref{WeakDiffFlame2} show the influence of vorticity on the flamelet stability near the extinction limit. The rotation of the flamelet due to vorticity causes a centrifugal effect on the counterflow velocity and thereby on the residence time in the vicinity of the reaction zone. $ K =0.275$ with values of $\omega_{\kappa} =0, 0.5,$ and $1.0$ are examined and reported here.

Figure \ref{WeakDiffFlame1}  shows that, without rotation and also with $\omega_{\kappa} = 0.5$, there is negligible reaction rate and heat release, essentially yielding extinction. Fuel and oxidizer just diffuse and mix without significant exothermic reaction. Further increase of the  rotational rate with $\omega_{\kappa} = 1.0$, however, yields a strong flame with a narrow reaction zone.
The heat release causes a decrease of the density in the vicinity of the flame. A plane still exists in the rotating reference frame where two different mixtures come together in a direction aligned with the scalar gradient while turning into the $z$ direction aligned with the vorticity. The expanding gas can cause a flow reversal of the inward flow from the $\xi$-direction, orthogonal to the scalar gradient, as shown in Figure \ref{WeakDiffFlame2}. The increased rotational rate produces the centrifugal acceleration that inhibits radially inward flow of the heavier gas and allows the expansion and velocity reversal for the lighter, hotter gas. Note that the development of negative values for $f_1'$ means that with the value $S_1 \leq 0$, the direction of the velocity component $u_{\xi}$  becomes radially outward, i.e., $u_{\xi} > 0 $ for $\xi > 0$ and $u_{\xi} < 0 $ for
$\xi < 0$.

Clearly, the combination of fluid rotation, variable density, and three-dimensional structure have major consequences for flamelet behavior. The specific mechanism is not immediately obvious but can be inferred from the results. Figure \ref{WeakDiffFlame2}b and  \ref{WeakDiffFlame2}c indicate that the  strong swirl causes the reversal of the $u_{\xi}$ velocity and an increase in the $w$ velocity, both  now being outward flows from the combustion zone. However, the fractional decrease in density implied by Figure \ref{WeakDiffFlame1}a is significantly larger than the fractional increase in outward velocity. So, the outward mass flow rate is reduced which is consistent with the reduction of the inward mass flow rate when the inward flow ceases to come from both the $\xi$ and $\chi$ directions and is limited to only the $\chi$ direction. Thereby, an increase in residence time of the flow in the reaction domain is allowed. This increase in vorticity and thereby in swirl rate changes the flammability limit.

\newpage

\begin{figure}[thbp]
  \centering
 \subfigure[enthalpy, $h/h_{\infty}$ ]{
  \includegraphics[height = 4.6cm, width=0.45\linewidth]{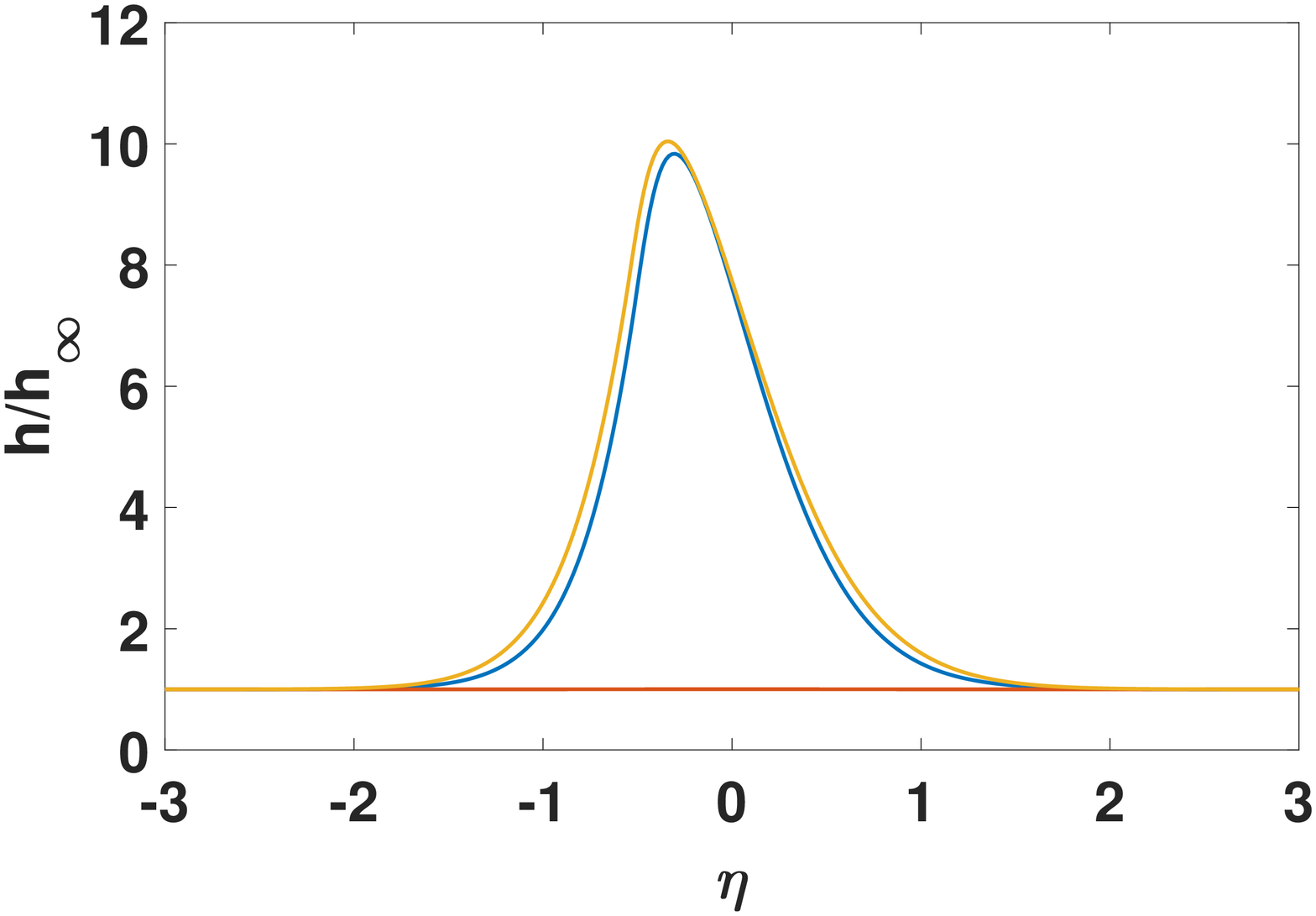}}
  \subfigure[fuel mass fraction, $Y_F$]{
  \includegraphics[height = 4.6cm, width=0.45\linewidth]{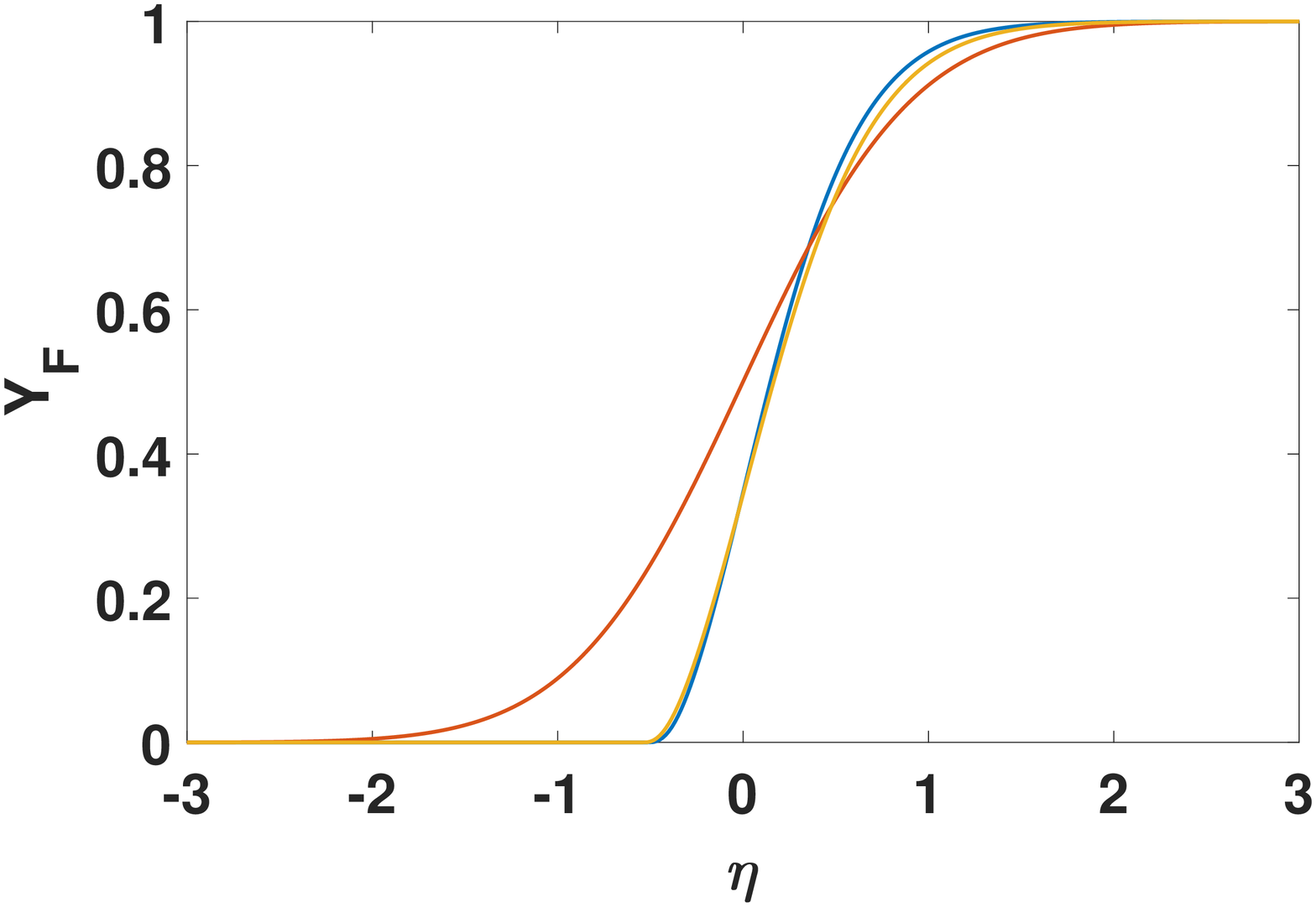}}     \\
  \vspace{0.2cm}
  \subfigure[ mass ratio x oxygen mass fraction, $\nu Y_O$]{
  \includegraphics[height = 4.6cm, width=0.45\linewidth]{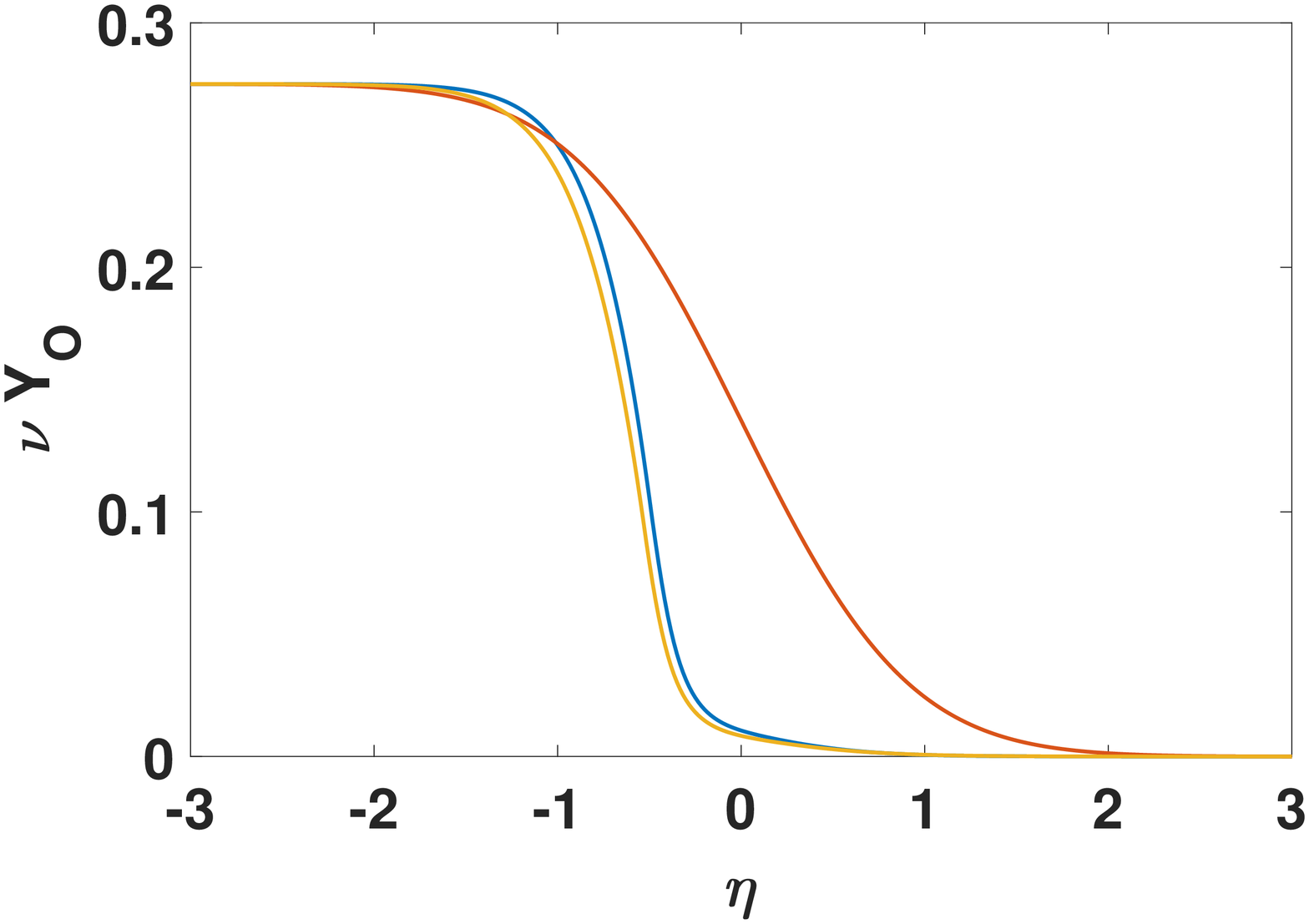}}
  \subfigure[integral of reaction rate, $\int \dot{\omega}_F d \eta$]{
  \includegraphics[height = 4.6cm, width=0.45\linewidth]{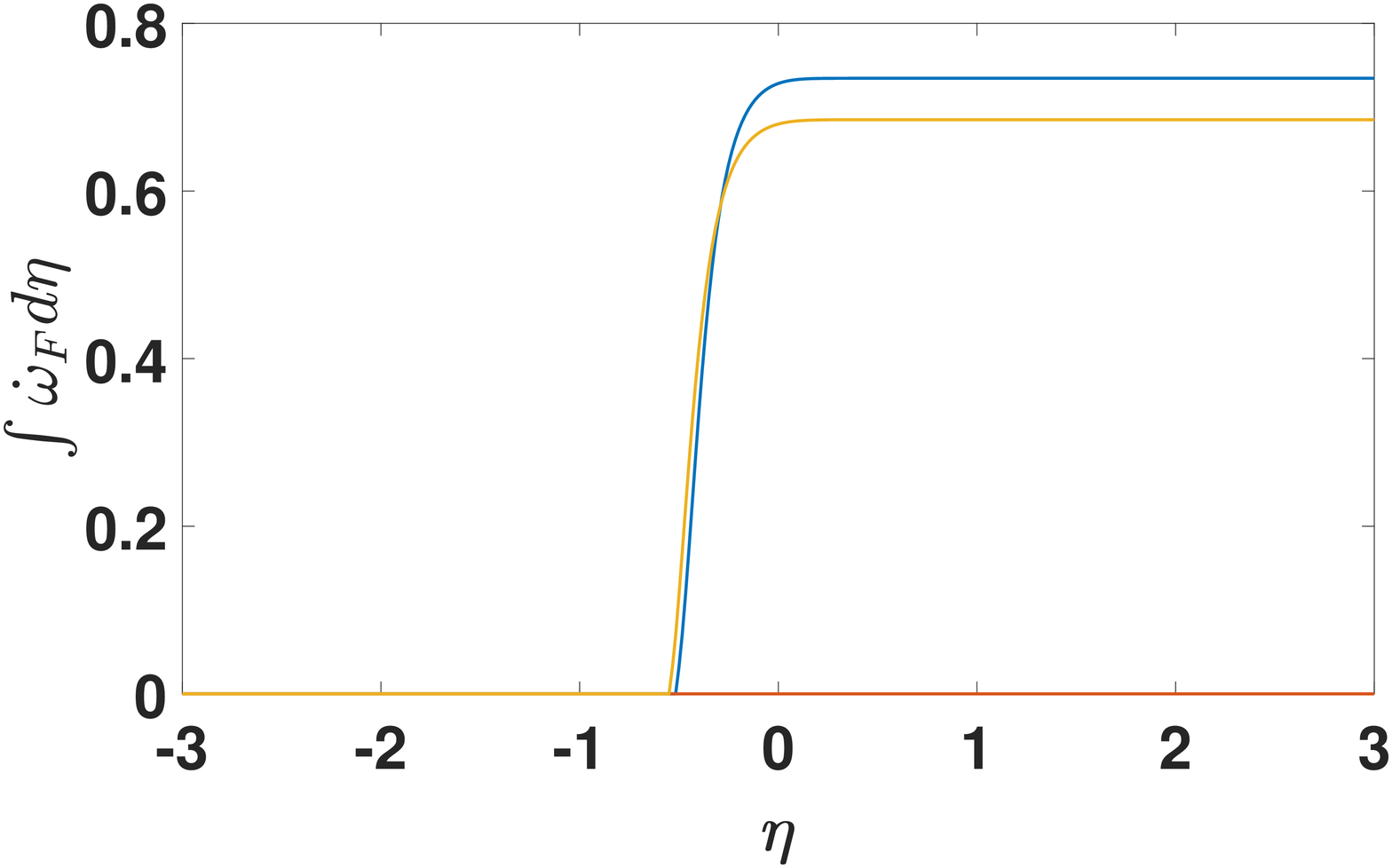}}     \\
  \vspace{-0.1cm}
  \caption{Effects of normal strain on scalar properties for diffusion flame. \\
  $K= 0.275; \;  \omega_{\kappa} = 1.0$.  \;  \; $S_1 =-1.25,  \; S_2 = 2.25$: \; red, \; no flame.
  \; $S_1 =-1.00,  \; S_2 = 2.00$:  \;   blue,  \; flame.  \; $S_1 =-0.75,  \; S_2 = 1.75$:  \;  \;  orange, \; flame. }
  \label{Strain1}
  \end{figure}
\begin{figure}[thbp]
  \centering
 \subfigure[mass flux per area, $f=\rho u_{\chi}$]{
  \includegraphics[height = 4.6cm, width=0.45\linewidth]{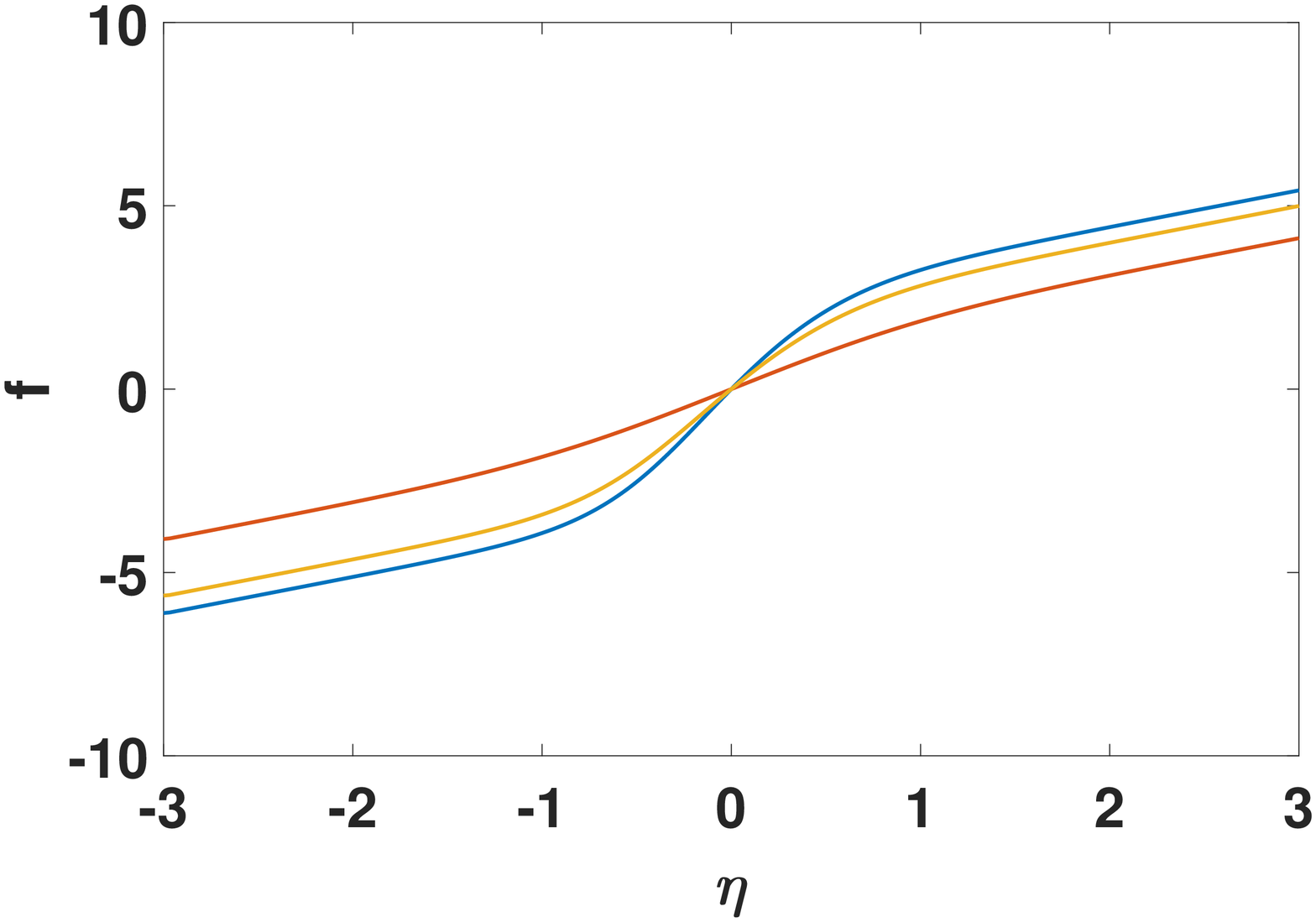}}
  \subfigure[velocity component, $f_1' = u_{\xi}/(S_1 \xi)$]{
  \includegraphics[height = 4.6cm, width=0.45\linewidth]{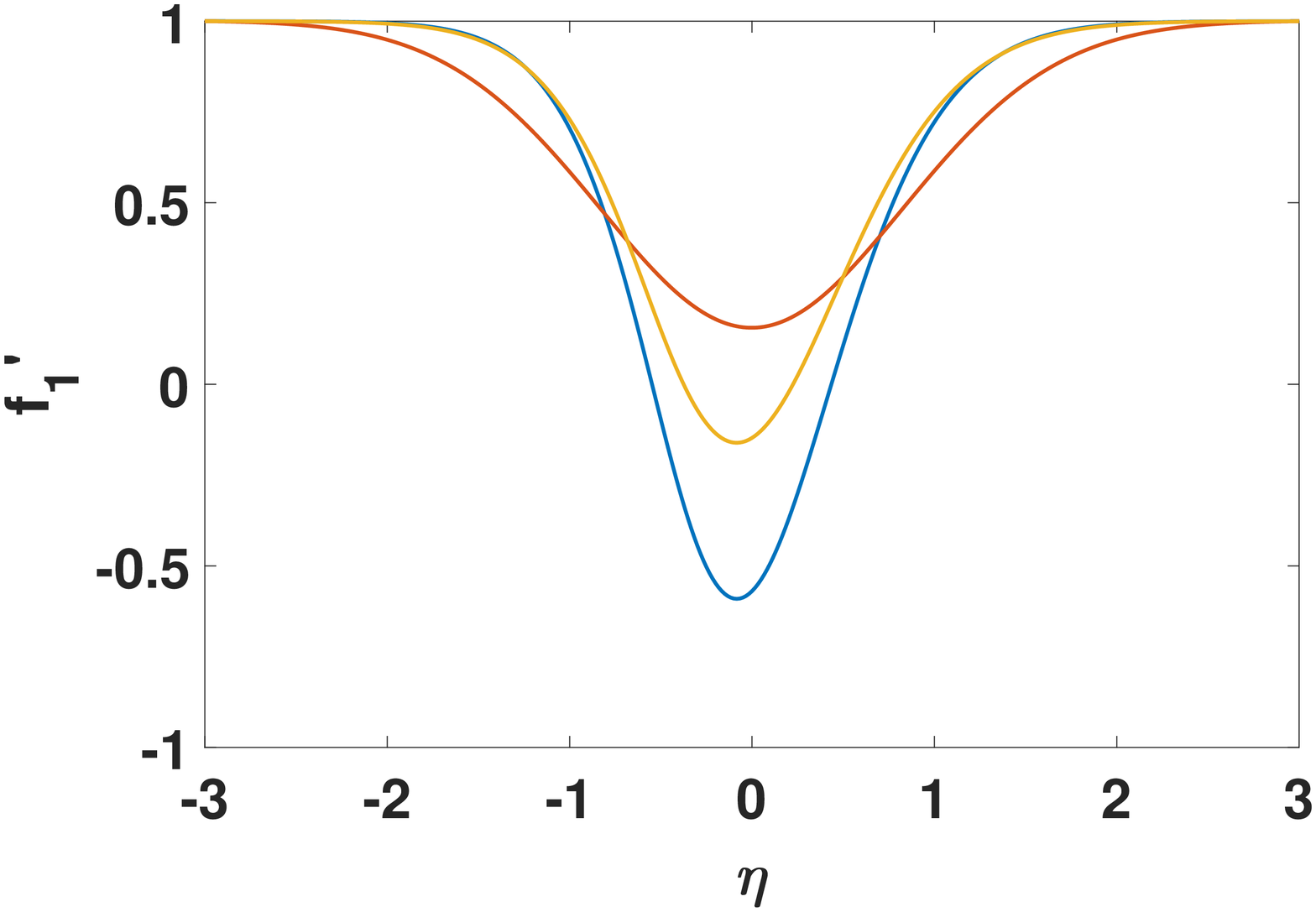}}     \\
  \vspace{0.2cm}
  \subfigure[velocity component, $f_2' = w/(S_2z)$]{
  \includegraphics[height = 4.6cm, width=0.45\linewidth]{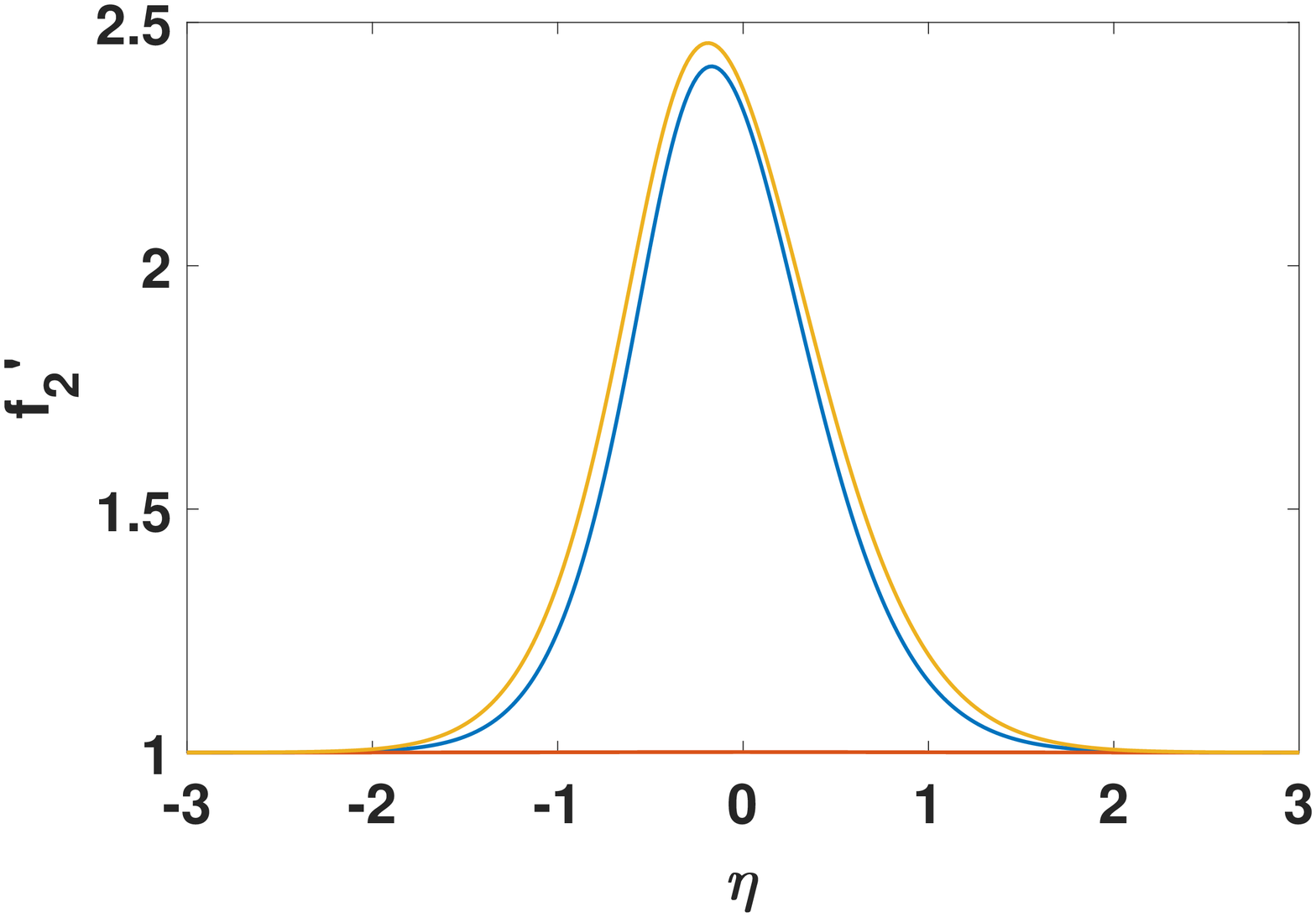}}
  \subfigure[velocity component, $u_{\chi}$]{
  \includegraphics[height = 4.6cm, width=0.45\linewidth]{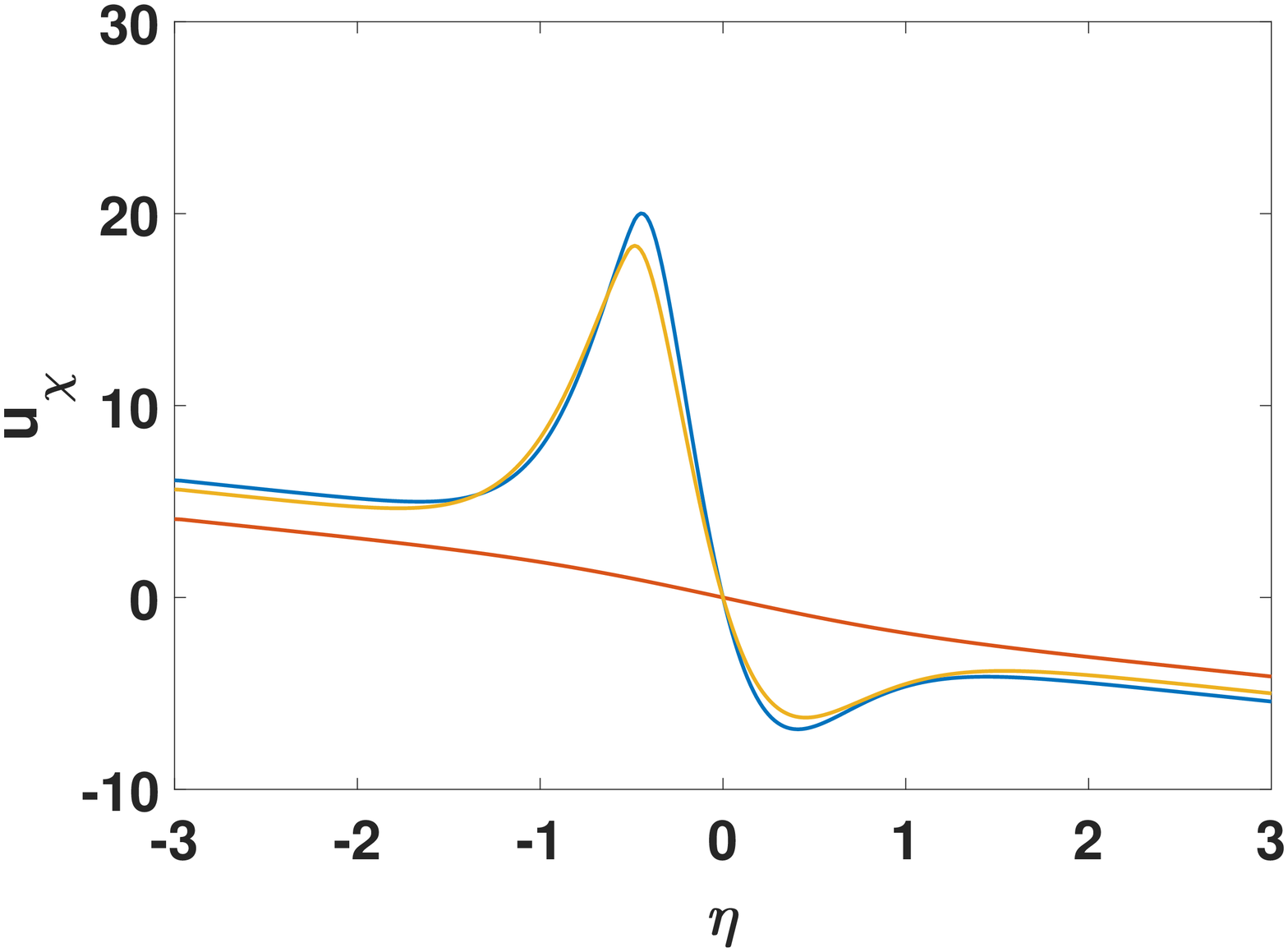}}     \\
  \vspace{-0.1cm}
  \caption{Effects of normal strain on flow properties for diffusion flame. \\    $K= 0.275; \;  \omega_{\kappa} = 1.0$.  \;  \; $S_1 =-1.25,  \; S_2 = 2.25$: \; red, \; no flame.
  \; $S_1 =-1.00,  \; S_2 = 2.00 $ :  \;   blue, \; flame,\; flow reversal.  \; $S_1 =-0.75,  \; S_2 = 1.75 $:  \;  \;  orange, \; flame, \; flow reversal.}
  \label{Strain2}
\end{figure}

  In Figures \ref{Strain1} and \ref{Strain2}, the effect of the imposed ambient normal strain rates is examined for a situation with $\omega_{\kappa} = 1$. Flame extinction results in this example with an increase of the magnitude of the $\xi$ normal strain rate $S_1$ beyond  the value of the $\chi$ normal strain rate. So, $S_1 + S_2 =1 < |S_1| =1.25$ yields no flame while a strong flame is established for the two cases where   $S_1 + S_2 =1 \geq |S_1|$.   Modest decreases in the integrated reaction rate, the mass flux $f$,  and the amount of flow reversal occur with a reduction of imposed strain in the $\xi$ direction from the base case where $S_1 = -1.0$; simultaneously,  a modest increase in peak temperature and enthalpy occurs. Apparently, the reduced mass flux and associated increase in residence time allows for a slightly greater temperature rise although the reaction  rate is slightly reduced. On the other hand, the increase in the magnitude of $S_1$ from the base case would, if density were reduced because of an established flame, yield too low a residence time to hold a flame; so, no flame occurs.

\begin{figure}[thbp]
  \centering
 \subfigure[enthalpy, $h/h_{\infty}$ ]{
  \includegraphics[height = 4.6cm, width=0.45\linewidth]{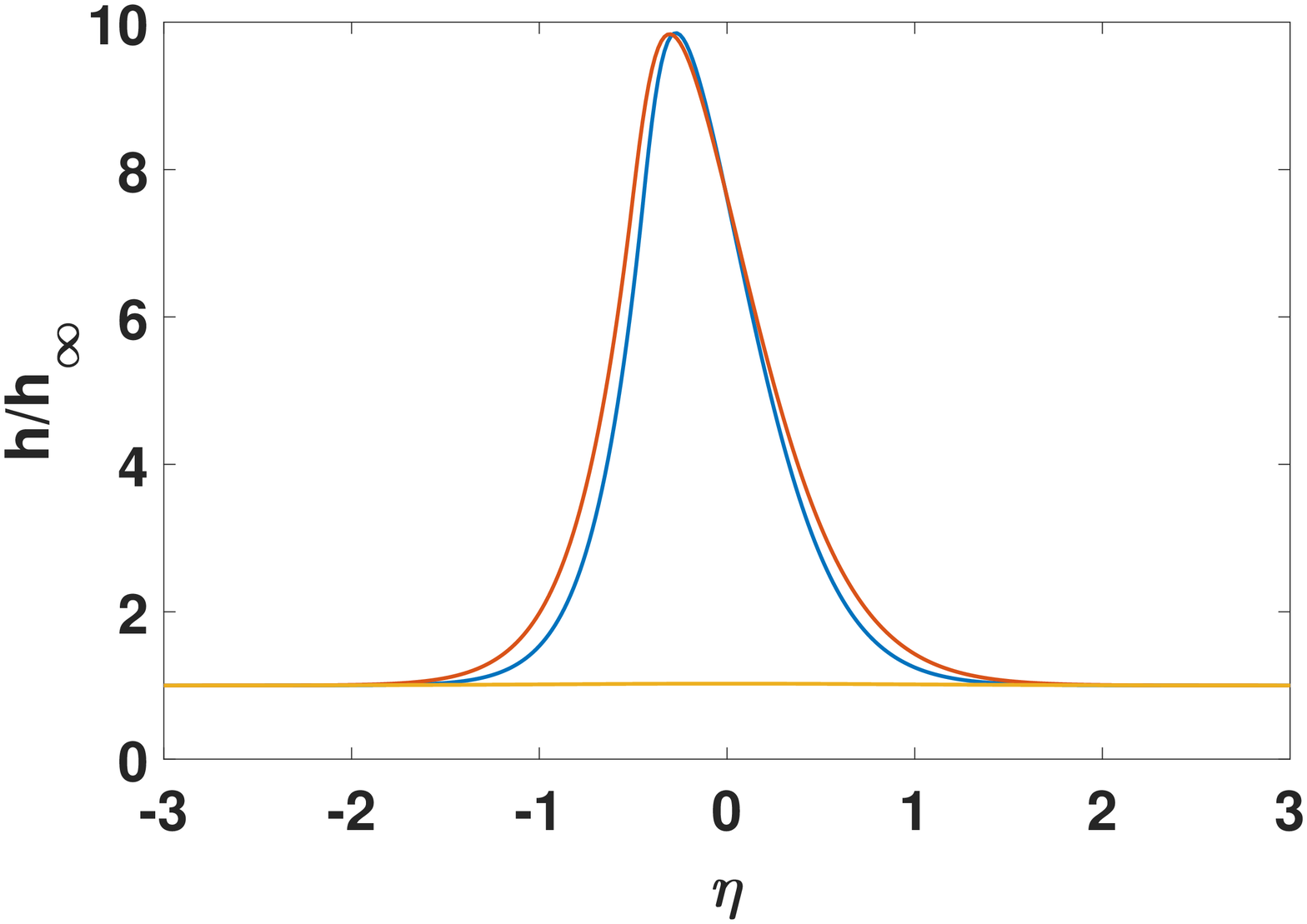}}
  \subfigure[fuel mass fraction, $Y_F$]{
  \includegraphics[height = 4.6cm, width=0.45\linewidth]{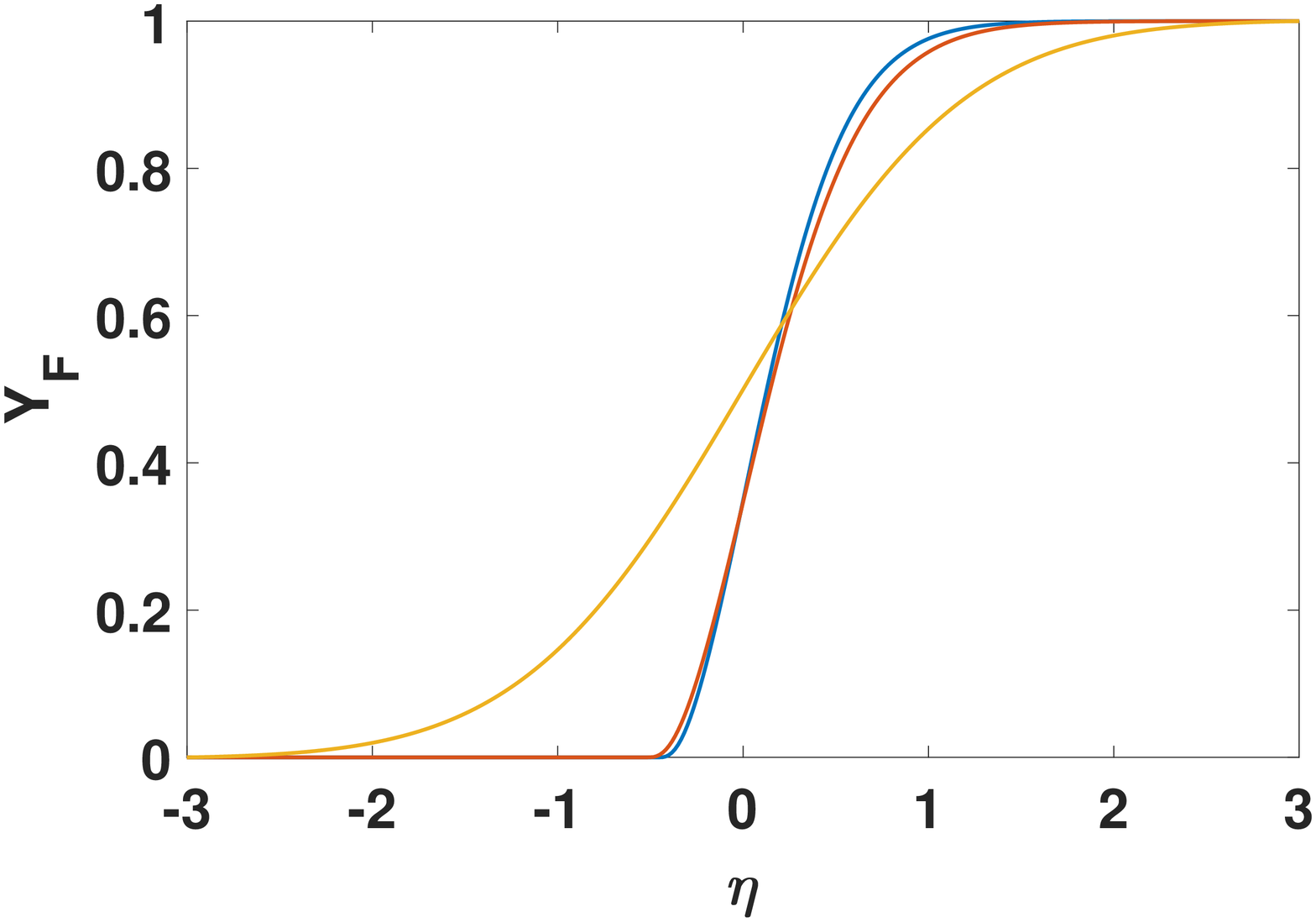}}     \\
  \vspace{0.2cm}
  \subfigure[ mass ratio x oxygen mass fraction, $\nu Y_O$]{
  \includegraphics[height = 4.6cm, width=0.45\linewidth]{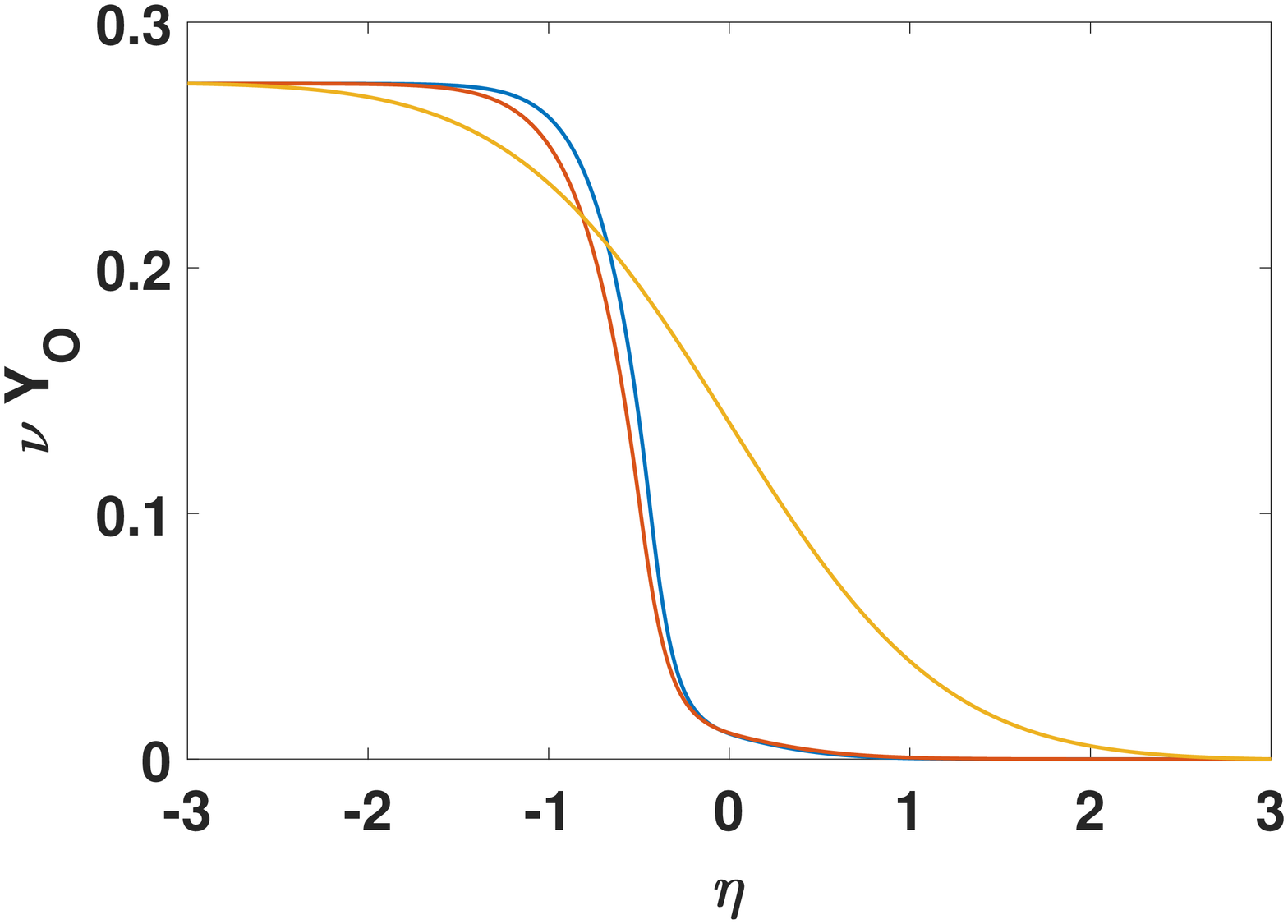}}
  \subfigure[integral of reaction rate, $\int \dot{\omega}_F d \eta$]{
  \includegraphics[height = 4.6cm, width=0.45\linewidth]{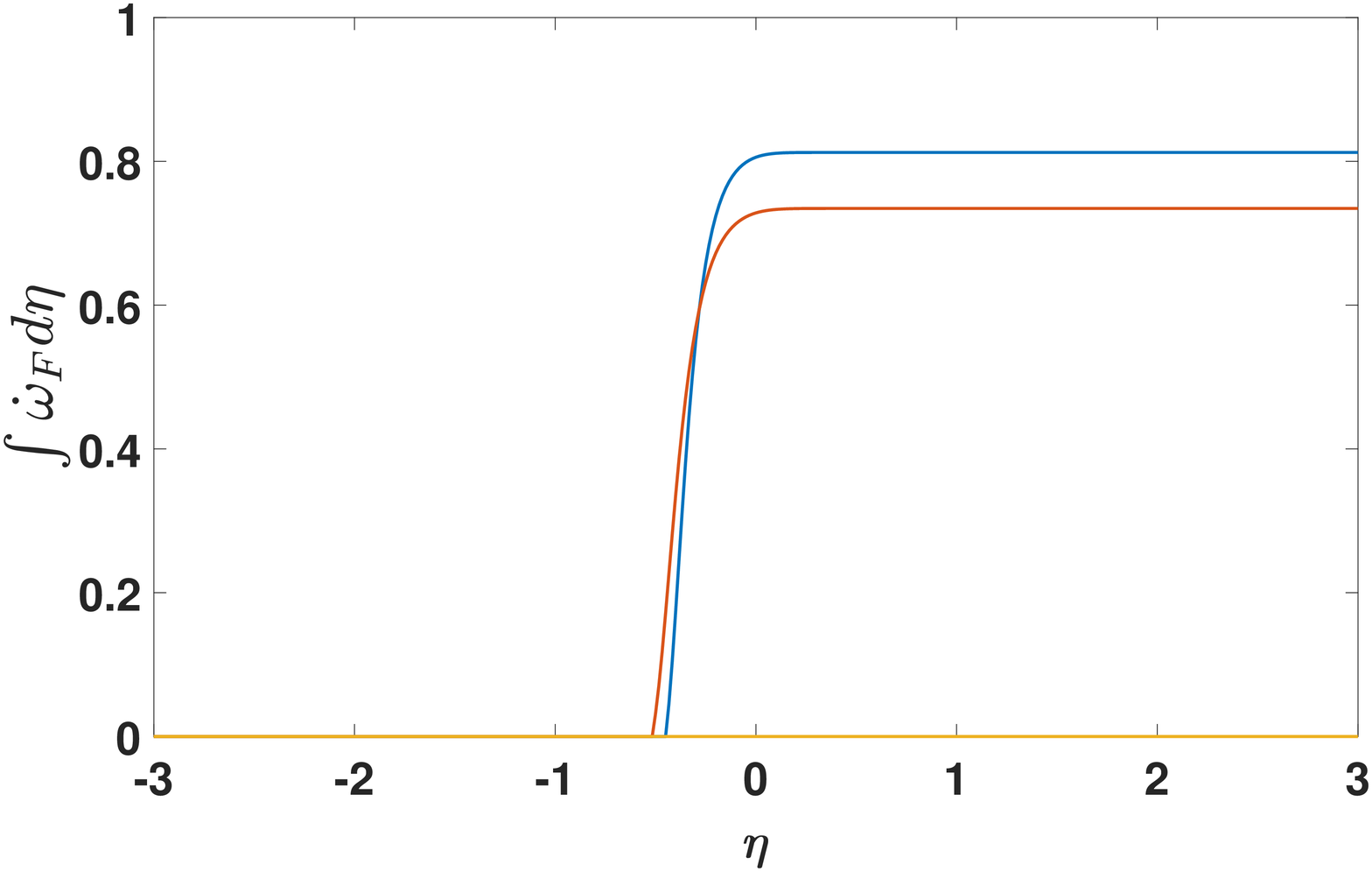}}     \\
  \vspace{-0.1cm}
  \caption{Effects of Prandtl number on scalar properties for diffusion flame. \\    $K= 0.275; \;  \omega_{\kappa} = 1.0; \; S_1 =-1.00;  \; S_2 = 2.00. $ \;  \; $Pr = 1.3$: \; blue, \; flame.
  \; $Pr =1.0 $:  \;   red,  \; flame.  \; $Pr =0.7$:  \;  \;  orange, \; no flame. }
  \label{Prandtl1}
  \end{figure}

\begin{figure}[thbp]
  \centering
 \subfigure[mass flux per area, $f=\rho u_{\chi}$ ]{
  \includegraphics[height = 4.6cm, width=0.45\linewidth]{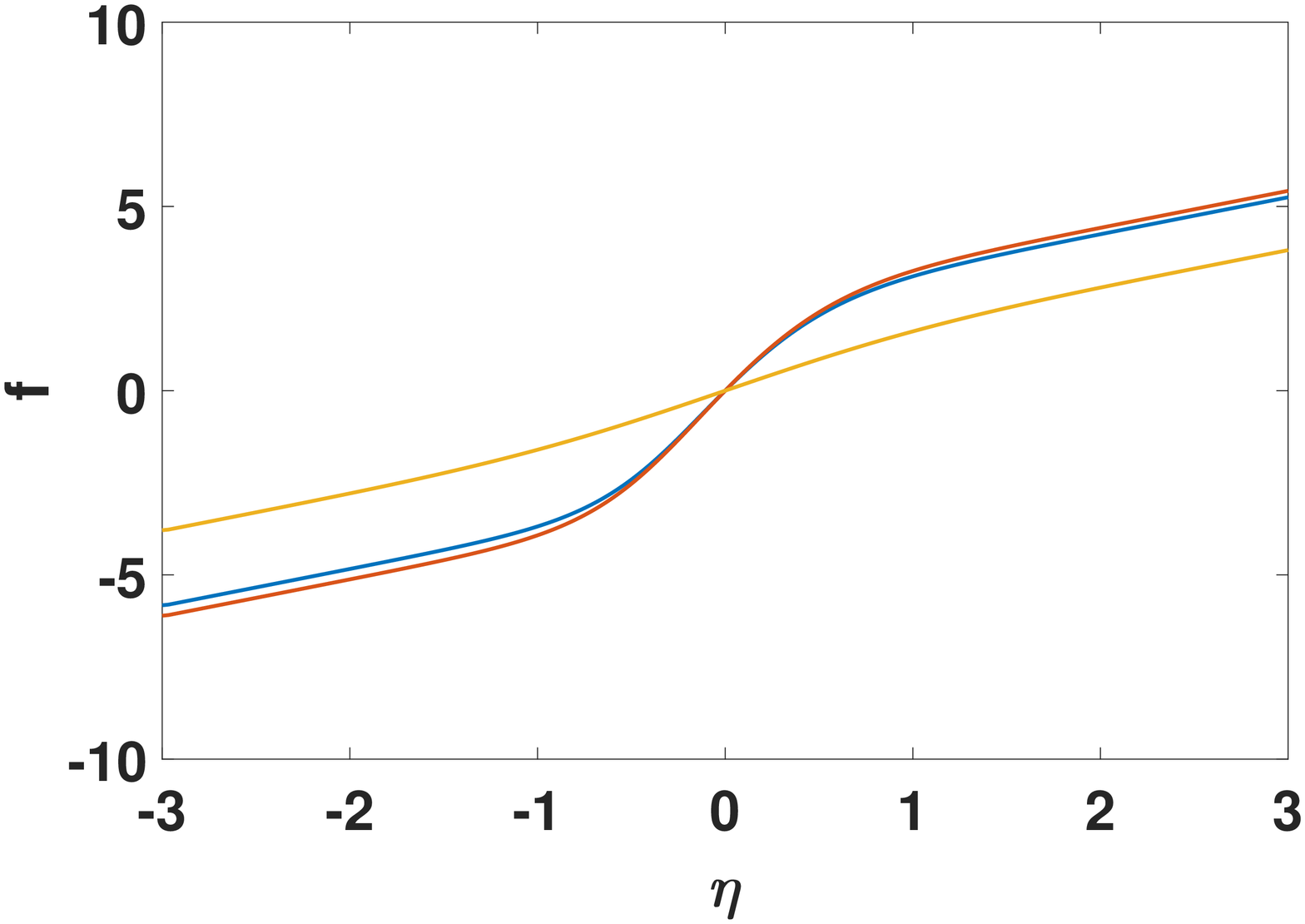}}
  \subfigure[velocity component, $f_1' = u_{\xi}/(S_1 \xi)$]{
  \includegraphics[height = 4.6cm, width=0.45\linewidth]{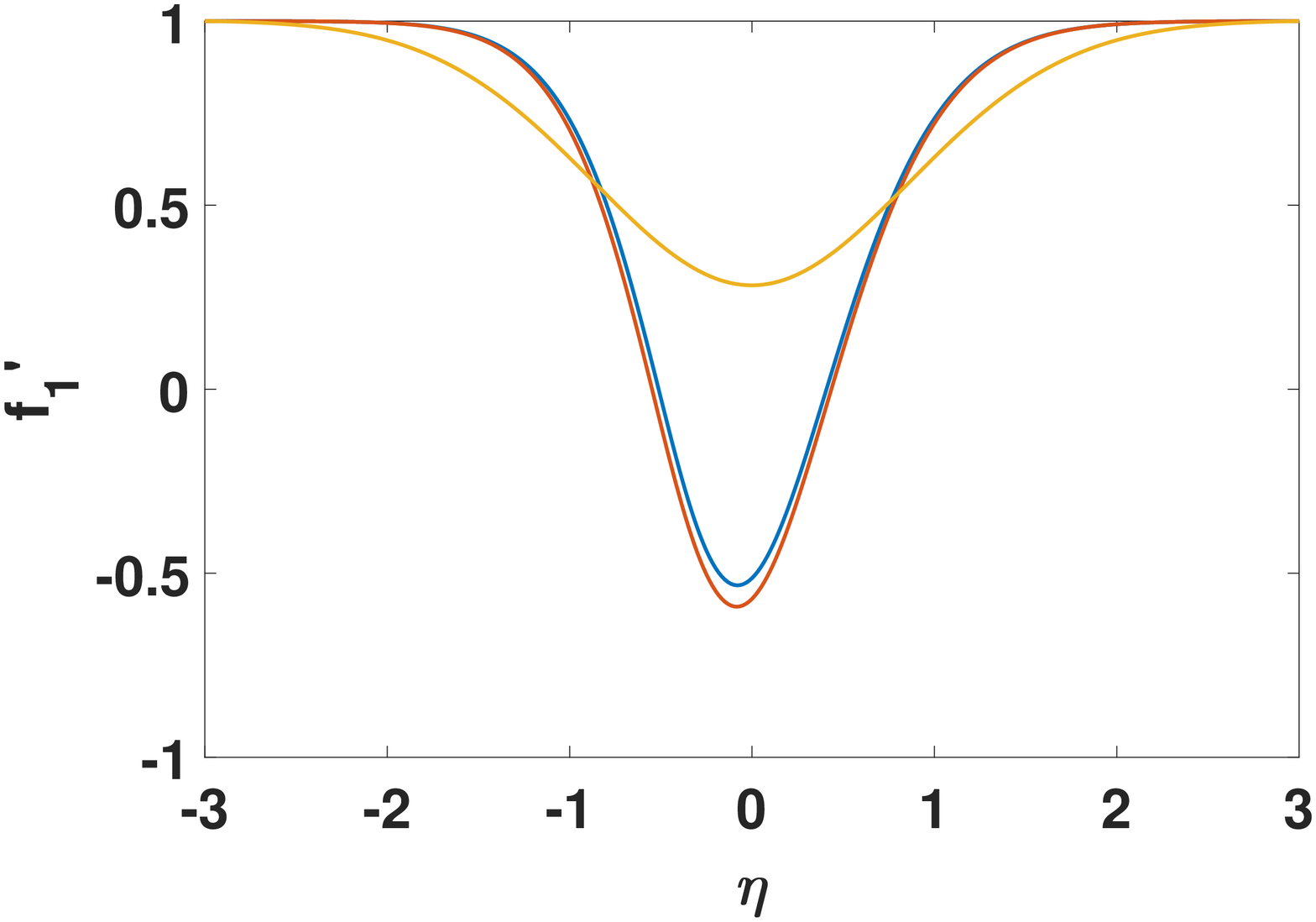}}     \\
  \vspace{0.2cm}
  \subfigure[ velocity component, $f_2' = w/(S_2z)$]{
  \includegraphics[height = 4.6cm, width=0.45\linewidth]{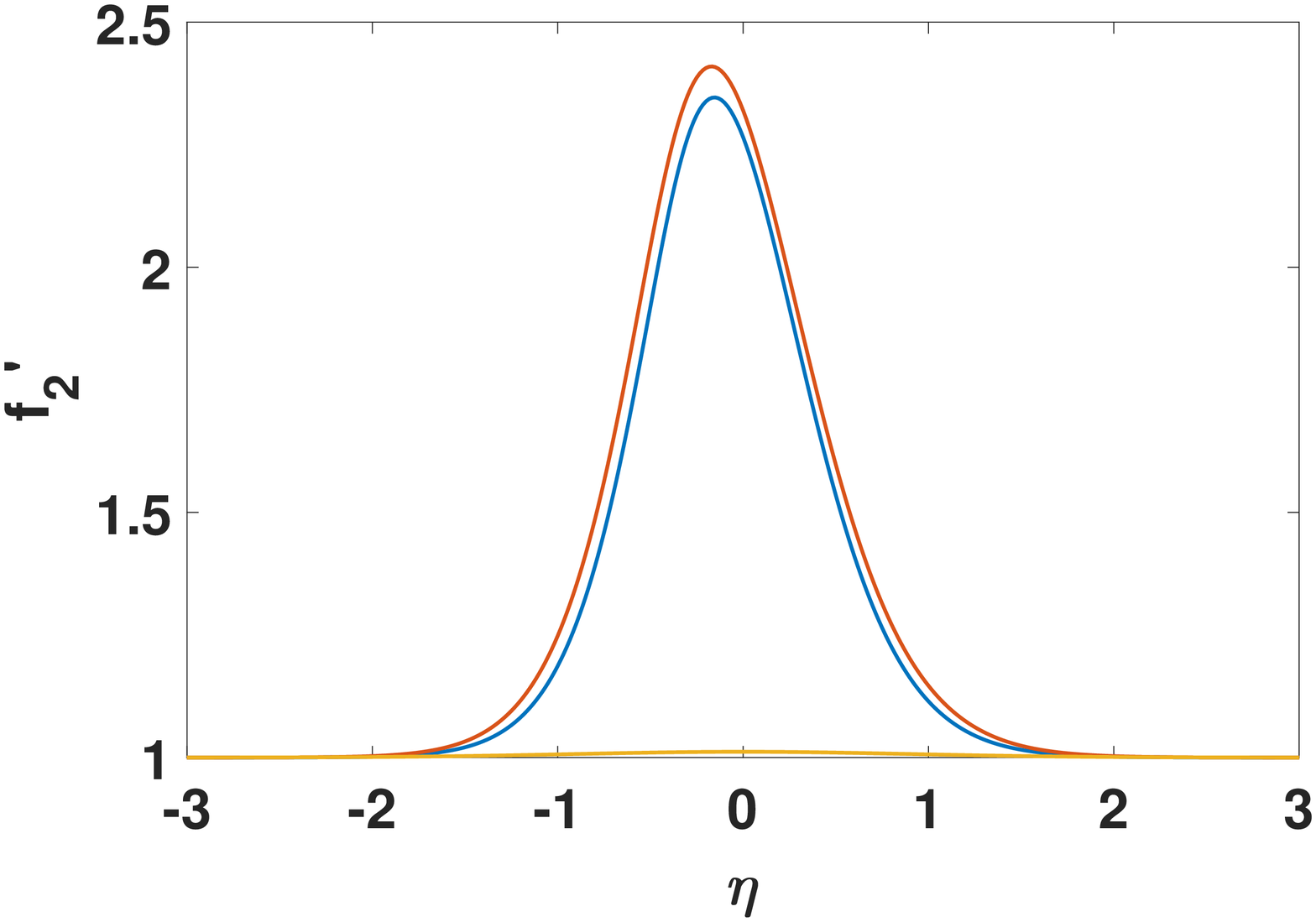}}
  \subfigure[velocity component, $u_{\chi}$]{
  \includegraphics[height = 4.6cm, width=0.45\linewidth]{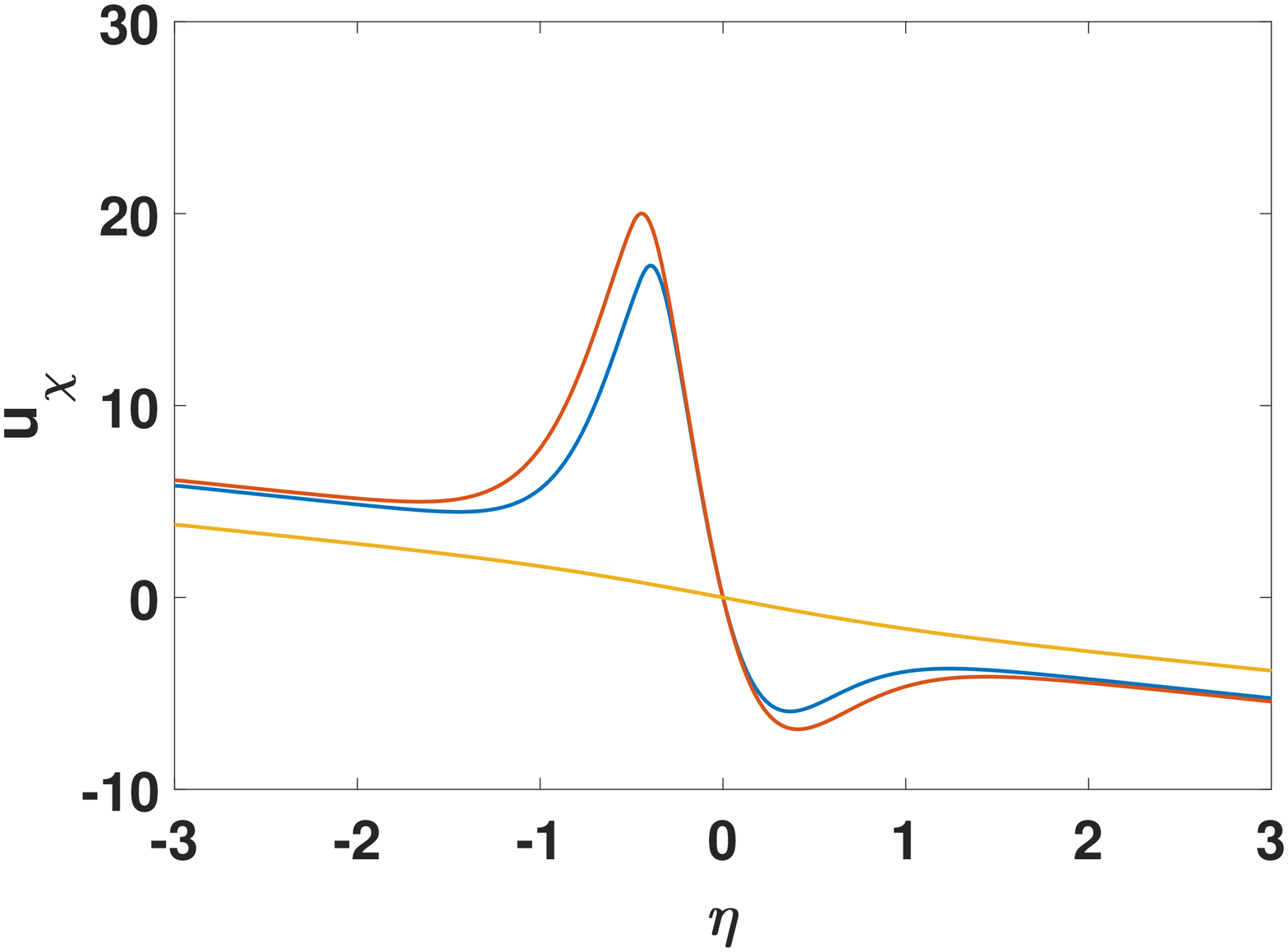}}     \\
  \vspace{-0.1cm}
  \caption{Effects of Prandtl number on flow properties for diffusion flame. \\    $K= 0.275; \;  \omega_{\kappa} = 1.0; \; S_1 =-1.00;  \; S_2 = 2.00. $ \;  \; $Pr = 1.3$: \; blue, \; flame, \; flow reversal.
  \; $Pr =1.0$:  \;   red,  \; flame,\; flow reversal.  \; $Pr =0.7$:  \;  \;  orange, \; no flame, \; no flow reversal.}
  \label{Prandtl2}
\end{figure}

The sensitivity to thermal and mass diffusivities is shown in Figures \ref{Prandtl1} and \ref{Prandtl2}. These diffusivities increase as Prandtl number $Pr$ decreases. Thus, higher $Pr$ results in thinner diffusion layers as shown in the figures However, when the diffusivity is too large, heat is carried away over too large a domain to maintain a flame.
\newpage

\subsection{Premixed Flamelet Calculations}\label{PremixedFlame}

\begin{figure}[thbp]
  \centering
 \subfigure[enthalpy, $h/h_{\infty}$ ]{
  \includegraphics[height = 4.6cm, width=0.45\linewidth]{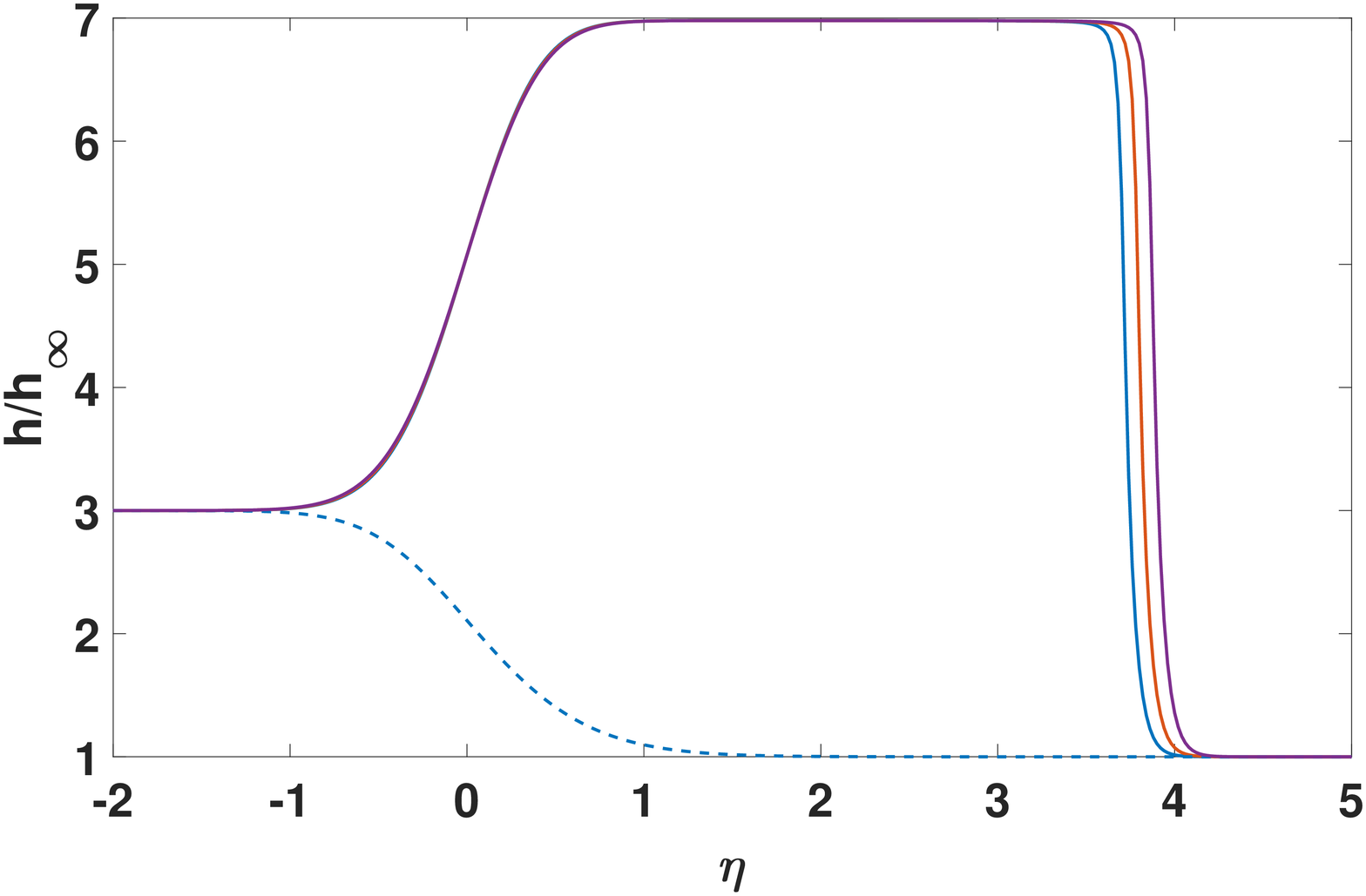}}
  \subfigure[fuel mass fraction, $Y_F$]{
  \includegraphics[height = 4.6cm, width=0.45\linewidth]{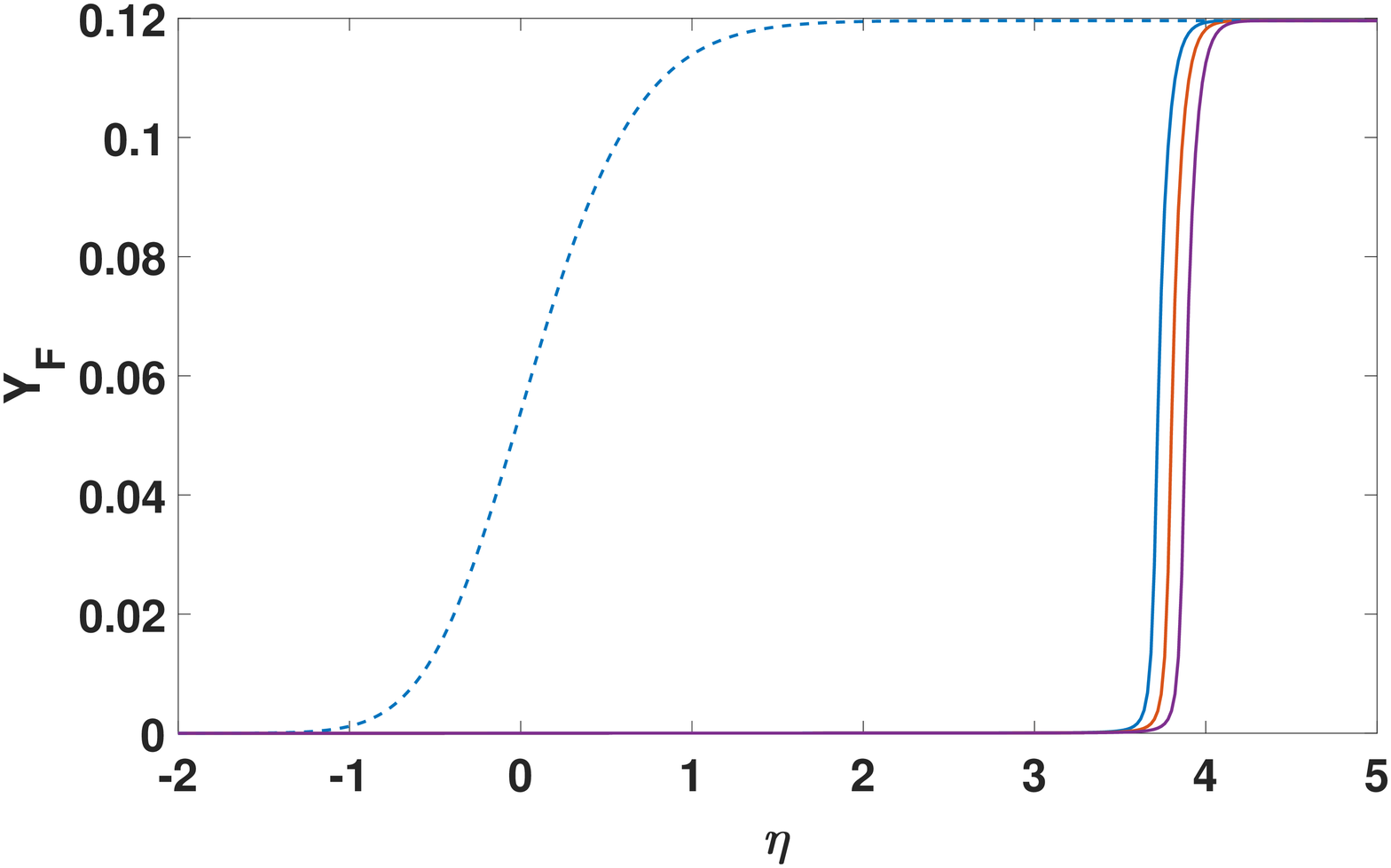}}     \\
  \vspace{0.2cm}
  \subfigure[ mass ratio x oxygen mass fraction, $\nu Y_O$]{
  \includegraphics[height = 4.6cm, width=0.45\linewidth]{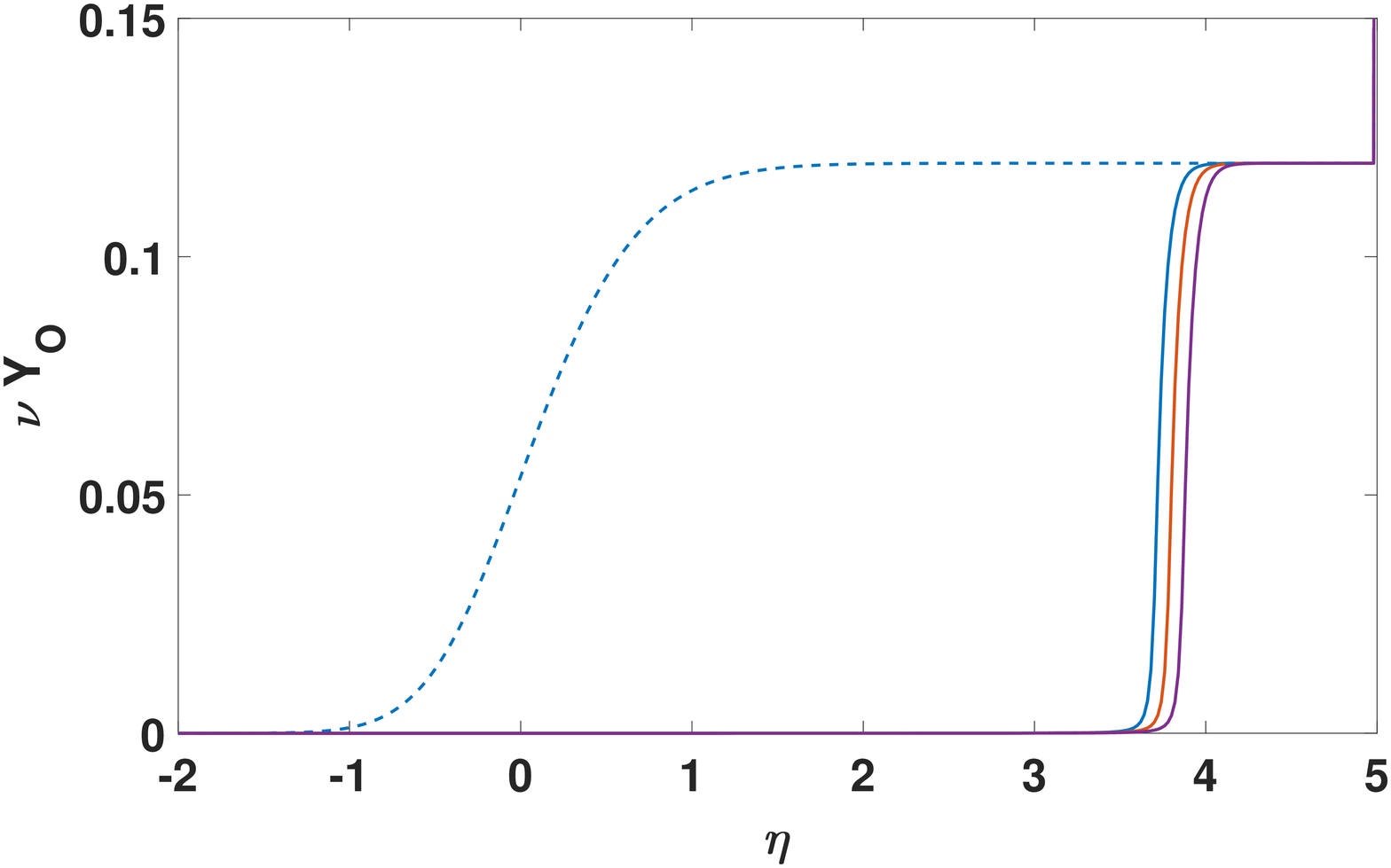}}
  \subfigure[integral of reaction rate, $\int \dot{\omega}_F d \eta$]{
  \includegraphics[height = 4.6cm, width=0.45\linewidth]{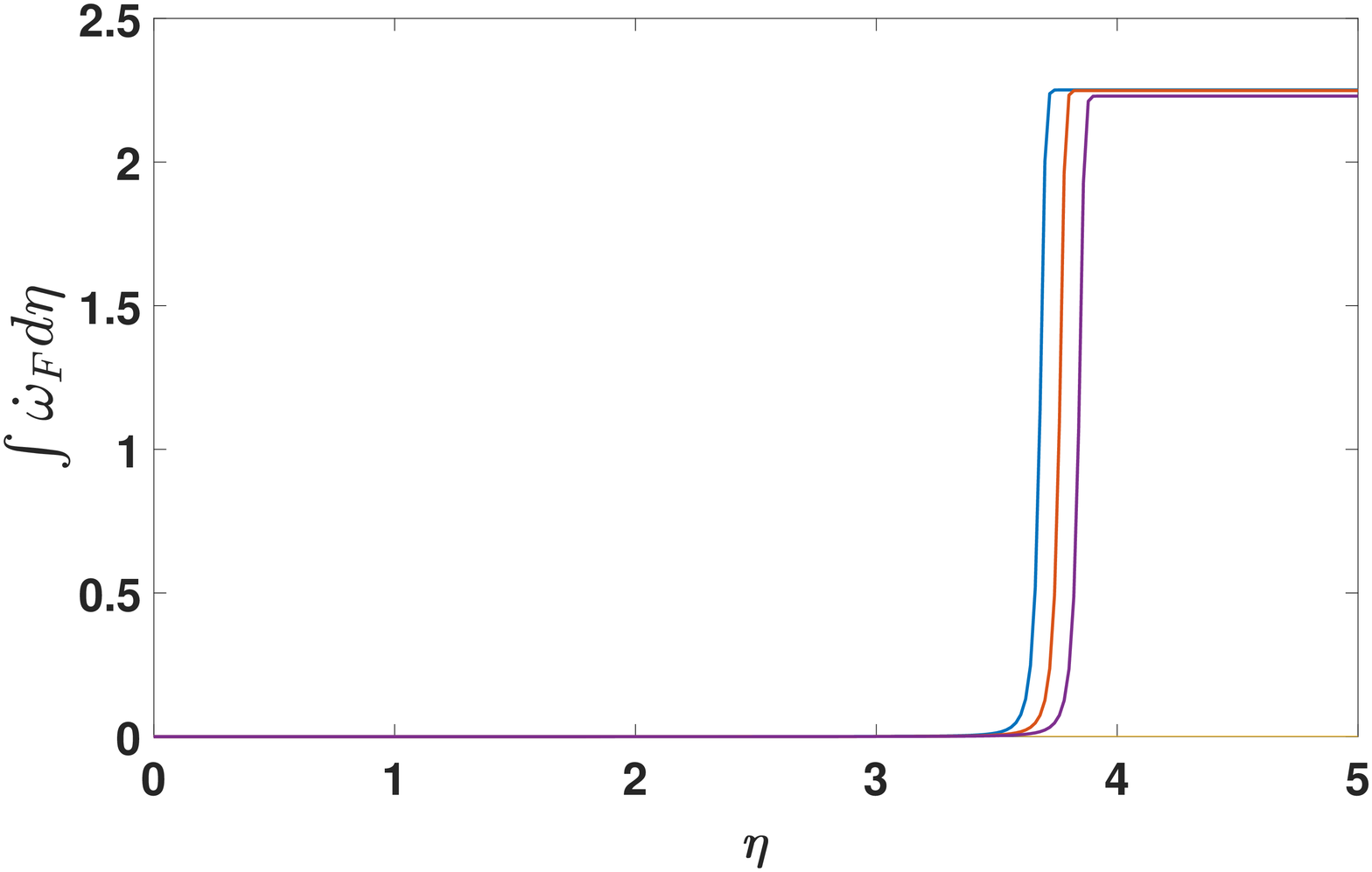}}     \\
  \vspace{-0.1cm}
  \caption{Scalar properties for premixed flame with varying vorticity. \\ $S_1 =-1.00,  \; S_2 = 2.00$. \;   $K = 3.00,\; \omega_{\kappa} =0 $, \; solid blue, flame; \; $K=3.00, \; \omega_{\kappa} =1.0$,\; red, flame; \;  $K=2.95, \; \omega_{\kappa} =1.0$,\; dashed blue, no flame; \;$K =2.95, \; \omega_{\kappa} =1.5$, \; purple, flame.}
  \label{PreFlame1}
\end{figure}

\begin{figure}[thbp]
  \centering
 \subfigure[mass flux per area, $f=\rho u_{\chi}$ ]{
 \includegraphics[height = 4.8cm, width=0.40\linewidth]{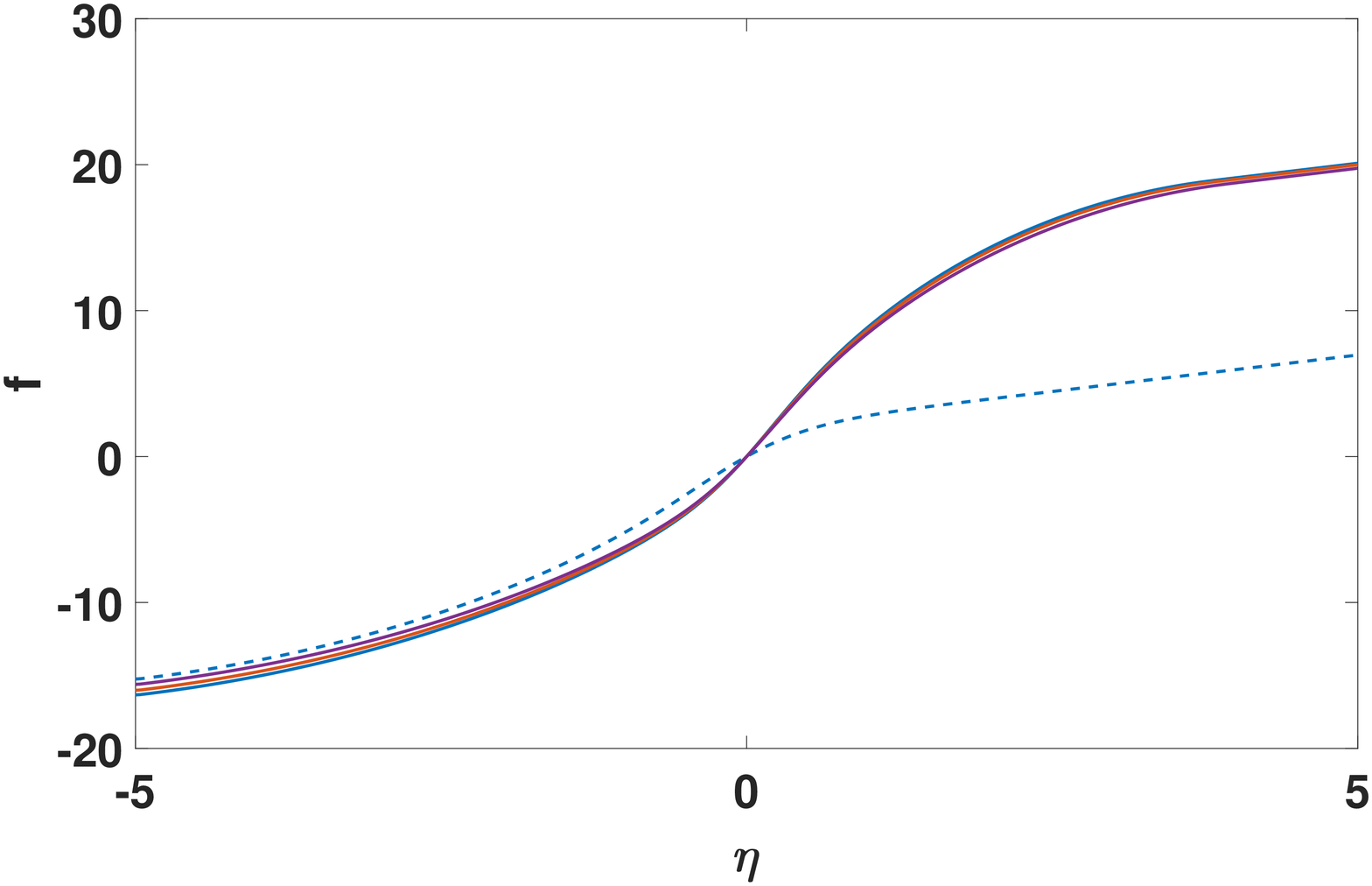}}
 \subfigure[velocity component, $f_1' = u_{\xi}/(S_1 \xi)$]{
  \includegraphics[height = 4.8cm, width=0.45\linewidth]{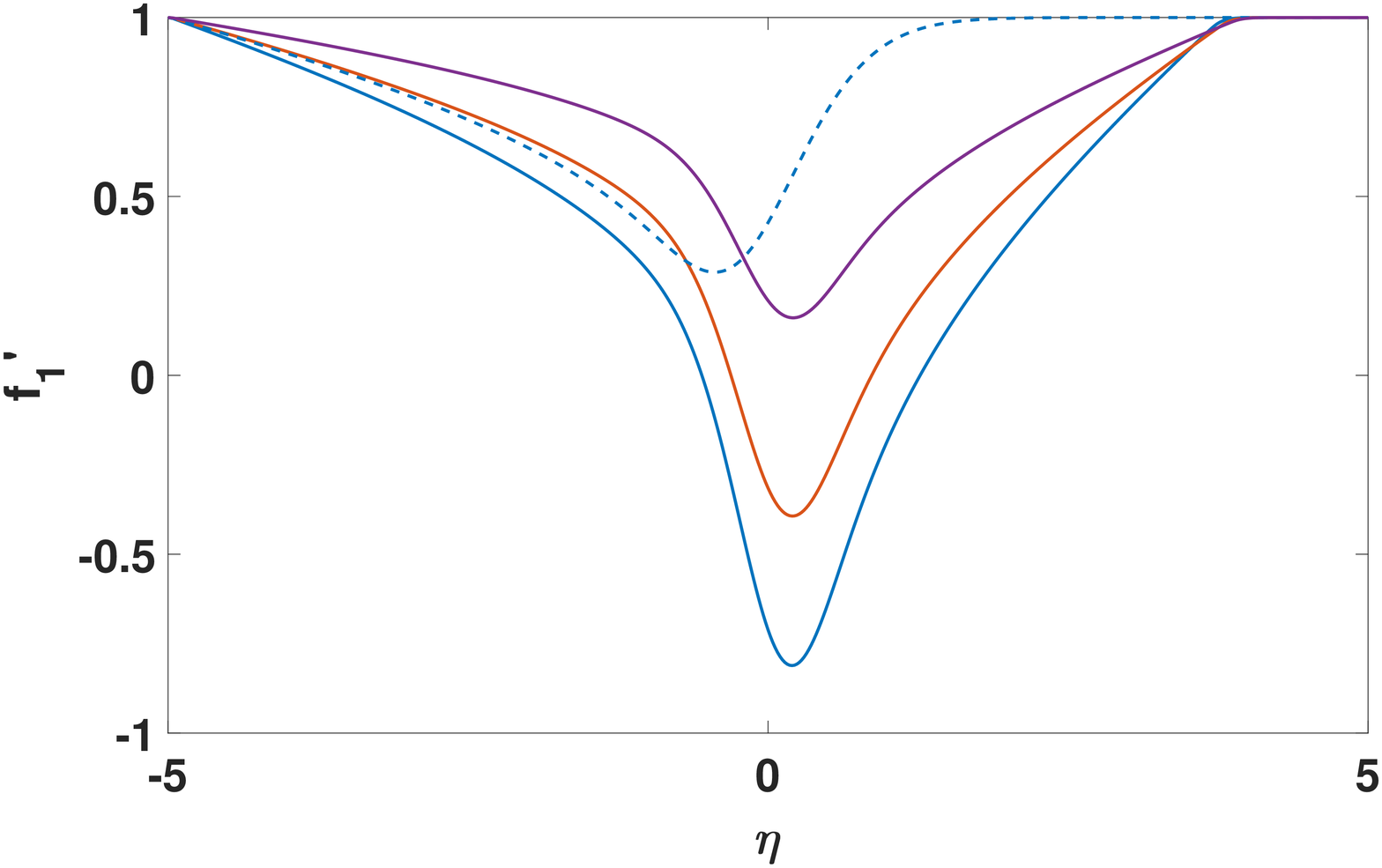}}     \\
  \vspace{0.2cm}
  \subfigure[ velocity component, $f_2' = w/(S_2z)$]{
  \includegraphics[height = 4.8cm, width=0.45\linewidth]{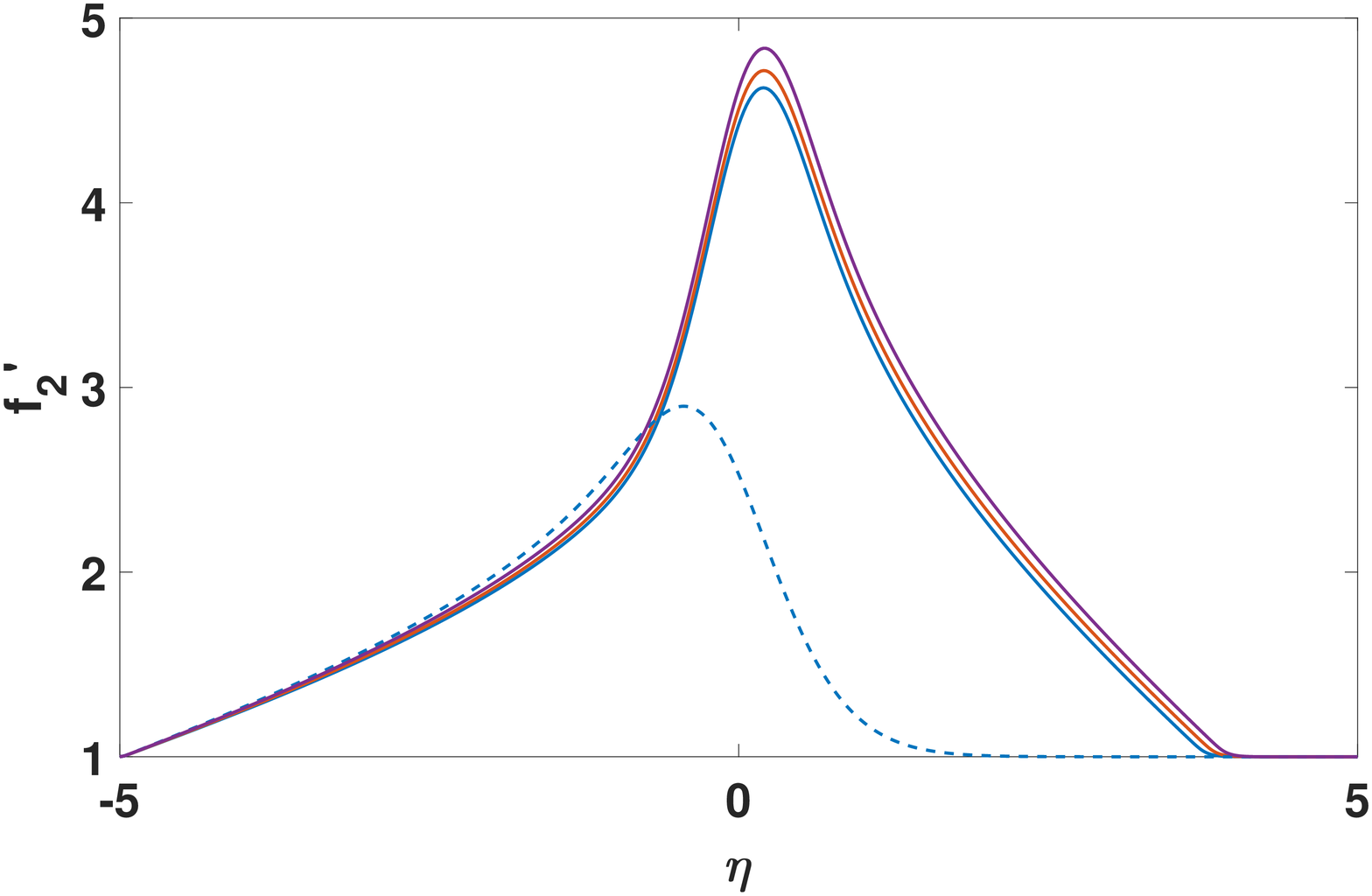}}
  \subfigure[velocity component, $u_{\chi}$]{
  \includegraphics[height = 4.8cm, width=0.45\linewidth]{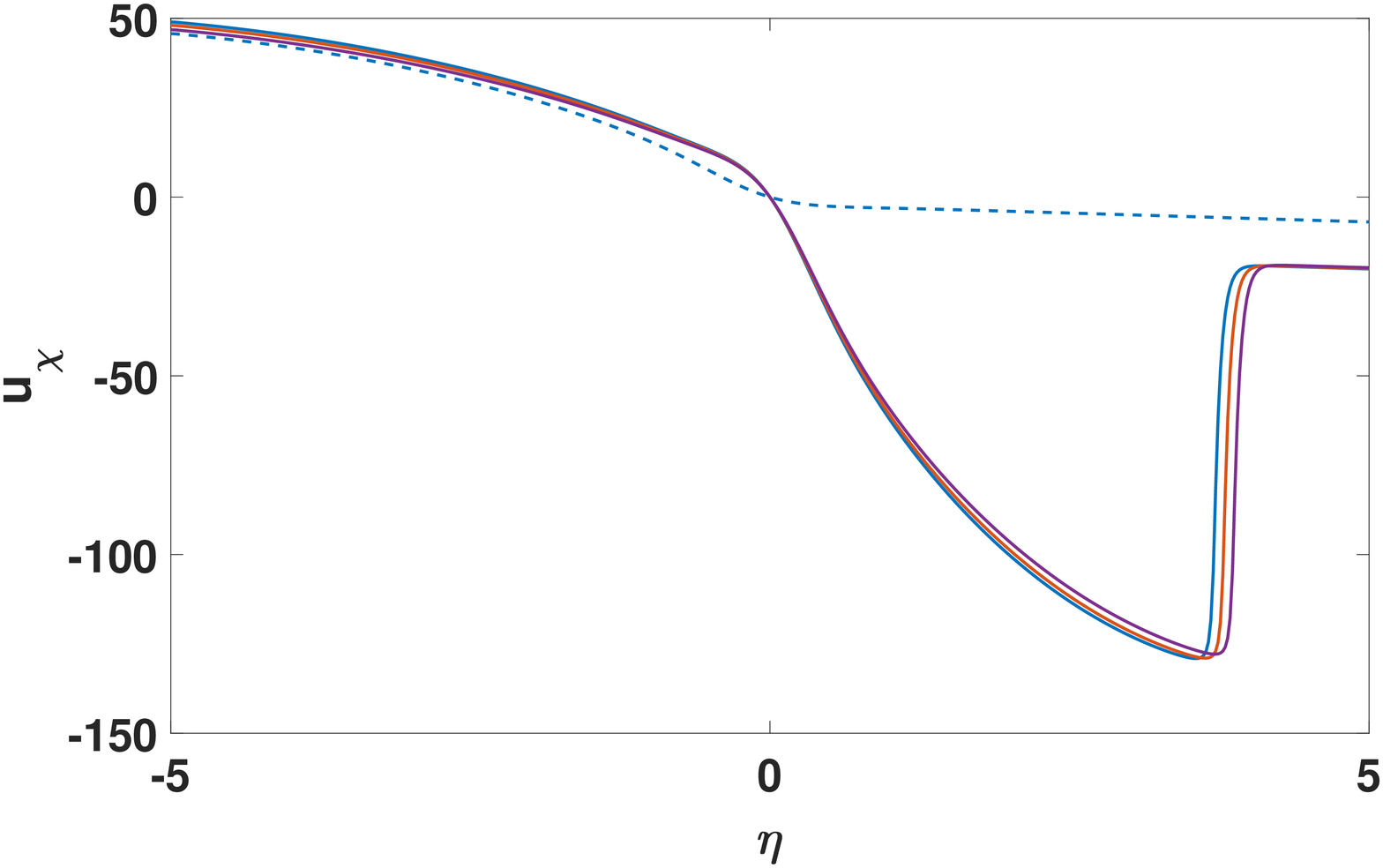}}     \\
  \vspace{0.2cm}
   \caption{Velocity behavior for premixed flame with varying vorticity . \\ $S_1 =-1.00, \; S_2 = 2.00$. \;   $K = 3.00,\; \omega_{\kappa} =0 $, \; solid blue, flame, flow reversal; \; $K=3.00, \; \omega_{\kappa} =1.0$,\; red, flame,  less flow reversal; \;  $K=2.95, \; \omega_{\kappa} =1.0$,\; dashed blue, no flame; \; $K =2.95, \; \omega_{\kappa} =1.5$, \; purple, flame, no flow reversal.}
  \label{PreFlame2}
\end{figure}

The analysis can apply to a situation where the inward swirling fluid has opposing streams of a combustible mixture and a hot inert gas (likely combustion products). A premixed flame can be established. The vorticity and centrifugal motion can have consequence, especially near a flammability limit. Figures \ref{PreFlame1} and \ref{PreFlame2} have some interesting results.
At a $Da$ value of three times the reference value, i.e., $K = 3.00$, a strong premixed flame is shown to exist with or without  a vorticity field. Application of  swirl through the vorticity results in the same peak enthalpy or temperature and flame speed (in the $\chi$ direction).  The $\xi$ component of velocity is seen to reverse direction with or without imposed vorticity. However, the increase in rotational rate and  centrifugal acceleration decreases the magnitude of the reversal. The premixed flame moves slightly farther upstream as measured by the $\eta$ value as swirl is applied. It likely occurs because, with less reversal in the $\xi$ direction, the expansion in the $\eta$ direction is enhanced.
\newpage
A slight decrease in $Da$ to a situation where $K = 2.95$ results in extinction with vorticity in the range up to $\omega_{\kappa} = 1.0$.  With further increase of the centrifugal acceleration through the increase of vorticity to the value $\omega_{\kappa} = 1.5$, a strong premixed flame is created. It has the same flame speed and peak temperature. However, there is no flow reversal and the flame stands further upstream than the flames with higher $K$ value.
\newpage

\subsection{Multi-branched Flamelet Calculations}\label{MultiFlame}

\begin{figure}[thbp]
  \centering
 \subfigure[enthalpy, $h/h_{\infty}$ ]{
  \includegraphics[height = 4.6cm, width=0.45\linewidth]{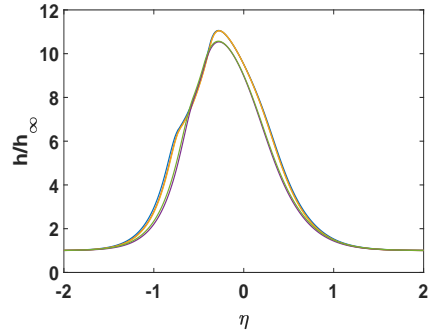}}
  \subfigure[fuel mass fraction, $Y_F$]{
  \includegraphics[height = 4.6cm, width=0.45\linewidth]{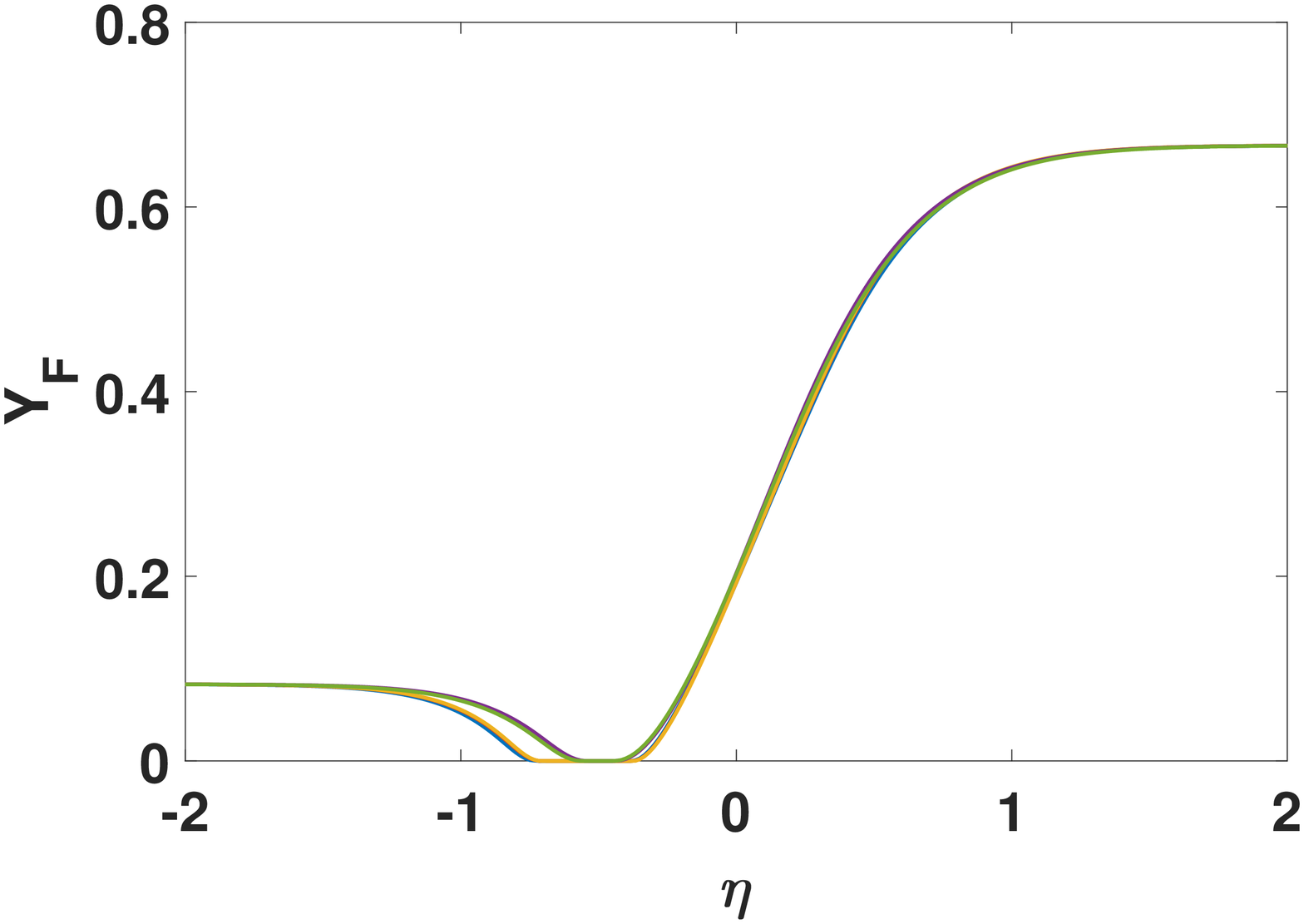}}     \\
  \vspace{0.1cm}
  \subfigure[ mass ratio x oxygen mass fraction, $\nu Y_O$]{
  \includegraphics[height = 4.6cm, width=0.45\linewidth]{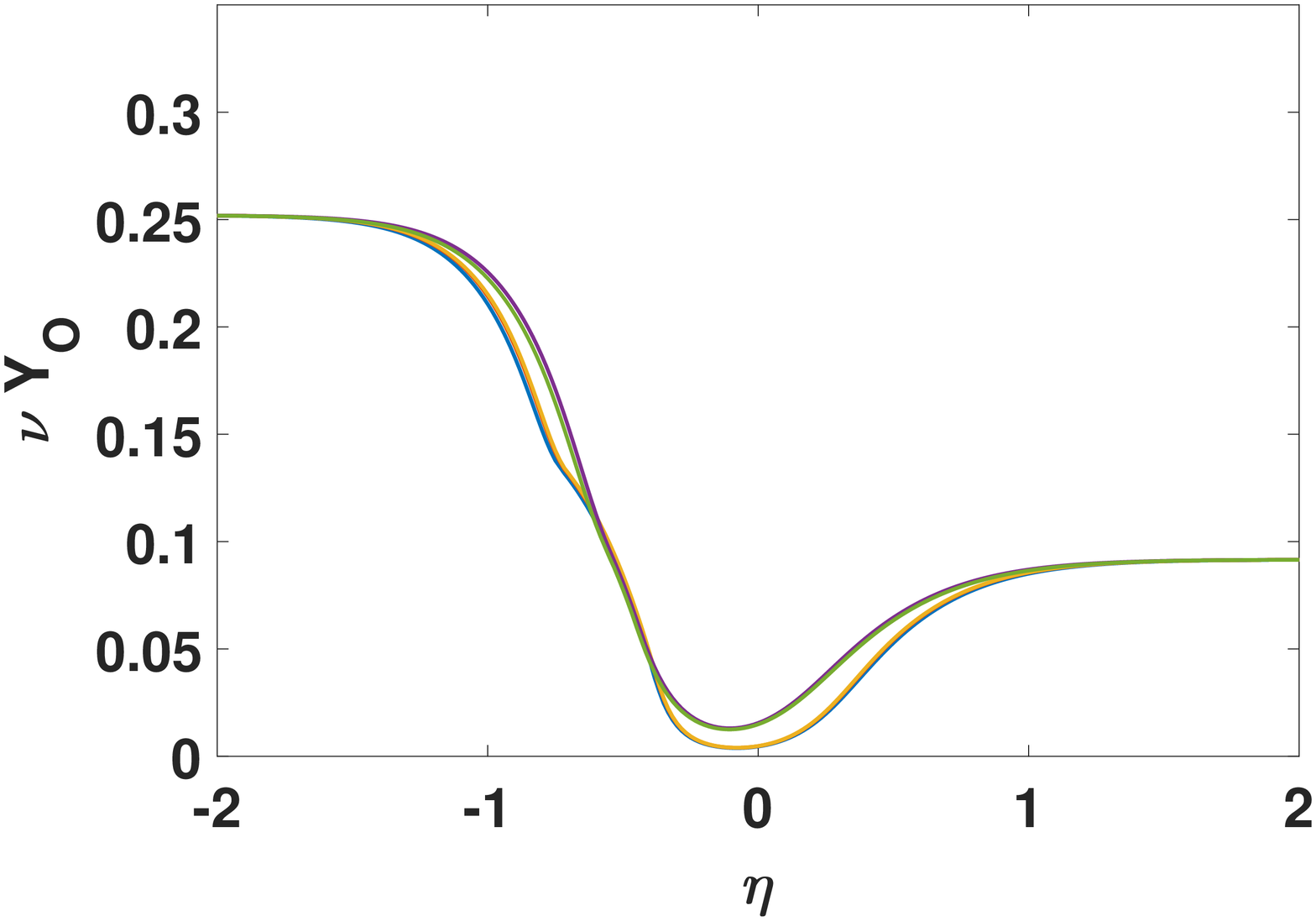}}
  \subfigure[integral of reaction rate, $\int \dot{\omega}_F d \eta$]{
  \includegraphics[height = 4.6cm, width=0.45\linewidth]{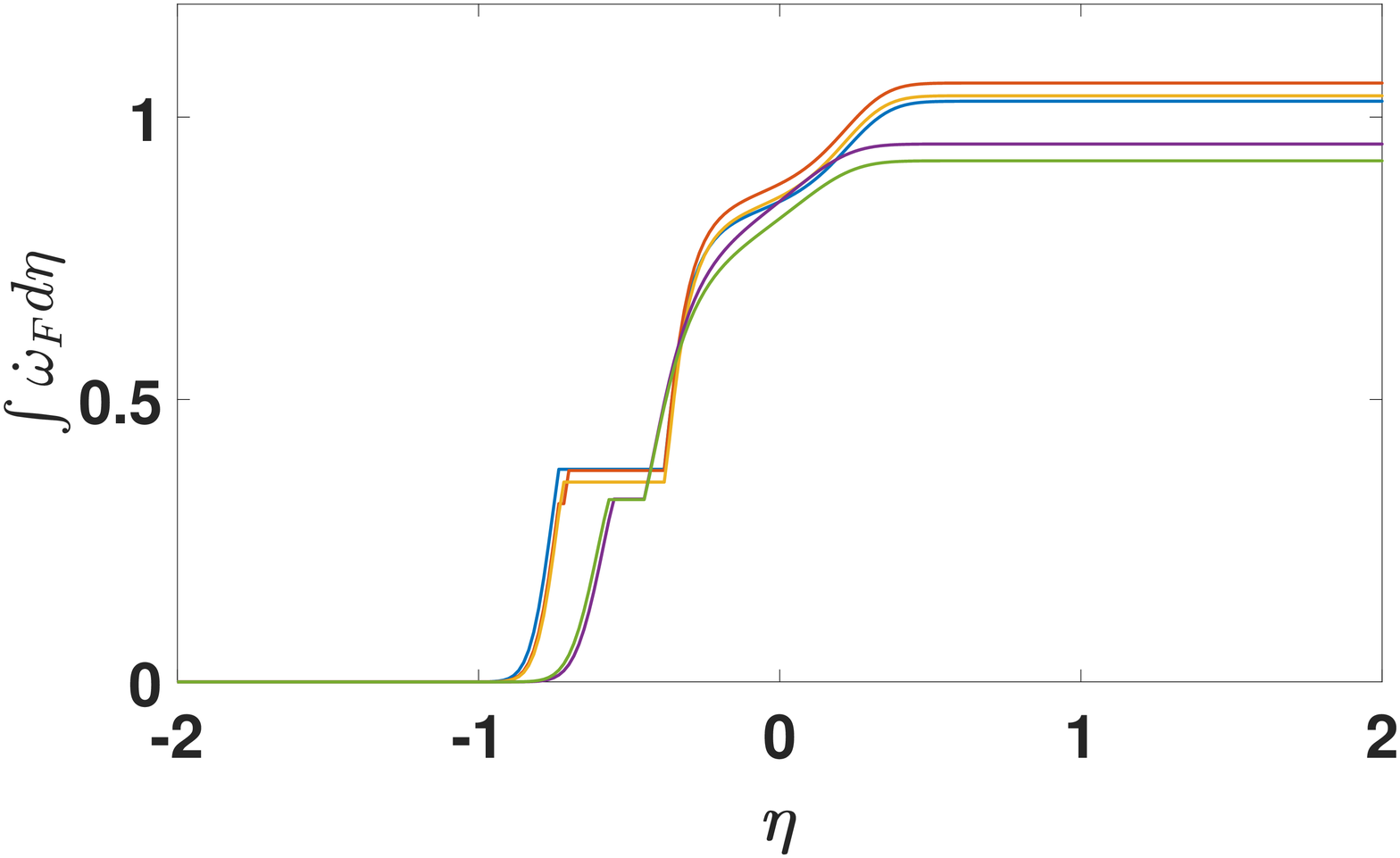}}     \\
  \subfigure[ enthalpy,  $h/h_{\infty}$, versus conserved scalar, $\Sigma$]{
  \includegraphics[height = 4.6cm, width=0.45\linewidth]{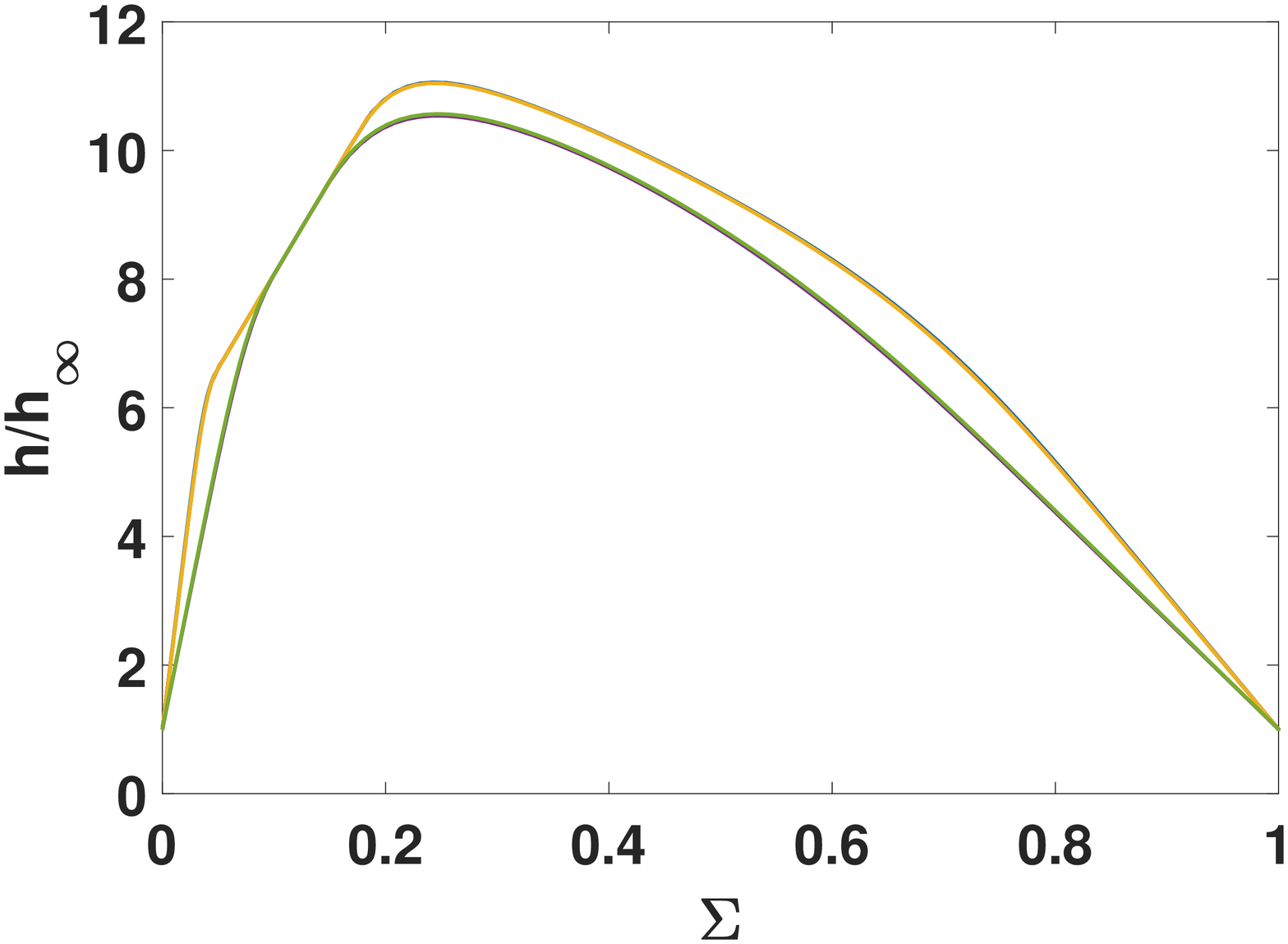}}
  \vspace{-0.2cm}
  \caption{Scalar properties for multibranched flame with varying Damk\"{o}hler number and vorticity. \; $S_1 =-1.00,  \; S_2 = 2.00$. \;  $K =1.00$:   \; $\omega_{\kappa} = 1.0$,  blue; \; $\omega_{\kappa} = 0.50$, red; \; $\omega_{\kappa} = 0$, orange.  \; \; $K =0.300$:   \; $\omega_{\kappa} = 1.0$,  green; \; $\omega_{\kappa} = 0$, purple.}
  \label{f1}
\end{figure}

\begin{figure}[thbp]
  \centering
 \subfigure[mass flux per area, $f=\rho u_{\chi}$ ]{
 \includegraphics[height = 4.8cm, width=0.40\linewidth]{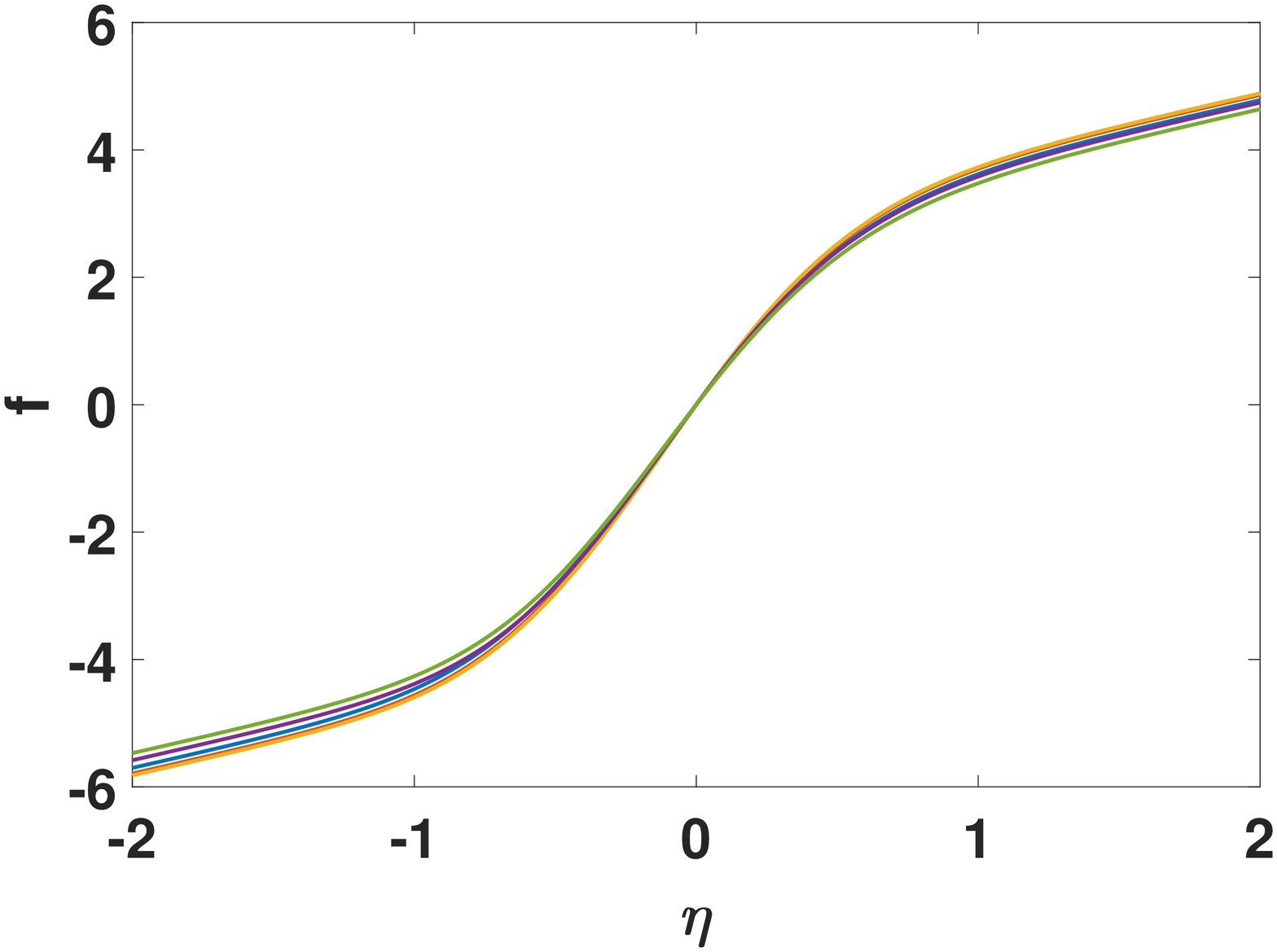}}
 \subfigure[velocity component, $f_1' = u_{\xi}/(S_1 \xi)$]{
  \includegraphics[height = 4.8cm, width=0.45\linewidth]{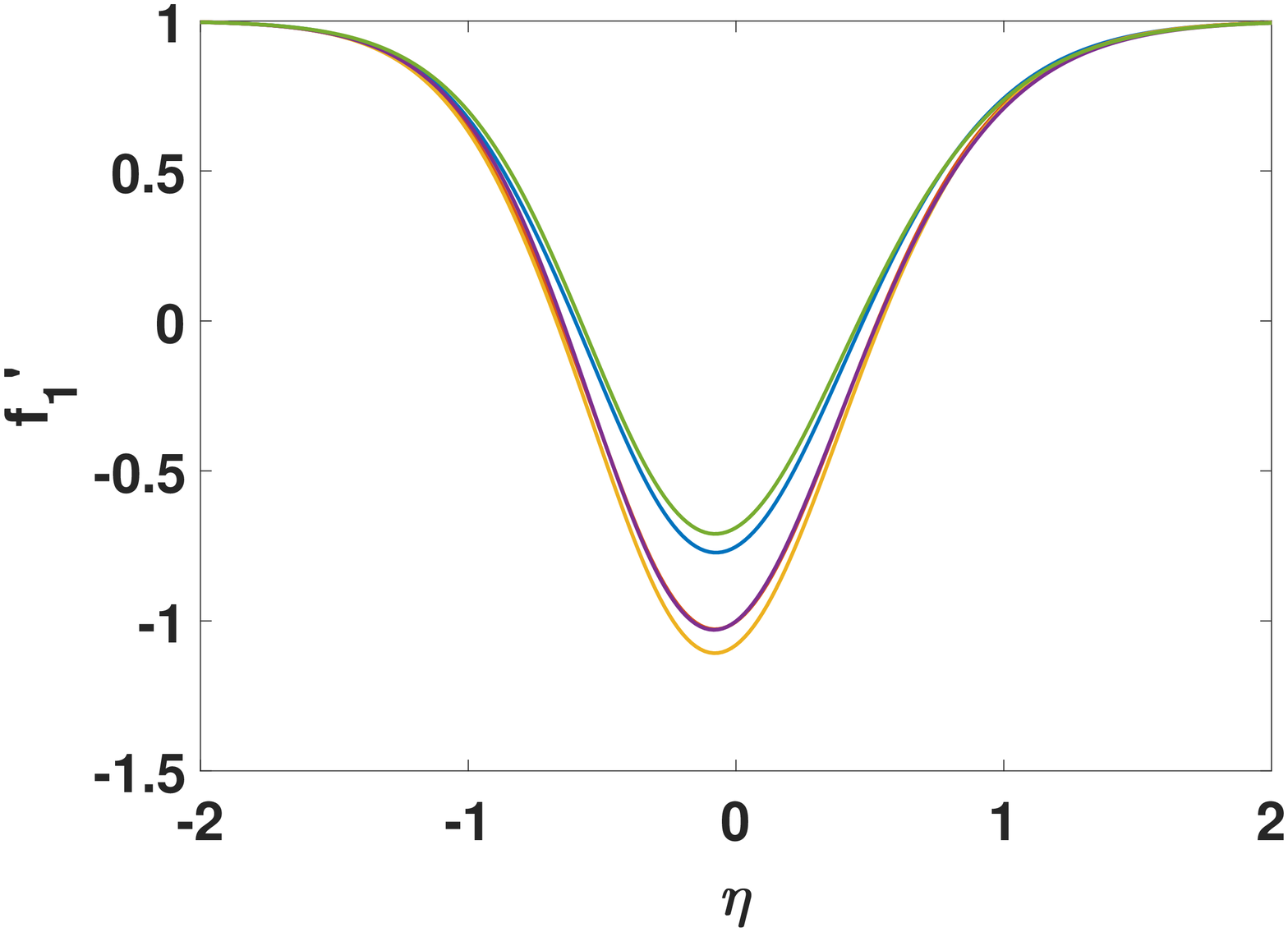}}     \\
  \vspace{0.2cm}
  \subfigure[ velocity component, $f_2' = w/(S_2z)$]{
  \includegraphics[height = 4.8cm, width=0.45\linewidth]{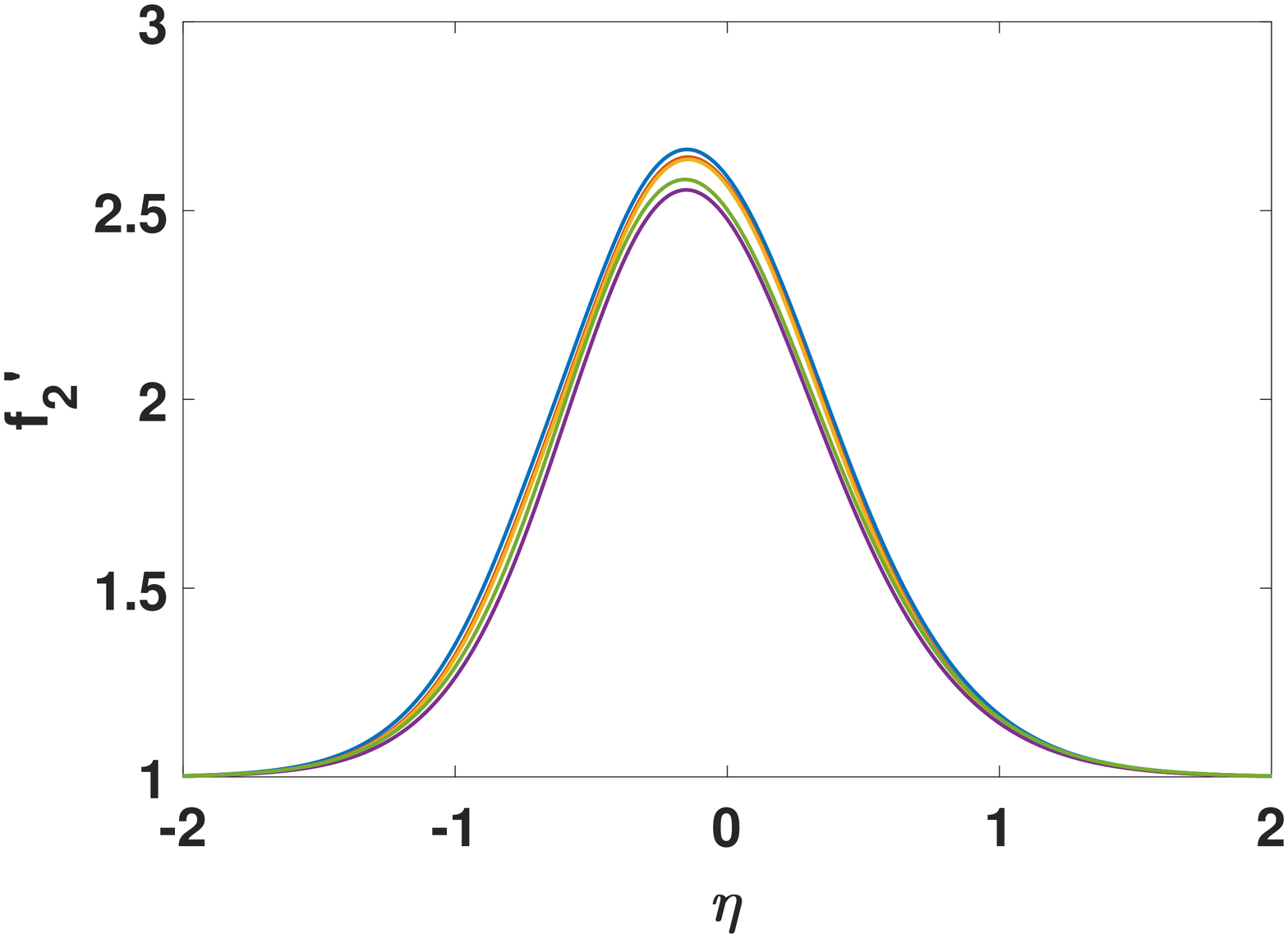}}
  \subfigure[velocity component, $u_{\chi}$]{
  \includegraphics[height = 4.8cm, width=0.45\linewidth]{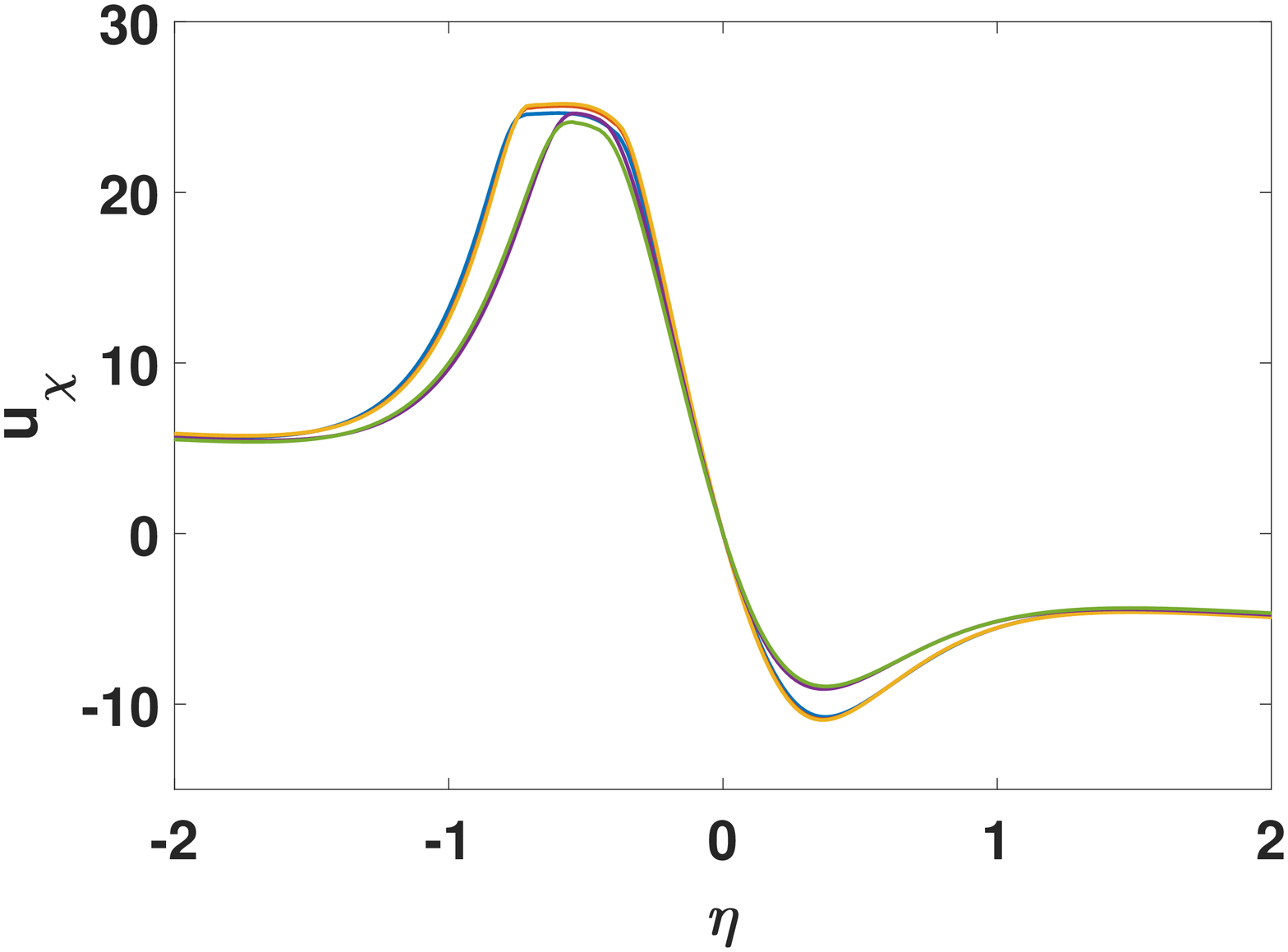}}     \\
  \vspace{0.2cm}
   \caption{Velocity behavior for multibranched flame with varying Damk\"{o}hler number and vorticity. \; $S_1 =-1.00,  \; S_2 = 2.00$. \;  $K =1.00$:   \; $\omega_{\kappa} = 1.0$,  blue \; ; \; $\omega_{\kappa} = 0.50$, red  ; \; $\omega_{\kappa} = 0$, orange.  \; \; $K =0.300$:   \; $\omega_{\kappa} = 1.0$,  green; \; $\omega_{\kappa} = 0$, purple.}
  \label{f2}
\end{figure}

\begin{figure}[thbp]
  \centering
 \subfigure[enthalpy, $h/h_{\infty}$ ]{
  \includegraphics[height = 4.6cm, width=0.45\linewidth]{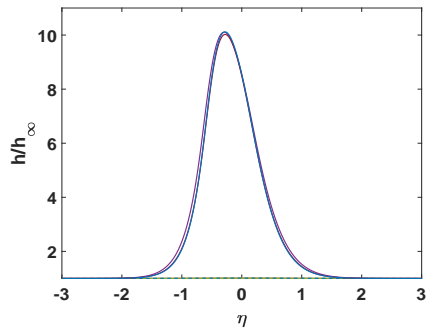}}
  \subfigure[fuel mass fraction, $Y_F$]{
  \includegraphics[height = 4.6cm, width=0.45\linewidth]{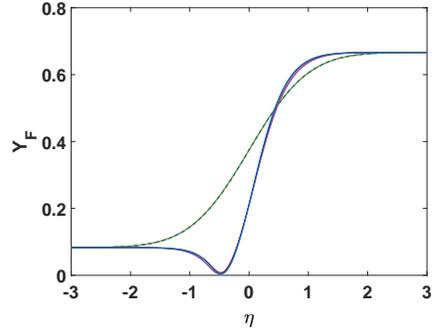}}     \\
  \vspace{0.1cm}
  \subfigure[ mass ratio x oxygen mass fraction, $\nu Y_O$]{
  \includegraphics[height = 4.6cm, width=0.45\linewidth]{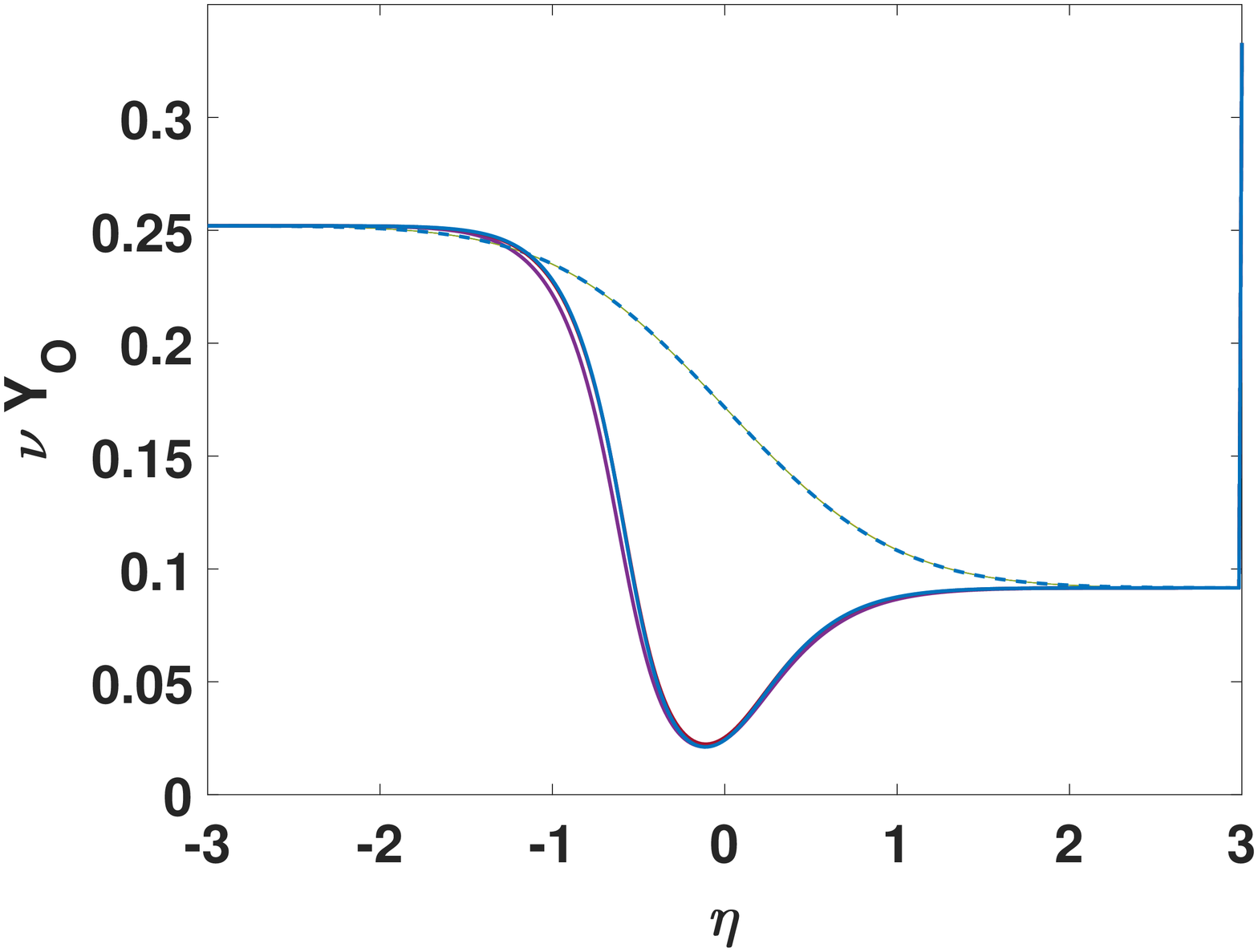}}
  \subfigure[integral of reaction rate, $\int \dot{\omega}_F d \eta$]{
  \includegraphics[height = 4.6cm, width=0.45\linewidth]{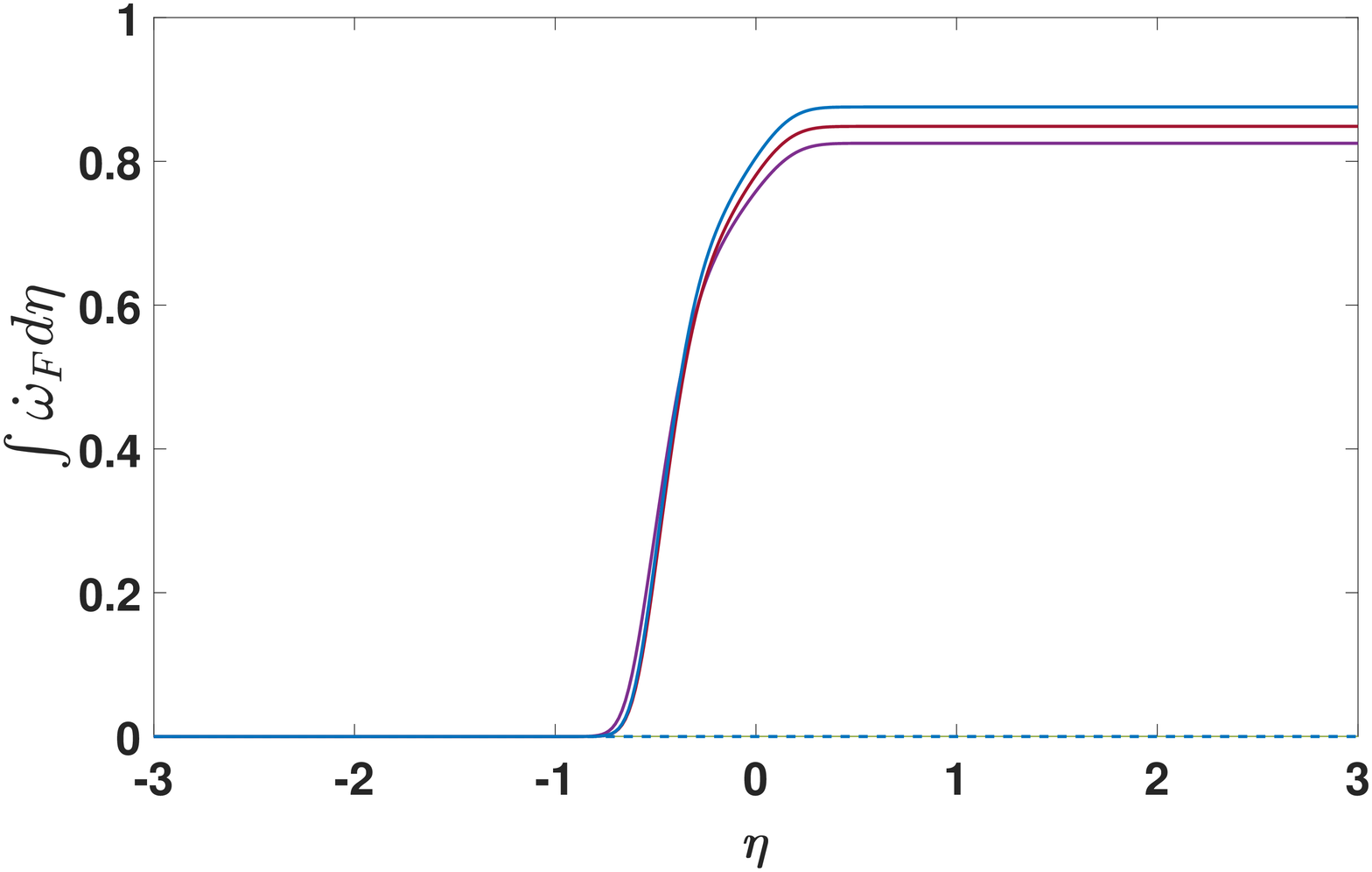}}     \\
  \subfigure[ enthalpy,  $h/h_{\infty}$, versus conserved scalar, $\Sigma$]{
  \includegraphics[height = 4.6cm, width=0.45\linewidth]{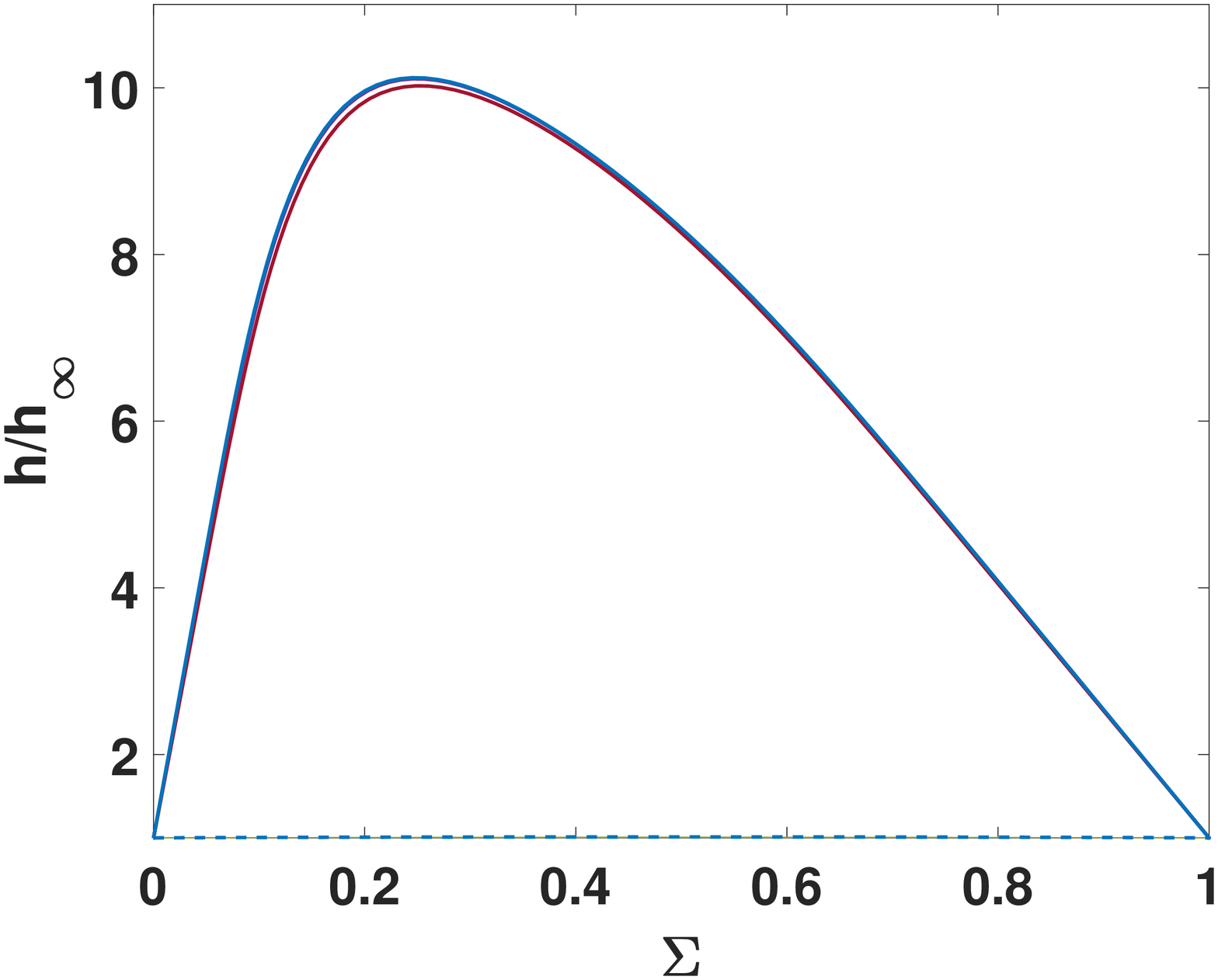}}
  \vspace{-0.2cm}
  \caption{Scalar properties for multibranched flame with varying vorticity. \;\; $S_1 =-1.00,  \; S_2 = 2.00$. \;  Cases with a strong flame: $K = 0.180,\; \omega_{\kappa} =0 $, \; solid blue; \; $K=0.170, \; \omega_{\kappa} = 0.75$,\; red; \;  $K=0.160, \; \omega_{\kappa} =1.50$,\; purple. Other curves show no flame (extinction) and all fall on the dashed blue line: \;$K =0.170, \; \omega_{\kappa} =0.50$;   \;$K =0.160, \; \omega_{\kappa} =1.00$.}
  \label{MultiFlame1}
\end{figure}

\begin{figure}[thbp]
  \centering
 \subfigure[mass flux per area, $f=\rho u_{\chi}$ ]{
 \includegraphics[height = 4.8cm, width=0.40\linewidth]{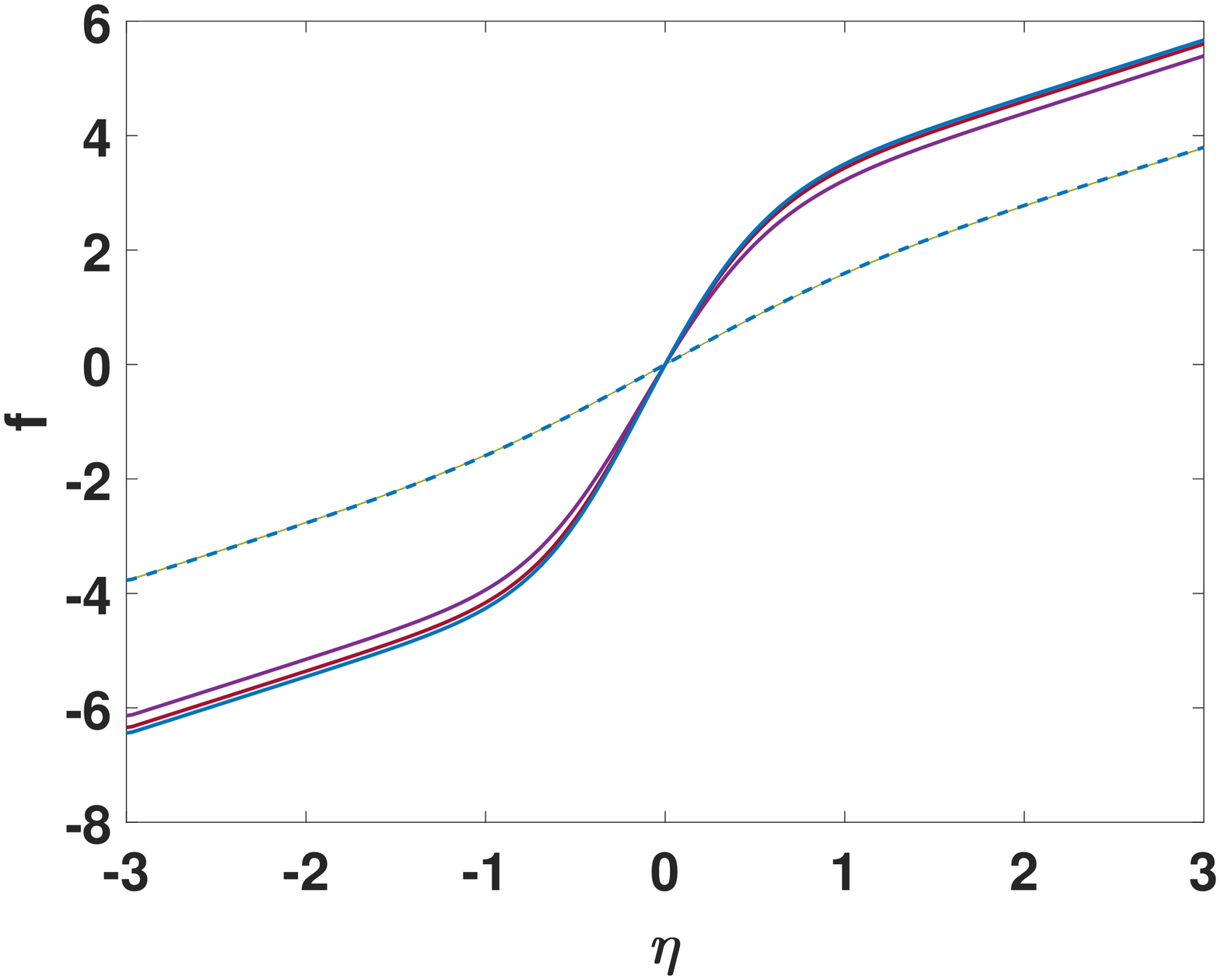}}
 \subfigure[velocity component, $f_1' = u_{\xi}/(S_1 \xi)$]{
  \includegraphics[height = 4.8cm, width=0.45\linewidth]{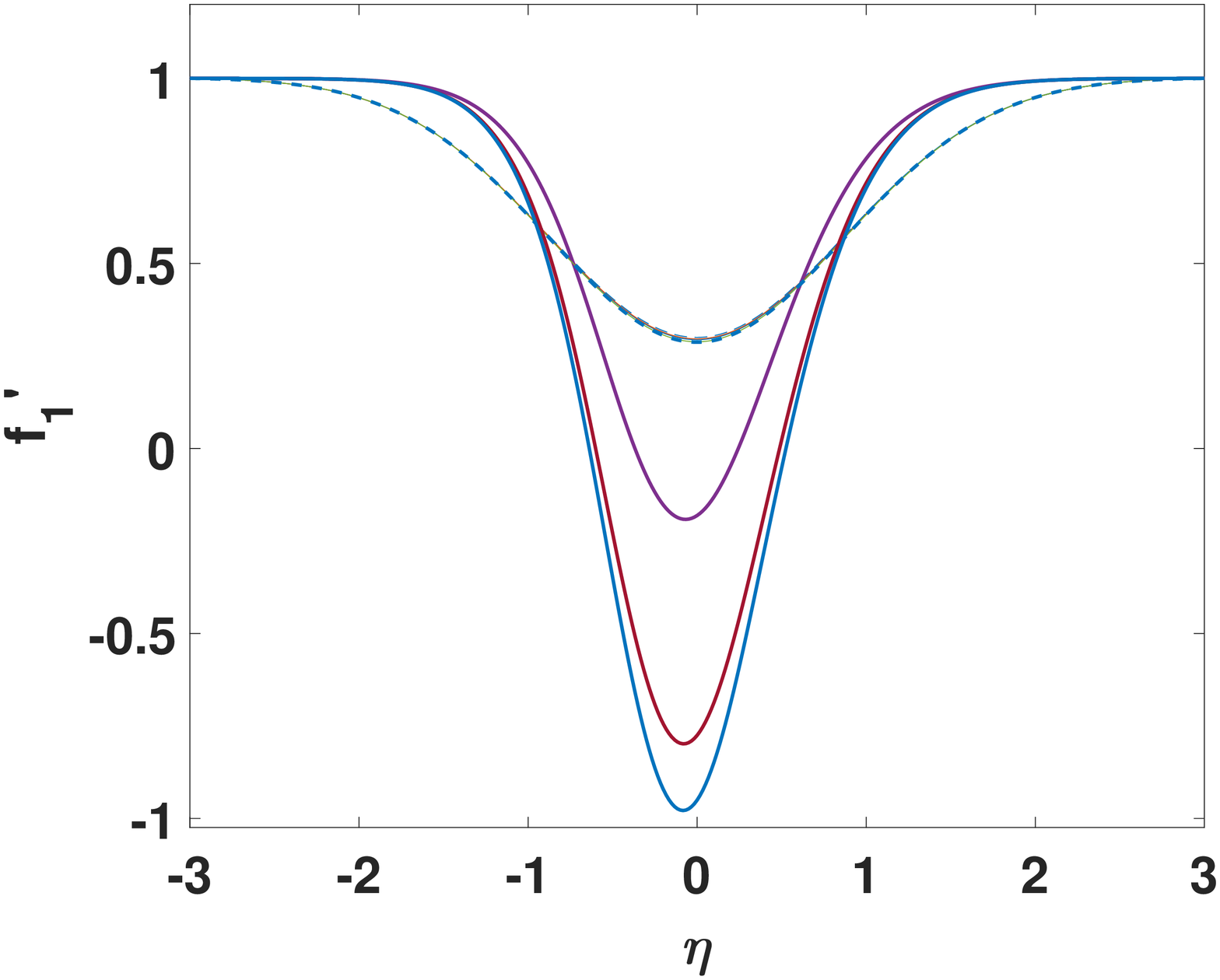}}     \\
  \vspace{0.2cm}
  \subfigure[ velocity component, $f_2' = w/(S_2z)$]{
  \includegraphics[height = 4.8cm, width=0.45\linewidth]{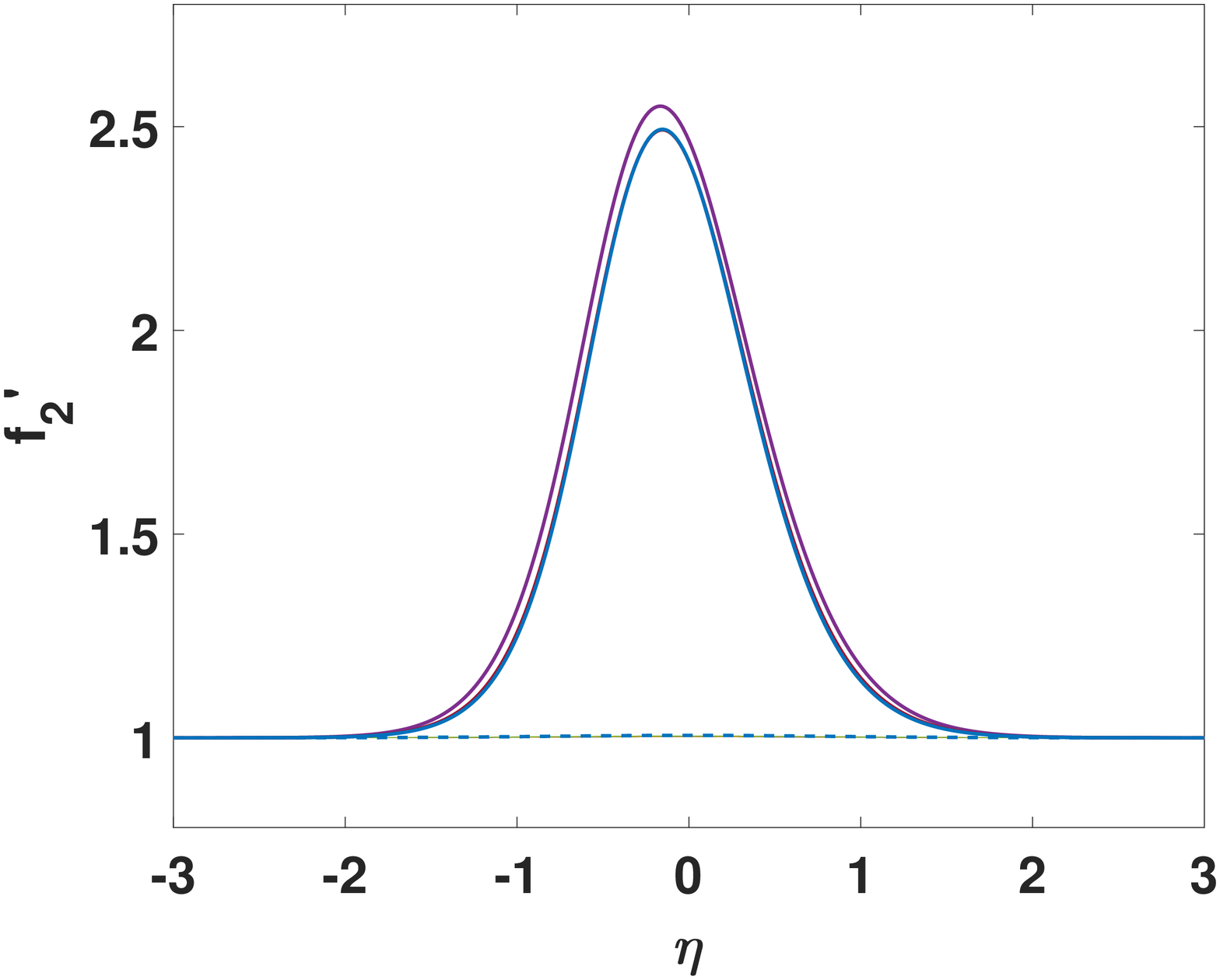}}
  \subfigure[velocity component, $u_{\chi}$]{
  \includegraphics[height = 4.8cm, width=0.45\linewidth]{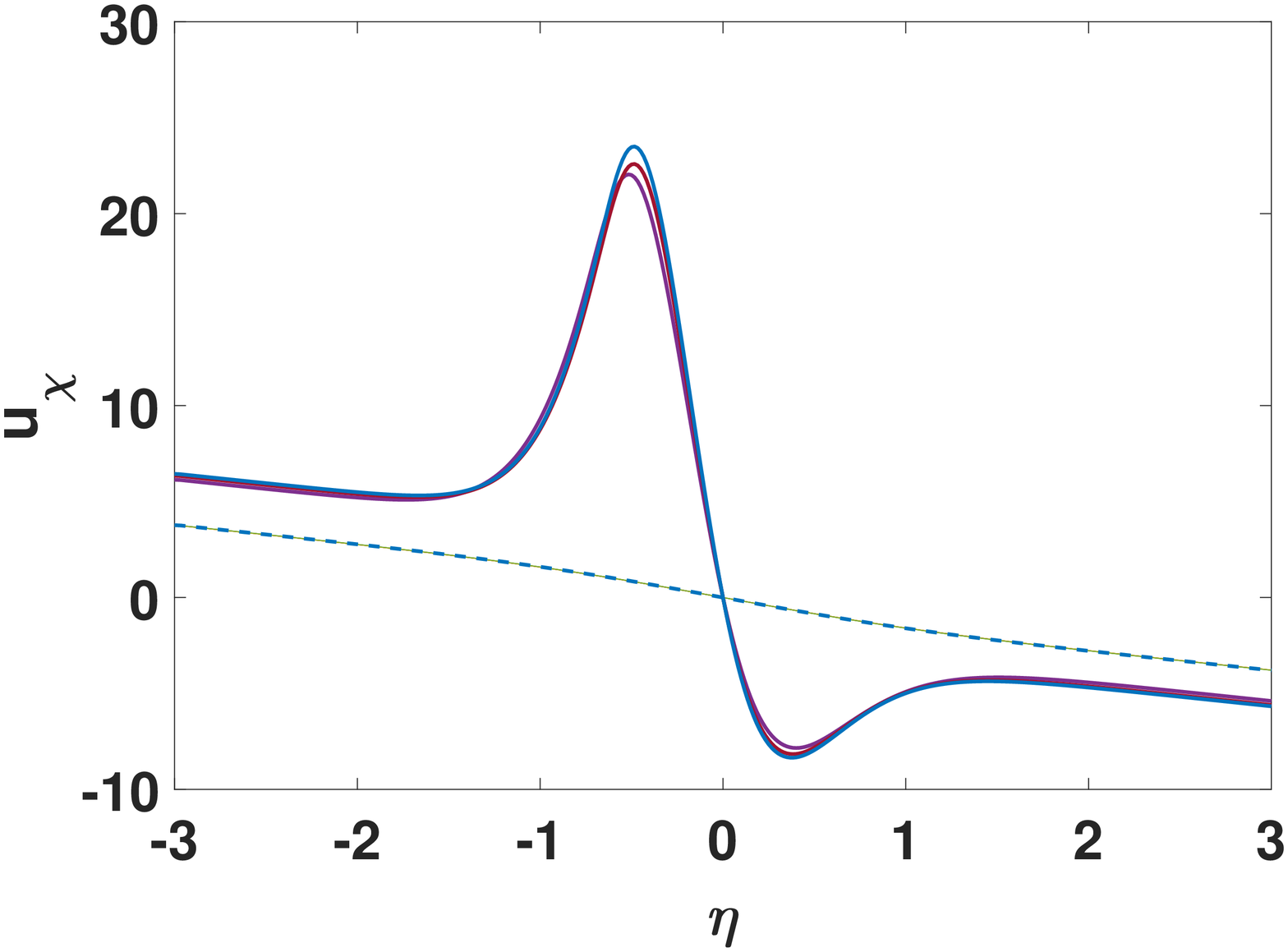}}     \\
  \vspace{0.2cm}
   \caption{Velocity behavior with varying vorticity for multibranched flame. \; $S_1 =-1.00 ;  \; S_2 = 2.00$. \;  Cases with a strong flame: $K = 0.180,\; \omega_{\kappa} =0 $, \; solid blue; \; $K=0.170, \; \omega_{\kappa} = 0.75$,\; red; \;  $K=0.160, \; \omega_{\kappa} =1.50$,\; purple. Other curves show no flame (extinction) and all fall on the dashed blue line: \;$K =0.170, \; \omega_{\kappa} =0.50$;   \;$K =0.160, \; \omega_{\kappa} =1.00$.}
  \label{MultiFlame2}
\end{figure}

Recent works have addressed the structure of multi-branched flamelets with a central diffusion flame and one or two premixed flames. The premixed flames can be fuel rich or fuel lean. They might be driven by heat transfer from the stronger diffusion flame. With three flames, the diffusion flame is centered between the two premixed flames and has the highest temperature. See \cite{Sirignano2021,Sirignano2020,Sirignano2022} which address three general configurations: a stagnation flow, counterflow imposed on a shear layer,and a rotational flamelet.  Here, we examine the multi-flame structure and behavior for the inward-swirling flamelet. Figures \ref{f1} and \ref{f2} show the computational results for relatively high values of $Da$. Later, Figures \ref{MultiFlame1} and \ref{MultiFlame2} will address the structure and behavior for values of $Da$ near the flammability limits. The effects of vorticity are especially of interest.

It is sometimes convenient to present the flamelet scalar variables as functions of a conserved scalar instead of as a function of the spatial coordinate. For our simple, one-step kinetics calculations, conserved scalars are formed by defining the Shvab Zel'dovich variables $\alpha \equiv Y_F - \nu Y_O$ and $ \beta \equiv h + \nu Y_O\tilde{Q}$ where $\tilde{Q}$ is the fuel heating value normalized by  $h^*(\infty)$.
	Then, Equation (\ref{ODEs3}) yields
	\begin{eqnarray}
		\alpha''  +Pr(S_1 f_1 + S_2f_2) \alpha'  &=& 0 \nonumber \\
		\beta''  +Pr(S_1 f_1 + S_2f_2) \beta'  &=& 0 \nonumber \\
		\label{ODEs4}
	\end{eqnarray}
	In the above relations,  the required constants are $S_1, S_2 = 1 - S_1, \rho_{-\infty},\nu, \tilde{Q}, Pr$. Solutions are coupled to the simultaneous solutions of Equations \ref{ODEs}.  A normalized conserved scalar $\Sigma$ varying between the values of $0$ and $1$ can be formed as shown by Sirignano.
\begin{eqnarray}
\Sigma \equiv   \frac{\alpha- \alpha_{-\infty}  }{\alpha_{\infty} - \alpha_{-\infty}   }
 = \frac{\beta- \beta_{-\infty}  }{\beta_{\infty} - \beta_{-\infty}   }
\label{Sigma}
\end{eqnarray}	
The use of 	plots of scalar variables as a function of $\Sigma$ are helpful in identifying the location of reaction zones in the flame structure.

In Figures \ref{MultiFlame1} and \ref{MultiFlame2}, results are shown for a configuration with a fuel-rich mixture at $\eta = \infty$ and a fuel-lean situation at $\eta = - \infty$. The fuel-rich mixture exists with $Y_F = 2/3$ and $Y_O = 1/3$
 flows inward on one side of the  swirling flame and a fuel-lean mixture with
$Y_F =1/12$ and $Y_O = 11/12$ flows inward from the other side.

For a sufficiently high value of $Da$, a strong diffusion flame and a weak, fuel-rich premixed can co-exist without a rotational flow.  See the case with $K =0.180$ and $\omega_{\kappa} =0$ in the figures. The weak premixed flame is indicated by the region of negative second derivative for enthalpy on the right-side of Figure \ref{MultiFlame1}e. (A negative second derivative also exists on the right-side of Figure \ref{MultiFlame1}a but  is more difficult to detect.) In Figure \ref{MultiFlame1}d,  the diffusion flame contributes to the region with the largest first derivative (the reaction rate which is the integrand) while the premixed flame contributes in the region where the first derivative is still positive but smaller. The existence of multiple flames is also indicated by consumption of oxygen to the right of the diffusion flame in \ref{MultiFlame1}c.  The weak premixed flame is driven by heat diffusion from the strong diffusion flame. More information about these multi-flame structures are provided by \cite{Sirignano2021}.

For the reduced values of Damk\"{o}hler number $Da$, rotation is needed to produce a flame. For example, as the same figures show with $K=0.170$, a strong flame appears with $\omega_{\kappa} = 0.75$ but there is no flame development possible with $\omega_{\kappa} \leq 0.5$. Similarly, for still smaller $Da$, even greater rotational rate is needed; for $K=0.160, \omega_{\kappa} = 1.50$ produces a strong flame,  while for $\omega_{\kappa} \leq 1.00$, no flame is sustained with any vorticity value.

Figure  \ref{MultiFlame2}b indicates that a strong flame will cause flow reversal in the $\xi$ direction due to gas expansion.

\newpage

\section{Conclusions}  \label{conclusions}

A new flamelet model is developed to treat a range of flame structures in a steady, stretched, three-dimensional vortex. Non-premixed flames, premixed flames, and multi-branched flames are addressed through a unified theory. The creation of a contrived parameter such as a progress variable is avoided. Four nondimensional  parameters are contolling: the imposed, normalized compressive strain rate $S_1$; the imposed,normalized vorticity $\omega_{\kappa}$; the Damk\"{o}hler number $Da$; and the Prandtl number $Pr$ which equals the Schmidt number $Sc$ here. The effects of these quantities are shown in the computational results. While this new theory is established for multi-step oxidation chemistry,  a simple example of one-step, propane-oxygen kinetics is considered.

The effects of the inward swirl inherent to the stretched vortex are shown to have significant effects, especially in modifying the flammability limits. Variable density is shown to have a critical role since the centrifugal force created through the vorticity has impact in that case.

For any of the flame structures, the increased vorticity can move the flammability limit to lower $Da$ values.  Higher $Da$ (for proper ambient mixtures) makes multi-branched flames more likely. Heat from the diffusion flame can drive the premixed flames. The distribution of the normal compressive strain between the directions for incoming swirling flow can affect the results.  The variation of $Pr$ within the expected range can have some  effect on the flammability limit.

For future studies, several issues are important. The computations should be extended to cases with detailed chemical kinetics, detailed transport models, and improved equations of state. Coupling of the flamelet model should be made with a RANS or LES analysis for a practical, reacting, mixing,  shear flow. Direct numerical simulations of reacting flows that give improved correlations of resolved-scale velocity gradients with the smallest-scale velocity gradients would be helpful.

\section*{Acknowledgements}

The effort was supported by AFOSR through Award  FA9550-18-1-0392 managed by Dr. Mitat Birkan.

\section*{Declaration of Interests.} The author reports no conflict of interest.
\newpage
\bibliographystyle{elsart-harv}
\bibliography {3DCompress}

\end{document}